%% file: main.tex
\documentclass[leqno,11pt]{article}

\usepackage[letterpaper]{geometry}
\usepackage{ltexpprt}

\usepackage[english]{babel}
\usepackage{mathtools}
\renewcommand{\epsilon}{\varepsilon}
\input{headers}
\usepackage{algpseudocode}

\title{Parallel Approximate Maximum Flows in Near-Linear Work and Polylogarithmic Depth}

\newcommand*\samethanks[1][\value{footnote}]{\footnotemark[#1]}
\newcommand{\ourinfo}{Research supported in part by NSF awards CCF-1934876 and CCF-2008305}

\author{Arpit Agarwal\thanks{Columbia University. \texttt{email:~arpit.agarwal@columbia.edu}}
    \and
Sanjeev Khanna\thanks{University of Pennsylvania. \texttt{email:~\{sanjeev,huanli,pprath\}@cis.upenn.edu}. \ourinfo.}
    \and
    Huan Li\samethanks
    \and
    Prathamesh Patil\samethanks
    \and
    Chen Wang\thanks{Rutgers University. \texttt{email:~wc497@cs.rutgers.edu}, supported in part by NSF CAREER Grant CCF-2047061, a gift from Google Research, and a Rutgers Research Fulcrum Award.}
    \and
    Nathan White\thanks{University of Pennsylvania.  \texttt{email:~nathanlw@cis.upenn.edu}. \ourinfo, and partially supported by the Department of Defense (DoD) through the National Defense Science and Engineering Graduate (NDSEG) Fellowship Program.}
    \and
    Peilin Zhong\thanks{Google Research. \texttt{email:~peilinz@google.com}}
}
\date{\today}

\begin{document}
\maketitle

\renewcommand{\nonl}{}
\newcommand{\algline}{\noindent\rule{\textwidth}{0.4pt}}

\begin{abstract}
    We present a parallel algorithm for the $(1-\epsilon)$-approximate maximum flow problem in capacitated, undirected graphs with $n$ vertices and $m$ edges, achieving O($\epsilon^{-3}\polylog n$) depth and $O(m \epsilon^{-3} \polylog n)$
    work in the \pram~model.
    Although near-linear time sequential algorithms for this problem have been known for almost a decade, no parallel algorithms that simultaneously achieved polylogarithmic depth and near-linear work were known. 

    At the heart of our result is a polylogarithmic depth, near-linear work recursive algorithm for computing congestion approximators. Our algorithm involves a recursive step to obtain a low-quality congestion approximator followed by a ``boosting’’ step to improve its quality which prevents a multiplicative blow-up in error. Similar to Peng [SODA’16], our boosting step builds upon the hierarchical decomposition scheme of R\"acke, Shah, and T\"aubig [SODA’14].
    A direct implementation of this approach, however,
    leads only to an algorithm with $n^{o(1)}$ depth and
    $m^{1+o(1)}$ work.
    To get around this,
    we introduce a new hierarchical decomposition scheme,
    in which we only need to solve maximum flows on subgraphs
    obtained by \textit{contracting} vertices,
    as opposed to vertex-induced subgraphs
    used in R\"acke, Shah, and T\"aubig [SODA’14].
    This in particular enables us to directly extract congestion approximators for the subgraphs from a congestion approximator for the entire graph, thereby avoiding additional recursion on those subgraphs. Along the way, we also develop a parallel flow-decomposition algorithm that is crucial to achieving polylogarithmic depth and may be of independent interest.
    
    We extend our results to related graph problems such as sparsest and balanced sparsest cuts, fair and isolating cuts, approximate Gomory-Hu trees, and hierarchical clustering. All algorithms achieve polylogarithmic depth and near-linear work.

    Finally, our \pram~results also imply the first polylogarithmic round, near-linear total space \mpc~algorithms for approximate undirected maximum flows, as well as all its aforementioned applications in the fully scalable regime where the local machine memory is $O(n^{\delta})$ for any constant $\delta>0$.

\end{abstract}

\newpage
{\hypersetup{hidelinks}\tableofcontents}
\newpage

\input{intro}

\section{Technical Overview.} \label{sec:tech_new}
\input{tech}

\section{Other Related Work.}
\label{sec:relate-work}

\input{Related}

\section{Preliminaries.}
\label{sec:prelim}
\input{pre}

\section{Recursive Computation of a Low-Quality Congestion Approximator.}
\label{sec:size_reductions}

\input{size-reductions}

\input{rst-on-trees}

\section{New Framework for Congestion Approximator Computation.}
\label{sec:new_congestion_approx}

\input{newRST14}

\section{Extraction of Congestion Approximators.}
\label{sec:extraction}
\label{sec:oracle}

\input{oracle}

\section{Parallel Flow Decomposition by Shortcutting.}
\label{sec:decomposition}

\input{shortcut}

\section{Applications.}
\label{sec:apps}
In this section, we discuss some notable applications of our parallel approximate max-flow algorithm. Namely, we show that our aforementioned result implies new or substantially improved parallel algorithms for (balanced) sparsest cuts, minimum-cost hierarchical clustering ~\cite{Dasgupta16}, fair-cuts \cite{Li+23} and approximate Gomory-Hu trees. The input instance to all these aforementioned problems are undirected, weighted graphs $G=(V,E,c)$ with positive edge-weights that are assumed to be polynomially-bounded (or alternatively, a polynomially bounded ratio of maximum to minimum edge weights).
\input{applications-sparsest-cut}

\input{applications-HC}

\input{applications-faircuts-GH}

\newcommand{\etalchar}[1]{$^{#1}$}

\appendix

\input{primitives}

\input{cut-matching}

\section{Parallel Implementation of Sherman's Algorithm.}

\input{Sherman}

\section{Computing Min Cut on Trees.}

\input{dp}

\section{Ensuring Polynomial Aspect Ratio.}

\input{poly-weight}

\end{document}

%% file: headers.tex
\usepackage{tikz}
\usetikzlibrary{arrows}
\usetikzlibrary{arrows.meta}
\usetikzlibrary{shapes}
\usetikzlibrary{backgrounds}
\usetikzlibrary{positioning}
\usetikzlibrary{decorations.markings}
\usetikzlibrary{patterns}
\usetikzlibrary{calc}
\usetikzlibrary{fit}
\usetikzlibrary{petri,decorations.markings}

\renewcommand{\top}{T}

\usepackage{fancybox}
\usepackage{enumitem}

\usepackage[noend]{algpseudocode}

\setlistdepth{9}

\newlist{myEnumerate}{enumerate}{5}
\setlist[myEnumerate,1]{label=\arabic*.}
\setlist[myEnumerate,2]{label=(\alph*)}
\setlist[myEnumerate,3]{label=\roman*.}
\setlist[myEnumerate,4]{label=\Alph*.}
\setlist[myEnumerate,5]{label=\Roman*.}

\usepackage{amsmath,amsfonts,amssymb,bm,xparse}
\usepackage{bm,xspace}
\usepackage{color,xcolor}
\usepackage[colorlinks=true,citecolor=blue]{hyperref}

\let\oldnl\nl%
\newcommand{\nonl}{\renewcommand{\nl}{\let\nl\oldnl}}%

\newtheorem{definition}{Definition}[section]

\newtheorem{remark}{Remark}[section]
\newtheorem{claim}{Claim}[section]

\newcommand{\hl}[1]{#1}

\newenvironment{fminipage}%
  {\begin{Sbox}\begin{minipage}}%
  {\end{minipage}\end{Sbox}\fbox{\TheSbox}}

\def\expec#1#2{{\mathbb{E}}_{#1}\left[ #2 \right]}

\def\setof#1{\left\{#1  \right\}}
\def\sizeof#1{\left|#1  \right|}
\def\deg{\textrm{deg}}

\def\union{\cup}
\def\intersect{\cap}
\def\Union{\bigcup}

\def\eps{\epsilon}

\def\norm#1{\left\| #1 \right\|}

\newcommand{\CS}[2]{{#1}(#2)}

\def\Acal{\mathcal{A}}

\def\Dcal{\mathcal{D}}

\def\Zcal{\mathcal{Z}}

\def\omegatil{\tilde{\omega}}

\def\Scal{\mathcal{S}}
\newcommand\Tcal{\mathcal{T}}
\newcommand\Pcal{\mathcal{P}}
\newcommand\Pcaltil{\tilde{\mathcal{P}}}

\newcommand{\Yg}{Y_{\mathrm{good}}}
\newcommand{\Yb}{Y_{\mathrm{bad}}}

\newcommand\rr{\boldsymbol{\mathit{r}}}

\newcommand\uu{\boldsymbol{\mathit{u}}}
\newcommand\vv{\boldsymbol{\mathit{v}}}

\newcommand{\card}[1]{\left\vert{#1}\right\vert}

\newcommand\MM{\mathrm{\textsf{M}}}

\newcommand\TT{\boldsymbol{\mathit{T}}}

\newcommand\lm{\mathsf{lmax}}

\DeclareMathOperator*{\argmin}{arg\,min}

\newcommand{\kh}[1]{\left(#1\right)}

\newcommand{\poly}{\mathrm{poly}}
\DeclareMathOperator{\polylog}{polylog}

\newcommand{\pram}{$\mathsf{PRAM}$\xspace}
\newcommand{\mpc}{$\mathsf{MPC}$\xspace}

\NewDocumentCommand{\ot}{g}{
    \IfNoValueTF{#1}{\tilde{O}}{\tilde{O}(#1)}
}
\DeclareMathOperator{\cost}{cost}
\DeclareMathOperator{\cut}{\texttt{cut}}
\DeclareMathOperator{\rcut}{\texttt{res\_cut}}
\newcommand{\algcomment}[1]{// \textcolor{gray}{#1}}
\newcommand{\hideref}[1]{{\hypersetup{hidelinks}\ref{#1}}}

\definecolor{RURed}{rgb}{0.95, 0.1, 0.1} %

\newcommand{\opt}{\ensuremath{\textsf{opt}}\xspace}

\usepackage{nameref}
\definecolor{ForestGreen}{rgb}{0.1333,0.5451,0.1333}
\definecolor{Red}{rgb}{0.9,0,0}
\usepackage[noabbrev,nameinlink]{cleveref}
\crefname{property}{property}{Property}
\creflabelformat{property}{(#1)#2#3}
\crefname{equation}{eq}{Eq}
\creflabelformat{equation}{(#1)#2#3}

\usepackage{tcolorbox}
\tcbuselibrary{skins,breakable}
\tcbset{enhanced jigsaw}
\newenvironment{tbox}{\begin{tcolorbox}[
    enlarge top by=5pt,
    enlarge bottom by=5pt,
     breakable,
     boxsep=0pt,
                  left=4pt,
                  right=4pt,
                  top=10pt,
                  arc=0pt,
                  boxrule=1pt,toprule=1pt,
                  colback=white
                  ]%
  }
{\end{tcolorbox}}

\newcommand{\paren}[1]{\ensuremath{\left(#1\right)}\xspace}

\newcommand{\PA}{\texttt{partition-A}}
\newcommand{\PAp}{\texttt{partition-A1}}
\newcommand{\PApp}{\texttt{partition-A2}}
\newcommand{\PB}{\texttt{partition-B}}

\newcommand{\PFD}{\texttt{flow-decomp}}
\newcommand{\RCA}{\texttt{hierarchical-decomp}}
\newcommand{\convb}{\texttt{convert-binary}}
\newcommand{\TPC}{\texttt{transform-paths-and-cycles}}
\newcommand{\polyweights}{\texttt{lower-aspect-ratio}}
\newcommand{\qual}{\log^{9}n}

\newcommand{\Fnew}{\ensuremath{F_{\mathrm{new}}}}

\usepackage{mdframed}

\newtheorem{mdresult}{Result}

\usepackage{bbm}

%% file: intro.tex
\section{Introduction.}
\label{sec:intro}

The \textit{maximum flow} problem, or equivalently, the \textit{minimum congestion flow} problem is one of the oldest and most well-studied combinatorial optimization problems that finds numerous applications across computer science and engineering. Formally, we are given a directed, capacitated flow network (graph) $G=(V,E,c)$ with $|V| = n$ vertices and $|E| = m$ edges with positive edge capacities $c\in \mathbb{R}^E$, along with a set of vertex demands $b\in \mathbb{R}^V$ specifying the desired excess flow at each vertex where $\sum_{v\in V} b_v = 0$. The objective is to find a flow $f\in \mathbb{R}^E$ that satisfies demands $b$, while minimizing the maximum congestion $|f_e|/c_e$ on any edge $e\in E$ in the network, i.e.
\begin{equation}
    \label{eq:mincongestioneq}
    \min~\|C^{-1}f\|_{\infty}~\textnormal{ s.t. } Bf = b ~\textit{(flow conservation constraints)};~f\geq 0,
\end{equation}
where $C\in \mathbb{R}^{E\times E}$ is a diagonal matrix of edge capacities with $C_{e,e} = c_e$, and $B\in \{-1,0,1\}^{V\times E}$ is the vertex-edge incidence matrix with $B_{v,e}$ being $1$ if $e=(u,v)$, $-1$ if $e=(v,u)$ and $0$ otherwise. For the special case where the edges are undirected, the formulation is identical except that the edges are oriented arbitrarily, and the non-negativity constraint on the flows is dropped; the sign of the flow on an edge specifies its direction relative to the edge orientation.

The maximum flow problem has a rich history, starting with the work of \cite{dantzig1951application} and \cite{ford1956maximal}, who first proposed algorithms with pseudo-polynomial running times of $O(mn^2 U)$ and $O(m^2U)$ respectively, for networks with maximum edge capacity $U$. Since then, substantial effort has been devoted to designing increasingly efficient algorithms for this problem. This has resulted in a sequence of exciting developments, with initial improvements coming from primarily combinatorial ideas such as shortest augmenting paths and blocking flows \cite{dinic1970algorithm,edmonds1972theoretical,karzanov1973finding,dinic1973metod,even1975network,boykov2004experimental}, push-relabeling \cite{goldberg1988new,goldberg2008partial,OrlinG21}, pseudo-flows \cite{hochbaum2008pseudoflow,chandran2009computational}, and capacity scaling \cite{ahuja1995capacity, goldberg1998beyond}, eventually culminating in the recent breakthrough $O(m^{1+o(1)})$ time result of \cite{chen2022maximum} achieved through a combination of second-order continuous optimization techniques (interior point methods) and efficient dynamic graph data-structures. Meanwhile, for the weaker objective of finding an approximately maximum flow in undirected graphs, even faster algorithms have been developed using simpler first-order continuous optimization techniques \cite{sherman2013nearly,Peng16,sherman2017area,sherman2017generalized}. In particular, \cite{Peng16} showed that in undirected graphs, a $(1-\eps)$-approximate maximum flow can be computed in just $O(m~\poly(1/\epsilon,\log n))$ time.

However, these aforementioned algorithmic results were largely developed in the sequential setting, and are not readily parallelizable. Moreover, the question of designing fast \emph{parallel} algorithms for max-flows has remained surprisingly under-explored.

In the context of parallel algorithms, several models of computation have been proposed over the years. Amongst them, the parallel random-access machine ($\mathsf{PRAM}$) is often considered as a standard, owing to its simplicity and its well-understood connections to other models of parallel computation. A generalization of the usual (sequential) $\mathsf{RAM}$ model, the \pram~is a synchronous, shared-memory, multi-processor model. Within \pram, there are several variants depending on how concurrent operations in the shared memory are handled: (in decreasing order of restrictiveness) exclusive-read-exclusive-write, concurrent-read-exclusive-write, and concurrent-read-concurrent-write. However, due to the low-level nature of the \pram, even the simplest algorithms designed for it involve tedious implementation details. In order to abstract away these specifics, we usually consider the equivalent work-depth paradigm, which measures the performance of a parallel algorithm using two parameters -- \emph{total work}, which is the total running time needed given only one processor, and \emph{depth}, which is the total parallel time given a maximal number of processors. Moreover, in this framework, the different variants of \pram~are equivalent up to polylogarithmic factors in work and depth, and therefore, can also be abstracted.

For the question of designing parallel algorithms for max flows, the literature is sparse. For finding exact max flows, early results of \cite{shiloach1982n2log,ramachandran1990parallel} achieved $\tilde{O}(n^2)$ depth  and $\tilde{O}(mn)$ work. More recently, ~\cite{peretz2022fast} improved the depth to $\tilde{O}(n)$ while retaining the same asymptotic work. For the weaker objective of finding an approximate max flow, the combined results of \cite{madry2016computing} and \cite{peng2014efficient} imply an algorithm with $\tilde{O}(\sqrt{m})$ depth and $\tilde{O}(m^{1.5})$ work at the cost of a $O(1/\poly(n))$ additive error in the flow value. A much earlier work of \cite{serna1991tight} also implies a $O(\poly(\log n,\log (1/\epsilon)))$ depth algorithm for finding a $(1-\epsilon)$-approximate max flow via a reduction to finding maximum matchings in bipartite graphs. However, this comes at the expense of an unspecified polynomial blow-up in work, a common issue with many of the early results for parallel algorithms. The question of whether it is possible to \textit{simultaneously} have a polylogarithmic depth and near-linear work approximate max-flow algorithm remained open. In this paper, we answer this in the affirmative for undirected graphs. Namely, we show the following main result.
{\begin{mdframed}[backgroundcolor=lightgray!40,topline=false,rightline=false,leftline=false,bottomline=false,innertopmargin=2pt]

\begin{theorem}
\label{rst:max-flow}
There is a randomized \pram~algorithm that given an undirected capacitated graph $G=(V,E,c)$, $s,t\in V$, and precision $\epsilon>0$, computes both a $(1-\eps)$-approximate $s$-$t$ maximum flow and a $(1+\eps)$-approximate $s$-$t$ minimum cut with high probability in $O\left(\epsilon^{-3}\polylog{n}\right)$ depth and $O\left(m\epsilon^{-3}\polylog{n}\right)$ total work.
\end{theorem}

\end{mdframed}}
A more recent model of parallel computation is the \emph{massively parallel computation} (\mpc) model, which is a common abstraction of many MapReduce-style computational frameworks (see i.e.~\cite{bks-mpc,AndoniNOY14,GoodrichSZ11}).
In this model, the input data is partitioned across multiple machines that are connected together through a communication network.
The computation proceeds in synchronous rounds where in each round, the machines can perform unlimited local computation on their local memory, but cannot communicate with other machines.
Between rounds, machines can communicate, so long as the total size of messages sent and received by any machine does not exceed the size of its local memory.
The performance of an \mpc~algorithm is measured by the number of rounds needed to complete the computation, the size of the local machine memory, as well as the total memory used across all machines.  
A simulation result of \cite{KarloffSV10,GoodrichSZ11} shows that any \pram~algorithm with $D$ depth and $W$ work can be simulated by an \mpc~algorithm with $O(D)$ rounds and $O(W)$ total memory even in the most stringent \emph{fully-scalable} regime\footnote{The dependence of the  the local memory parameter (the constant $\delta>0$) on the number of rounds and total memory is a multiplicative $O(1/\delta)$, and is usually omitted.} where the local memory size is $O(n^{\delta})$ for any constant $\delta > 0$. Using this simulation result, we obtain the following corollary of our main result.

{\begin{mdframed}[backgroundcolor=lightgray!40,topline=false,rightline=false,leftline=false,bottomline=false,innertopmargin=2pt]
\begin{corollary}[of Theorem \ref{rst:max-flow}]
    \label{rst:max-flow-mpc}
    There is a randomized \mpc~algorithm that given an undirected capacitated graph $G=(V,E,c)$, $s,t\in V$ and precision $\epsilon>0$, computes both a $(1-\eps)$-approximate $s$-$t$ maximum flow and a $(1+\eps)$-approximate $s$-$t$ minimum cut with high probability in $O\left(\epsilon^{-3}\polylog{n}\right)$ rounds, $O\left(m\epsilon^{-3}\polylog{n}\right)$ total memory and $O\left(n^{\delta}\right)$ local memory, for any constant $\delta > 0$.
\end{corollary}

\end{mdframed}}

Note that by standard reduction, our Results \ref{rst:max-flow} and \ref{rst:max-flow-mpc} extend to general vertex demands $b\in \mathbb{R}^{V}$ in the same form of Eq \ref{eq:mincongestioneq}.

\subsection{Applications.}
Our result for approximate max flows has  broad implications to other related graph problems, where it implies either new or substantially improved parallel algorithms (both \pram~and \mpc) for them. 

\medskip

\noindent
\textbf{Sparsest cut and balanced sparsest cut:}
The \textit{sparsest cut} problem is an important
subroutine for divide-and-conquer based approaches for several
graph problems and has many applications including
image segmentation, VLSI design, clustering and expander decomposition.
In this problem, given a weighted undirected graph $G=(V,E,c)$, the objective is to find a cut $(S, \bar{S})$ with minimum sparsity $\phi(S)$ which is defined as $\phi(S) := c(E(S, \bar{S}))/\min\{|S|,|\bar{S}|\}$, where $c(E(S,\bar{S}))$ is the total weight of the edges going across the cut.
The \textit{balanced sparsest cut} problem is a variant of this problem where there is an additional requirement
that the cut $(S, \bar{S})$ must be $\beta$-balanced, i.e.~$\min \{|S|, |\bar{S}|\} \geq \beta n$ for some given parameter $\beta \in (0,1/2)$.
Balanced sparse cuts are useful in applications where one wants the
divide-and-conquer tree to have low-depth.

The sparsest cut problem is NP-hard
and the best known approximation factor of $O(\sqrt{\log n})$ is achieved by an SDP-based algorithm due to Arora, Rao and Vazirani \cite{AroraRV09}.
However, this algorithm is highly sequential, and computationally expensive.
The most efficient algorithms for sparsest cut problem are
based on the cut-matching game framework of Khandekar, Rao and Vazirani \cite{KhandekarRV06}, which effectively reduces the sparsest cut problem to a poly-logarithmic number of
single commodity max-flow computations.
In this framework, a cut-player and a matching-player play an alternating game; in each round, the former produces a bisection $(S, \bar{S})$ of vertices, and the latter produces a perfect matching between $S$ and $\bar{S}$ that can be embedded in the underlying graph.
The game ends when either the cut player produces a bisection for which the matching player cannot find a perfect matching  (i.e. a sparse cut has been found), or the union of the perfect matchings produced thus far form an expander (i.e. an expander can be embedded in the underlying graph, certifying its expansion).
\cite{KhandekarRV06} showed that there is a cut strategy with runtime $T_{\text{cut}} = \tilde{O}(n)$ such that this game terminates in $\alpha = O(\log^2n)$ rounds, regardless of the matching player's strategy.
Furthermore, the matching player's strategy can be implemented using a max-flow
computation with the sources and sinks being the two sides of the bipartition.
In combination, this framework produces an $O(\alpha)$-approximation to sparsest cut with a
runtime of $O( \alpha \cdot(T_{\text{cut}} + T_{\text{flow}}))$, where
$\alpha$ is an upper bound on the number of rounds of this game, $T_{\text{cut}}$ is the time to implement cut player's strategy and $T_{\text{flow}}$ is the time to compute a single commodity max-flow.
This framework also generalizes to the problem of $\beta$-balanced
sparsest cut for any constant $\beta$
with the same approximation factor and  running time.

Nanongkai and Saranurak \cite{ns17} further showed that the matching player's
strategy can be implemented using
approximate max-flow computations rather than
exact max-flows.
Building on \cite{ns17}, we show that
our parallel algorithm for max-flows (Result~\ref{rst:max-flow}) can be
used to implement the matching player's strategy, yielding a $O(\polylog n)$ depth and $\tilde{O}(m)$ work
algorithm for the matching player.
Furthermore, we show that the cut player's strategy can also be implemented in
$O(\polylog n)$ depth and $\tilde{O}(m)$ work using our new \emph{parallel flow decomposition} algorithm (see \Cref{sec:ovflowdecomp} for a discussion).
In combination, this gives a $O(\polylog n)$ depth and $\tilde{O}(m)$ work algorithm for sparsest cuts that has an approximation factor of $O(\log^3 n)$.
These results also imply a bicriteria approximation algorithm for
$\beta$-balanced sparsest cut with the same depth and work.
Specifically, our algorithm finds a $(\beta/\log^2n)$-balanced cut whose sparsity is at most $O(\log^3 n)$ times the
sparsity of an optimal $\beta$-balanced cut.

\medskip

\noindent
\textbf{Minimum cost hierarchical clustering:}
Hierarchical clustering is a fundamental problem
in data analysis where the goal is
to recursively partition data into clusters which results in a rooted tree structure. This problem has wide-ranging applications across different domains including phylogenetics, social network analysis, and information retrieval.
While the study of hierarchical clustering
goes back several decades,
Dasgupta \cite{Dasgupta16} initiated its study from an optimization viewpoint. Specifically, he proposed a minimization objective for hierarchical clustering on similarity graphs that measures the cost of a hierarchy as the sum of costs of its internal splits, which in turn is measured as the total edge-weight across the split scaled by the size of the cluster at that split.
\cite{Dasgupta16} and subsequent work by \cite{Cohen-AddadKMM18, CharikarC17} showed
that the hierarchical clustering produced
by recursively splitting the graph using a $\alpha$-approximate sparsest cut subroutine
results in a hierarchy that is an $O(\alpha)$-approximation
for this minimization objective. Furthermore, using
an $\alpha$-approximate $\beta$-balanced sparsest cut
subroutine results in the same approximation with an additional property that the depth of the tree is $O(\log n)$ for $\beta = \Omega(1)$.
\cite{CharikarC17} also showed that a $(\alpha,\beta')$-bicriteria approximation oracle for $\beta$-balanced min-cut is sufficient to achieve a $O(\alpha/\beta')$-approximation for minimum cost hierarchical clustering.
As a consequence, using our parallel balanced min-cut algorithm as a subroutine, we get the first parallel algorithm that computes a tree that is a $O(\log^5 n)$-approximation to Dasgupta's objective with $\tilde{O}(m)$ work and ${O}(\polylog n)$ depth.

\medskip

\noindent
\textbf{Fair cuts and approximate Gomory-Hu trees:}
Li \textit{et al}.~\cite{Li+23} recently introduced the notion of fair
cuts for undirected graphs which are a ``robust'' generalization of  approximate min-cuts.
Specifically, for $\alpha\geq 1$, a $s$-$t$-cut is \textit{$\alpha$-fair} if there exists an $s$-$t$
flow that uses at least $1/\alpha$-fraction of the capacity
of every edge in the cut.
\cite{Li+23} showed that a near-linear time oracle 
for computing a fair cut is useful 
for obtaining near-linear time algorithms for several 
applications including computation of (approximate) all-pairs maxflow values (using 
approximate Gomory-Hu trees).
They also showed that, given an unweighted graph,  a fair cut can be 
computed in parallel using $m^{1+o(1)}$ work 
and $n^{o(1)}$ depth, 
implying a similar result for 
approximate Gomory-Hu trees.
Our results improve upon their results by giving parallel algorithms 
for fair cuts and approximate Gomory-Hu trees that have $O(m\,\polylog n)$ work 
and $O(\polylog n)$ depth.
The key technical tool that results in this improvement
is our $O(\polylog n)$ depth construction 
of a $O(\polylog n)$-\emph{congestion approximator}
(see \Cref{sec:overview_congestion_approx} for a discussion).

%% file: tech.tex
\interfootnotelinepenalty=10000

The starting point of our discussion is Sherman's framework~\cite{sherman2013nearly,sherman2017generalized},
which was initially proposed for computing fast approximate undirected max-flows in the \textit{sequential}
setting.
Given the optimization problem defined in Equation~\ref{eq:mincongestioneq},
let $\opt_G(b):=\min_{f:Bf=b} \|C^{-1}f\|_{\infty}$ be the minimum congestion
of any feasible flow routing given demands $b$ in the graph $G$.
Sherman showed that if we have a matrix $\Phi$ such that for \emph{any} demand vector $b$, $\|\Phi b\|_{\infty}$ is always an $\alpha$-approximation to $\opt(b)$, i.e., $\|\Phi b\|_{\infty}\leq \opt(b)\leq \alpha\cdot \|\Phi b\|_{\infty}$, then there is an iterative algorithm to compute a $(1+\varepsilon)$-approximate minimum congestion flow. %
The matrix $\Phi$ is called a \emph{$\alpha$-congestion approximator matrix},
and the algorithm requires $\poly(\alpha,\epsilon^{-1},\log n)$ iterations\footnote{To be precise, Sherman's algorithm upon termination routes a flow that \emph{almost} satisfies the desired demand $b$, and the negligible residual demand is such that it can be routed in the flow network with $1/\poly(m)$ congestion. We get a feasible flow by trivially routing this residual demand along a maximum spanning tree, which is known to admit an efficient parallel construction with
$\ot(m)$ work and $\polylog(n)$ depth~\cite{pram-mst}.} where each iteration requires in total $O(m)$-time algorithmic operations plus computing the matrix-vector products $\Phi x$ and $\Phi^{\top}y$ for some arbitrary vectors $x$ and $y$.

A line of work in the parallel transshipment ($\ell_1$ flow) literature ~\cite{andoni2020parallel,li2020faster,rozhovn2022undirected,zuzic2022universally} observed that Sherman's framework can be adapted to the \textit{parallel} computing regime.
In particular, they showed the following: given an $\alpha$-congestion approximator matrix $\Phi $, there is a \pram~algorithm which outputs a   $(1+\varepsilon)$-approximate minimum congestion flow. %
The depth of the algorithm is $\poly(\alpha,\epsilon^{-1},\log n)\cdot (\mathrm{depth}(\Phi x) + \mathrm{depth}(\Phi ^\top y))$, and the work is $\poly(\alpha,\epsilon^{-1},\log n)\cdot (\mathrm{work}(\Phi x) + \mathrm{work}(\Phi ^\top y) + O(m))$, where $\text{depth}(\cdot),\text{work}(\cdot)$ denote the depth and work required to compute the corresponding sub-problem respectively, and $\Phi x,\Phi ^{\top}y$ denote the sub-problems of multiplication between $\Phi $ and an arbitrary vector and the multiplication between $\Phi ^{\top}$ and an arbitrary vector respectively.
Therefore, the problem effectively reduces to
efficiently computing a good congestion approximator matrix $\Phi $
that also allows efficient computation of $\Phi x$ and $\Phi ^T y$.

R\"acke, Shah, and T\"aubig \cite{RackeST14} gave an efficient algorithm for building a \textit{tree-structured} congestion approximator. Specifically, they showed that given any graph $G=(V,E,c)$, one can construct an $O(\log n)$-depth tree $R=(V_R,E_R,c_R)$ supported on $V_R \supseteq V$
such that for any demand vector $b$, $\opt_R(b)\leq \opt_G(b)\leq O(\log^4 n)\cdot\opt_R(b)$.
A useful property of the tree $R$ is that the flow that goes from $u$ to the parent of $u$ is equal to the total demand within the subtree of $u$.
This implies that we can write $\opt_R(b)=\|\Phi b\|_{\infty}$ for a matrix $\Phi $, where each row of $\Phi $ corresponds to a node in $R$, such that the $u$-th entry of $\Phi b$ equals
the congestion of the tree edge connecting $u$ to its parent:
$$
(\Phi b)_u = \frac{\sum_{v\text{ in the subtree of }u} b_v}{c_R(u,\text{parent of }u)}.
$$
Furthermore, computing $\Phi x$ is equivalent to computing the sum of rescaled node weights in each subtree, and computing $\Phi ^{\top}y$ is equivalent to computing the sum of rescaled edge weights of the path from each node to the root.
Thus, both computations of $\Phi x$ and $\Phi ^{\top}y$ can be done in $\polylog(n)$ depth and $n\cdot\polylog(n)$ work using standard dynamic programming ideas.
Hence, our goal becomes to compute such a congestion approximator tree (or hereafter simply \textit{congestion approximator}) $R$.

The algorithm of \cite{RackeST14} constructs such a tree by performing a hierarchical decomposition of the graph.
At a high level, starting with the vertex set $V$, they recursively apply a subroutine to partition a given cluster $S$ of vertices
into smaller sub-clusters such that
the \textit{boundary edges} leaving each sub-cluster
are {\em well-linked},
in the sense that one can route a product multicommodity flow between them with low congestion.
However, this partitioning routine itself requires computing
$(1+\varepsilon)$-approximate minimum congestion flows on the induced subgraph $G[S]$,
which is the exact same problem (though on a subgraph)
we set out to solve by constructing a congestion approximator!

In the \textit{sequential setting}, this chicken-and-egg situation was resolved
by
Peng \cite{Peng16} using a clever recursive construction,
which we summarize below:
\begin{enumerate}
\item Given an input graph $G=(V,E,c)$, the goal is to output an $O(\log^4 n)$-congestion approximator tree $R=(V_R,E_R,c_R)$.
\item Simulate the algorithm of~\cite{RackeST14} on $G$.
\textbf{Every time} a $(1+\varepsilon)$-approximate minimum congestion flow computation is required on some \textbf{vertex-induced subgraph} $G[S]$ of $G$, do the following:
\begin{enumerate}
\item Compute a sparsifier $H$ of $G[S]$ such that for any demand $b$,
$\opt_{H}(b)\leq \opt_{G[S]}(b)\leq \polylog(n)\cdot\opt_{H}(b).$
In addition, $H$ can be decomposed into
a core graph $C$ and a forest $F$ such that
(i) $C$ only has $|S|/\polylog(n)$ nodes and edges, and
(ii) each connected component of $F$ has exactly one vertex in $C$.
\item \textbf{Recursively} compute an $O(\log^4 n)$-congestion approximator tree $R_{C}$ for the core graph $C$.
\item Let $R_{G[S]}=R_{C}\cup F$.
It is easy to show that by composing,
$R_{G[S]}$ is a $\polylog(n)$-congestion approximator tree for $G[S]$.
\item Compute a $(1+\varepsilon)$-approximate minimum congestion flow on $G[S]$ by
plugging $R_{G[S]}$ into Sherman's algorithm.
\end{enumerate}
\item Output an $O(\log^4 n)$-congestion approximator tree $R$ for $G$
obtained from simulating the algorithm of \cite{RackeST14}.
\end{enumerate}
Note that since $C$ has significantly smaller size than $G[S]$,
the computation of a congestion approximator tree for $C$ is much faster than for $G[S]$.
Leveraging this observation, and using a large enough $\polylog(n)$ factor in the size reduction of the sparsifier,
\cite{Peng16} showed that the overall running time of their algorithm can be bounded by $m\cdot\poly(\varepsilon^{-1},\log(n))$.

However, there are two major challenges to convert Peng's algorithm into a small depth \pram~algorithm.
The first challenge is that Peng's
algorithm has multiple recursive calls and the input to each recursive call depends on the output of previous recursive calls.
These dependent recursive calls result in long dependent computation paths and thus the algorithm has a large (i.e.~super-logarithmic) depth.
The second challenge comes from how the approximate maximum flows are used in~\cite{RackeST14}. In particular, given an approximate maximum $s$-$t$ flow,
\cite{RackeST14} crucially requires a decomposition of the flow into $s$-$t$ flow paths.

Our paper's {\em main technical contribution} is to address these two challenges.
To address the second challenge, we propose a novel efficient parallel flow decomposition routine
which we discuss in more detail later
in Section~\ref{sec:ovflowdecomp} of this overview.
With this routine, we are already able to implement the algorithm of~\cite{RackeST14}
with $m^{1+o(1)}$ work and $n^{o(1)}$ depth.
However, getting the depth down to $\polylog(n)$ leaves us with the considerably
more difficult task of addressing the first challenge.
To this end, we devise a variant of the algorithm in~\cite{RackeST14} for
constructing congestion approximators. While our new algorithm also computes a
hierarchical decomposition of the graph, it allows us to apply our partitioning routine
by running maximum flows on subgraphs obtained by \textit{contracting} vertices, as
opposed to recursing on vertex-induced subgraphs as done in \cite{RackeST14}. This in particular
allows us to \textit{extract} congestion approximators for the subgraphs from a given
congestion approximator of the entire graph, thus avoiding any additional recursive calls on these subgraphs.
As a result, we only have \textit{one} recursive call, whose goal is to compute a congestion
approximator of the core graph $C$ of the sparsifier, enabling us to obtain a depth
of $\polylog(n)$.

We adopt a similar recursive idea to that of Peng~\cite{Peng16}, with the key difference being
our approach involves only one recursive call.
Specifically,
our construction consists of a \textit{recursive step} to obtain a low-quality
(though still $\polylog(n)$-approximate) congestion approximator, and a \textit{boosting step}
to improve it to an $O(\qual)$-congestion approximator,
so as to avoid an accumulation of error over successive recursive calls.
We summarize our new construction as follows:
\begin{enumerate}
\item Given input $G=(V,E,c)$, the goal is to output an $O(\qual)$-congestion approximator tree $R=(V_R,E_R,c_R)$.
\item Compute a sparsifier $H$ of $G$ such that for any demand $b$,
$\opt_{H}(b)\leq \opt_{G}(b)\leq \polylog(n)\cdot\opt_{H}(b).$
In addition, $H$ can be decomposed into
a core graph $C$ and a forest $F$ such that
(i) $C$ only has $|S|/\polylog(n)$ nodes and edges, and
(ii) each connected component of $F$ has exactly one vertex in $C$. \label{line:j-tree}
\item (Recursive step) \textbf{Only one recursive call:} Recursively compute an $O(\qual)$-congestion approximator tree $R_{C}$ for the core graph $C$.
\item Let $R_G=R_{C}\cup F$.
Then $R_G$ is a $\polylog(n)$-congestion approximator tree for $G$.
\item (Boosting step) Simulate our variant of the algorithm in~\cite{RackeST14} on $G$ to find an $O(\qual)$-congestion approximator tree $R$,
where every time when $(1+\varepsilon)$-approximate minimum congestion flow computation is required on some
subgraph $G(S)$ obtained from $G$ by contracting
vertices $V\setminus S$ into a single supernode,
do the following: \label{line:new-rst-construction}
\begin{enumerate}
\item Extract from $R_G$ a $\polylog(n)$-congestion approximator tree $R_S$ for $G(S)$.
\item Compute a $(1+\varepsilon)$-approximate minimum congestion flow required on $G(S)$ by plugging $R_S$ into Sherman's algorithm.
\end{enumerate}
\end{enumerate}
Line~\ref{line:j-tree} involves constructing ultrasparsifiers \cite{SpielmanT14,KoutisMP14}, and subsequently transforming them into $j$-trees~\cite{madry10}.
Though not particularly technically challenging, we are the first to give efficient constructions of these objects in the \pram~model.
We refer the reader to Section~\ref{sec:size-reducution} for their parallel implementations.
Line~\ref{line:new-rst-construction},
namely our variant of the algorithm in~\cite{RackeST14},
is substantially more involved,
as we must open the blackbox of~\cite{RackeST14} entirely.
In the following section, we explain in more detail, our new construction of congestion approximators that is used in our boosting step.

\subsection{Overview of New Congestion Approximator Construction.}
\label{sec:overview_congestion_approx}

As mentioned above, the difficulty in parallelizing the construction of
congestion approximators in \cite{RackeST14} arises from the fact that we need to
solve (approximate) maximum flows on vertex-induced subgraphs,
which in turn requires congestion approximators for those subgraphs.
However, recursing on these subgraphs necessarily blows up the depth of our algorithm
to super-logarithmic.
There are two possibilities of resolving this:
(i) use a more aggressive size reduction than $\polylog(n)$ when computing sparsifiers, or (ii) reduce the number of dependent recursive calls.
Here, (i) can be immediately ruled out, since a more aggressive size reduction
implies a worse approximation for the sparsifier, which in turn results in a larger iteration count (and hence, depth) of Sherman's algorithm to compute
maximum flows.
This leaves us with (ii) as the only option.
Notice that having even just {\em two} dependent recursive calls would result in a depth
of $2^{O\left(\frac{\log n}{\log \log n}\right)} = n^{o(1)}$,
which makes (ii) even more imperative.

This raises the following question: can we reduce the number of dependent
recursive calls by \textit{reusing} congestion approximators?
While it indeed seems plausible to combine the congestion approximators for the subgraphs into one for the entire graph,
this does not suit the construction of \cite{RackeST14} -
in particular, the hierarchical decomposition tree there is computed
in a \textit{top-down manner}, and thus
we do not know on what subgraphs we want to run maximum flows until we are done with the partitioning at higher levels.
This suggests that we should instead consider \textit{extracting} congestion
approximators for the subgraphs \textit{from} that of the entire graph.
But unfortunately,
this is in general impossible due to the information loss due to congestion approximation (which in turn is a form of sparsification that preserves both cuts and flows).

Our solution here is to open the blackbox of \cite{RackeST14} entirely
and propose a new framework for computing a hierarchical decomposition. Our new framework
allows us to partition the graph by running (approximate) maximum flows
on \textit{contracted subgraphs}, each of which is obtained by contracting
a subset of vertices into a single \textit{supernode}, as opposed to vertex-induced subgraphs as in \cite{RackeST14}.
We show that for the contracted subgraphs we encounter in our construction,
we are actually able to extract congestion approximators for them
from a given congestion approximator of the entire graph by contracting its vertices\footnote{This is a vast simplification of our extraction process. In particular, since we need our congestion approximator to be a tree in order to run Sherman's algorithm, we can only do certain ``partial'' contractions that preserves the tree structure, which significantly complicates both our algorithm and its analysis. We refer the reader to Section~\ref{sec:oracle} for more detail.}, thereby avoiding \emph{any} additional recursive calls.
Crucially, the congestion approximators obtained for the contracted subgraphs
have size proportional to the size of the corresponding subgraphs which prevents any substantial blow up in the total work of our algorithm.

We next highlight some additional ideas needed to make our new construction work. Since we run maximum flows on contracted subgraphs,
the routing we get could use paths in the contracted subgraphs, which
are not necessary valid in the entire graph. Therefore we need a ``fixing'' step in our routing - namely, whenever we obtain a routing in a contracted subgraph, we have to then fix it so that it becomes a valid routing (with low congestion) in the entire graph, while routing the exact same demands\footnote{In a contracted subgraph, we only ever route demands that are supported entirely on the ``uncontracted'' vertices. Thus there is no ambiguity in saying ``same demands'' in contracted subgraphs vs. in the entire graph.}.
This fixing step is possible because our construction guarantees that
the edges incident on a contracted supernode (which we call \textit{boundary edges}) are always well-linked, in the sense that one could route in the entire graph a product multicommodity flow between them with low ($\polylog(n)$) congestion. Thus, we can convert all the flow paths passing through a supernode in a contracted subgraph into flow paths in the entire graph using the well-linkedness of these edges while blowing up the congestion by at most
$\polylog(n)$.

As stated thus far, it may seem that too many contracted subgraphs might end up using
a same edge in the entire graph to route and thus overcongesting the edge by too much. We address this issue by fixing the routing level-by-level in our hierarchical decomposition in a bottom-up manner; at each level, we only partially fix the routing in a contracted subgraph, in the sense that it becomes valid (i.e.~routes the exact same demand) in the slightly larger contracted subgraph obtained one level above.
We aim to maintain the invariant that in each contracted subgraph, the edges incident on the contracted supernode (i.e.~boundary edges) are never congested by a factor larger than $\polylog(n)$.
This invariant guarantees that in the (partial) fixing step, the edges inside the contracted supernodes do not get congested by a factor larger than $\polylog(n)$ either: any routing that utilizes these edges
has to go through the boundary edges in the first place, and the latter are guaranteed to have low congestion.

However, notice that naively performing this (partial) fixing operation level-by-level does not get us the desired invariant. This is because each time we use well-linkedness of the boundary edges to fix the routing, we get a multiplicative $\polylog(n)$ blowup in congestion, which accumulates across all $\Theta(\log n)$ levels to a very large value. To address this issue, we use a similar trick as in \cite{RackeST14} where, when solving maximum flows in each contracted subgraph, we lower the weights of the boundary edges by a large enough $\polylog(n)$ factor. Therefore, the actual congestion we get on these edges is in fact $\polylog(n)$ times smaller than what we get in our maximum flow routing, canceling out the multiplicative blowup caused by the fixing step. Moreover,
a $\polylog(n)$-congestion approximator for the original contracted subgraph remains a $\polylog(n)$-congestion approximator for the reweighted contracted subgraph, albeit with a worse $\polylog(n)$ approximation factor. Therefore we can still use Sherman's algorithm to solve maximum flows on the reweighted subgraph in $\polylog(n)$ depth.
Leveraging all these ideas, we get a parallel algorithm for computing an $O(\qual)$-congestion approximator.

\subsection{Overview of Our Parallel Flow Decomposition Routine.}\label{sec:ovflowdecomp}

We now summarize the ideas behind
our parallel flow decomposition routine that addresses our second
challenge.
Here we are given an $s$-$t$ flow represented by a flow network,
with each edge carrying some non-negative amount of flow,
and our goal is to decompose the flow into $s$-$t$ flow paths.
We will decompose the flow in an iterative manner where in each iteration, we \emph{shortcut} flow paths of length two in the flow network by replacing them with a single edge having the same flow value. We then repeat until (almost) all remaining edges directly go
from $s$ to $t$ (i.e. all $s$-$t$ paths have length $1$).

In order to achieve small depth, in every iteration, we need to find a large
(in capacity)
collection of edge-disjoint length two flow paths so that they can be shortcut in parallel without interfering with each other. We show that such a collection exists, and moreover, can be found efficiently by formulating the problem as $b$-matching instances. Specifically, for every (directed) edge $e=(u,v)$ that does not directly connect $s$ and $t$, we independently and randomly assign $e$ either as an ``outgoing" edge to its head vertex $u$, or as an ``incoming'' edge to its tail vertex $v$. Now taking a local view at any vertex $w$ in the flow network, the task of shortcutting reduces to a $b$-matching problem in a complete bipartite graph, where the two sides are vertices corresponding to incoming edges into $w$ and outgoing edges from $w$, respectively, with demands equal to their flow values.
This way, we never have to worry about an edge being matched more than once as it gets assigned
to exactly one of its two endpoints.

We show that in expectation over the random incoming/outgoing assignments of the edges, the total weight of the maximum $b$-matchings across all vertices must be at least a constant fraction of the total remaining flow that does not directly go from $s$ to $t$.
Moreover, algorithmically, in each $b$-matching instance,
we can utilize a natural greedy strategy to gather at least a constant fraction of the maximum matched value, which can further be implemented in parallel across all vertices.
This guarantees that in each iteration, the total $\ell_1$-norm of the
remaining flow shrinks by a constant factor,
and thus after $O(\log n)$ iterations we have decomposed
a $(1 - 1/\poly(n))$-fraction of the total flow into $s$-$t$ paths, which suffices
for our purpose since we only want approximate maximum flows.
At the end, we output the entire shortcutting history represented by
a DAG data structure from which we can recover our desired information.

\subsection{Algorithms.}\label{sec:ovalgs}
Lastly, we present all the key algorithms outlines in our overview. Our main algorithm is a recursive \pram~algorithm for computing a $O(\qual)$-congestion approximator in $O(\polylog n)$ depth and $\ot{m}$ total work.
We present our algorithms for the case where the ratio between the largest and smallest capacities in the graph is bounded by $\poly(n)$, but by the reduction shown in Section \ref{sec:poly-weights}, this is without loss of generality.

\begin{algorithm}{$\texttt{congestion-approx}(G)$}\\
    \label{alg:congestion-approx}\nonl\textbf{Input:} Graph $G=(V,E,c)$ with $n=|V|,m=|E|$ \\
    \nonl\textbf{Output:} A $O(\qual)$-congestion approximator for $G$. \\
    \textbf{Procedure:}
    \begin{algorithmic}[1]
        \State$G_s \gets \texttt{ultrasparsifier}(G,\kappa)$ with $\kappa=10\log^{4}n$.
        \algorithmiccomment{Section \ref{sec:ultrasparsifier}} \\
        $(\mathcal{C},\mathcal{E}) \gets \texttt{$j$-tree}(G_s)$ with $\mathcal{C}$ the core and $\mathcal{E}$ the envelope of the $j$-tree.
        \algorithmiccomment{Section \ref{sec:jtree}} \\
        \nonl\algcomment{This is the only recursive call to \texttt{congestion-approx}} \\
        $R_{\mathcal{C}} \gets \texttt{congestion-approx}(\mathcal{C})$. \\
        \nonl\algcomment{$R_{\mathcal{C}}\cup \mathcal{E}$ is an $O(\log^{13}n)$-congestion approximator for $G$} \\
        $R' \gets \texttt{tree-hierarchical-decomp}(R_{\mathcal{C}}\cup \mathcal{E})$ \algorithmiccomment{Section \ref{sec:rst-on-trees}} \\
        \nonl\algcomment{From the call to \texttt{tree-hierarchical-decomp}, $R'$ is a binary tree with $O(\log n)$ depth} \label{line:improve-ca}\\
        $R \gets \texttt{improve-CA}(G,R')$ \algorithmiccomment{Algorithm \ref{alg:improve-ca}} \\
        \Return $R$
    \end{algorithmic}
\end{algorithm}

Our major technical contributions are in the new framework for computing congestion approximators, the  implementation specifics of which are described in the \texttt{improve-CA} subroutine and the details of which are discussed in Section~\ref{sec:new-rst}.
\begin{algorithm}{$\texttt{improve-CA}(G,R)$}\\
    \noindent\label{alg:improve-ca}\nonl\textbf{Input:} Graph $G=(V,E,c)$ and $O(\log^{d}n)$-congestion approximator $R$ for $G$, with $9 < d \leq 30$. \\
    \nonl\textbf{Output:} An $O(\qual)$-congestion approximator for $G$.\\
    \textbf{Procedure:}
    \begin{algorithmic}[1]
        \State $A_1,A_2 \gets \PA(G)$ \algorithmiccomment{Section \ref{sec:parta}} \\
        $B_1,B_2 \gets \PB(G, A_1, A_2)$ \algorithmiccomment{Section \ref{sec:partb}} \\
        \nonl\algcomment{Claim \hideref{claim:well-linked-part} guarantees the edges leaving each $Z_i$ are $O(1/\log^{9}n)$-well-linked} \\
        Set $Z_1 = A_1 \cap B_1, Z_2 = A_2 \cap B_1, Z_3 = A_1 \cap B_2, Z_4 = A_2 \cap B_2$.
        \ForAll{$Z_i$}
            \State \nonl\algcomment{$R_i$ has size $\ot{|Z_i|}$ by Lemma \hideref{lem:ca-contraction}}
            \State $R_i\gets \texttt{ca-contraction}(R,Z_i)$ \algorithmiccomment{Section \ref{sec:ca-contraction}}
            \State\nonl\algcomment{$G(Z_i)$ is a contracted subgraph (Definition \hideref{def:contract-subgraph}), reweighted as in Section \hideref{sec:rc-ca}}
            \State$T_i \gets \texttt{improve-CA}(G(Z_i), R_i)$
        \EndFor
        \State Set $R'$ as the tree formed by the $T_i$, using lines \ref{line:combine1} and \ref{line:combine2} of Algorithm \ref{alg:congestion-sapprox-details}.
        \State \Return $R'$
    \end{algorithmic}
\end{algorithm}

\begin{theorem}
    \label{thm:alg1}
    Given an undirected capacitated graph $G=(V,E,c)$ with $n$ nodes and $m$ edges, Algorithm \ref{alg:congestion-approx} is a $\polylog n$-depth, $O(m~\polylog n)$ total work \pram~algorithm which outputs a hierarchical decomposition tree that, with high probability, is a $O(\qual)$-congestion approximator for $G$.
\end{theorem}
\begin{proof}
    The depth, work and correctness of all subroutines besides \texttt{improve-CA} follow from the proofs in their respective sections.
    Specifically, these sections show that $R'$ passed to \texttt{improve-CA} is a $O(\polylog n)$-congestion approximator for $G$ which is a binary tree with depth $O(\log n)$, and that $R'$ can be computed using $O(\polylog n)$ depth and $\ot{m}$ work.
    Then, the correctness of \texttt{improve-CA} follows from Theorem \ref{thm:RCA}, as does the depth and work of all but the \texttt{ca-contraction} calls.
    To bound the work from contracting the congestion approximators, by Lemma \ref{lem:parta} we have that $A_1,A_2$, $B_1,B_2$ are both partitionings of $V$, and so $\bigcup_i Z_i = V$.
    Let $G(Z_i)$ be as defined in Definition \ref{def:contract-subgraph}, and properly modifying the capacities of some edges as in Section \ref{sec:rc-ca}.
    As the $Z_i$'s are a partitioning of nodes $V(G)$, each edge in $G$ can be present in a most two contracted graphs $G(Z_i)$.
    Thus, $\sum_{i}|E(G(Z_i))| = O(m)$ as well.
    Constructing all contracted graphs $G(Z_i)$ can therefore be done in $O(m)$ work (and $O(1)$ depth), and so by Lemma \ref{lem:ca-contraction}, computing the contracted congestion approximators for all $G(Z_i)$ takes work $\ot{m}$ and depth $O(\log n)$.
    The total work and depth of the algorithm then follow from Theorem \ref{thm:RCA}.
\end{proof}

\begin{proof}(Theorem~\ref{rst:max-flow})
Using the congestion approximator from Theorem~\ref{thm:alg1}, we can run the parallel version of Sherman's algorithm (Theorem~\ref{thm:Sherman-parallel}) to get both a $(1-\eps)$-approximate maximum $s$-$t$ flow and a $(1+\eps)$-approximate minimum $s$-$t$ cut.
\end{proof}

\subsection{Organization.}
We first discuss additional related work in 
Section \ref{sec:relate-work} and then preliminaries
in Section \ref{sec:prelim}.
We give details of our 
construction of congestion approximators 
in three subsequent sections. Firstly, Section \ref{sec:size_reductions} 
presents the construction of low-quality
congestion approximators (outlined in Algorithm \ref{alg:congestion-approx}).
Secondly, Section \ref{sec:new_congestion_approx}
presents the boosting step where the quality 
of the congestion approximator is improved (outlined in Algorithm \ref{alg:improve-ca}).
Finally, Section \ref{sec:extraction} presents 
details on how we extract congestion approximators 
for contracted subgraphs and utilize them within Sherman's framework \cite{sherman2013nearly}. 
We give details of our parallel flow decomposition
algorithm in Section \ref{sec:decomposition}.
We finally discuss applications of our approximate max-flows result in Section \ref{sec:apps}.

%% file: Related.tex
In addition to the related work we covered in the introduction,
parallel maximum flow has also been studied for restricted classes of graphs. 
For example, when the flow network is a DAG with depth $r$, \cite{cohen95} gives a \pram algorithm with near linear work and $\poly(r,\log n)$ depth.
Another problem related to the parallel maximum flow is parallel global minimum cut. 
The global minimum cut problem has been studied by~\cite{karger1996new,karger2000minimum,geissmann2018parallel}.
The current best known algorithm~\cite{geissmann2018parallel} has $\polylog(n)$ depth and $\tilde{O}(m)$ work.

In the sequential setting, the maximum flow problem on \textit{directed} graphs has been extensively studied over the past $\sim$70 years. 
Ford and Fulkerson gave the first maximum flow algorithm using the idea of augmenting paths~\cite{ford1956maximal}.
Since then, there has been a long line of work giving increasingly more computationally efficient algorithms for this problem; a partial list includes~\cite{dinic1970algorithm,edmonds1972theoretical,karzanov1973finding,dinic1973metod,even1975network,garbow1985scaling,goldfarb1988computational,goldberg1988new,goldberg1998beyond,boykov2004experimental,goldberg2008partial,daitch2008faster,hochbaum2008pseudoflow,chandran2009computational,lee2014path,goldberg2015faster,madry2016computing,liu2020faster,kathuria2022unit,chen2022maximum}.
When supplies and demands are polynomially bounded integrals, \cite{chen2022maximum} provides the current fastest maximum flow algorithm which achieves a running time of $m^{1+o(1)}$ by leveraging interior point methods, which are powerful second-order continuous optimization methods. 
When considering \emph{undirected} maximum flow problem, continuous optimization techniques have also been used to obtain fast $(1-\epsilon)$-approximate maximum flow algorithms including, e.g., \cite{christiano2011electrical,kelner2012faster,lee2013new,sherman2013nearly,kelner2014almost,Peng16,sherman2017area,sherman2017generalized}. Instead of using IPMs, these algorithms use simpler first-order methods.

In the parallel computation setting, flow problems other than maximum flow/minimum congestion flow have also been widely studied.
As we have discussed in previous sections, the goal of maximum flow/minimum congestion flow problem is to minimize the $\ell_{\infty}$ norm of the (rescaled) flow vector while satisfying the flow conservation constraints.
Alternatively, if the objective is to minimize the $\ell_1$ norm of the (rescaled) flow vector while satisfying the same constraints, the problem is known as the uncapacitated minimum cost flow problem. 
This problem can be seen as a generalization of the shortest path problem.
There is a line of work of parallel undirected uncapacitated minimum cost flow algorithms that are based on Sherman's framework~\cite{sherman2017generalized} as well. %
In particular, \cite{andoni2020parallel,li2020faster,rozhovn2022undirected} provide efficient procedures to compute a $\polylog(n)$-approximate $\ell_1$-oblivious routing scheme ($\ell_1$ preconditioner), and thus they are able to apply Sherman's framework~\cite{sherman2017generalized} to compute a $(1-\epsilon)$-approximate uncapacitated minimum cost flow for undirected graphs using $\polylog(n)$ depth and $\tilde{O}(m)$ work.
If the goal is to minimize the $\ell_2$ norm of the (rescaled) flow vector while satisfying the flow conservation constraints, the problem is known as the electric flow problem. 
The efficient parallel electric flow ($\ell_2$-flow) algorithm essentially reduces to an efficient (approximate) solver for  symmetric, diagonally dominant (SDD) linear systems due to \cite{peng2014efficient}.
In particular, \cite{peng2014efficient} provides a parallel algorithm to solve SDD systems with $n$ dimensions and $m$ non-zero entries in $\polylog(n)$ depth and $\tilde{O}(m)$ work, and thus the electric flow problem can be solved in the same complexity. 
Interestingly, by plugging the parallel SDD system solver into the framework of \cite{madry2016computing} to deal with the computation of each Newton step in the IPMs, it implicitly gives a parallel maximum flow algorithm for directed graphs with $\tilde{O}(\sqrt{m})$ depth and $\tilde{O}(m^{1.5})$ work (and with $1/\poly(n)$ additive error).

Finally, we also note that a recent paper by Forster et al.~\cite{Forster21} used a relevant technique in~\cite{cohen95}
to round fractional flows to integral ones in their distributed maximum flow algorithms. However, we emphasize that the scheme of~\cite{cohen95} is for flow rounding rather than flow decomposition, i.e. rounding fractional flows to integer flows, as opposed to finding the actual 
$s$-$t$ flow paths. The latter is crucial for our application in cut matching games, as the matching player needs to compute a (fractional) matching between the sources and sinks from a given 
$s$-$t$ flow.

%% file: pre.tex
\paragraph{Graphs.}
We represent an $n$-vertex, $m$-edge undirected, capacitated graph as $G = (V,E,c)$ with $V$ being the vertices (nodes),
$E$ being the edges, and $c$ being the edge capacities (weights).
We also use $V(G)$ and $E(G)$ to denote the vertices and edges of the graph $G$,
respectively.
We will interchangeably write the capacity of an edge $e$ as
$c(e)$ or $c_e$.
Sometimes the same edge $e$ appears in two different graphs with different capacities,
thus we write $c^H(e)$ or $c^H_e$ to denote the weight of $e$ in graph $H$ to avoid
ambiguity.
In Section \ref{sec:poly-weights}, we show how to transform a graph with arbitrary capacities into one with polynomial aspect ratio while only reducing the maximum $s$-$t$ flow by a $(1-\varepsilon)$ factor, for any $s,t\in G$ and $\varepsilon>0$.
Thus, throughout the paper we assume that graphs have $\poly(n/\varepsilon)$ aspect ratio.

We usually need to solve maximum flows by adding a source and a sink,
for which we set up the following notation.

\begin{definition}[Graphs with Sources and Sinks]
  Given a graph $H$ with two (not necessarily disjoint) vertex subsets $V_1,V_2\subseteq V(H)$,
  and a parameter $c_u \geq 0$ for each $u\in V_1$,
  and a parameter $d_u \geq 0$ for each $u\in V_2$,
  we write $H_{st}$ to denote the graph obtained from $H$ by doing the following:
  \begin{enumerate}
    \item Add a vertex $s$, and connect $s$ to each $u\in V_1$ with capacity $c_u$;
    \item Add a vertex $t$, and connect $t$ to each $u\in V_2$ with capacity $d_u$.
  \end{enumerate}
\end{definition}

\paragraph{Congestion Approximators and Sherman's Algorithm.} Given a graph $G = (V,E,c)$ and a demand vector $b$,
let $\opt_G(b)=\min_{f:Bf=b} \|C^{-1}f\|_{\infty}$ be the optimal congestion
of any flow routing a given demand vector $b$ in the graph $G$,
with $B$ being the edge-vertex incidence matrix of $G$.

We first define congestion approximators. Note that in this paper, we \textit{always} consider congestion approximators that are trees, which are readily plugged into Sherman's algorithm.

\begin{definition}[Congestion Approximators]
    Given a graph $G = (V,E,c)$,
    a tree $R = (V_R,E_R,c_R)$ with $V\subseteq V_R$
    is an $\alpha$-congestion approximator of $G$
    for some $\alpha \geq 1$ if for any demand vector $b$, we have
    \begin{align*}
        \opt_R(b') \leq \opt_G(b) \leq \alpha\cdot \opt_R(b') ,
    \end{align*}
    where $b'$ is obtained by appending $b$ with zeros on entries $V_R\setminus V$.
\end{definition}

We next state the parallel version of Sherman's algorithm.
While we believe it is known that
Sherman's algorithm can be implemented
with low depth \emph{modulo the computation of a good congestion approximator},
we give a discussion of such an implementation in Appendix \ref{sec:shermans}
for completeness.

\begin{theorem}[Sherman's algorithm~\cite{sherman2013nearly}, Parallel Version]\label{thm:Sherman-parallel}
    There is a \pram\ algorithm that,
    given a graph $G = (V,E,c)$, an $\alpha$-congestion approximator
    $R = (V_R,E_R,c_R)$ of $G$,
    a source $s\in V$ and a sink $t\in V$,
    and an $\eps \in (0,1)$,
    computes a $(1 - \eps)$-approximate maximum $s$-$t$ flow as well as
    a $(1+\eps)$-approximate minimum $s$-$t$ cut in $G$.
    The total work of the algorithm is $O((|E| + |E_R|) \alpha^2 \eps^{-3} \polylog(n))$, and the depth of the algorithm is
    $O(\alpha^{2} \eps^{-3} \polylog(n))$.
\end{theorem}

\paragraph{Contracted Subgraphs.}
We give the definition of contracted subgraphs below.
We use $X$ to denote the subset of vertices that we contract,
and use $S = V(G)\setminus X$ to denote the remaining vertices.

\begin{definition}[Contracted Subgraphs]
  \label{def:contract-subgraph}
  Given a graph $G = (V,E,c)$ and a
  subset of vertices $S\subseteq V(G)$,
  we denote by $G(S)$ the graph obtained from $G$
  by contracting $X := V\setminus S$ into a single supernode $u_{X}$,
    deleting any resulting self-loops,
    and keeping all resulting parallel edges.
\end{definition}

We now define reweighted contracted subgraphs which will be useful
in our new congestion approximator construction in Section \ref{sec:new-rst}.

\begin{definition}[Reweighted Contracted Subgraphs]
  Given a graph $G = (V,E,c)$ and a pair
  $\Pcal = (S,\omega)$,
  where $S\subseteq V(G)$ is a subset of vertices,
  and $\omega : E(S,V(G)\setminus S) \to (0,1]$
  is a real-valued {\em reweighting function},
  we denote by $\CS{G}{\Pcal}$ the graph obtained from $G$
  by doing the following operations: %
  \begin{enumerate}
    \item Contract $X := V\setminus S$ into a single supernode $u_{X}$,
    deleting any resulting self-loops,
    and keeping all resulting parallel edges;
    \item Scale the weight of each (parallel) edge $e$ incident on
    $u_{X}$ by a factor of $\omega(e)$.
  \end{enumerate}
  We call $u_{X}$ the \textit{contracted vertex}, and
  refer to other vertices (a.k.a.~$S$) in $\CS{G}{\Pcal}$ as
  \textit{uncontracted vertices}.
  We refer to the edges in $G(\Pcal)$ that
  are incident on $u_X$ the {\em boundary edges}.
\end{definition}

\paragraph{Subdivision Graphs and Well-Linked Edges.} We recall the definition of subdivision graphs from~\cite{RackeST14}.

\begin{definition}[Subdivision Graphs]
  \label{def:subdivision}
  Given a capacitated graph $G = (V,E,c)$,
  we write $G' = (V',E',c')$ to denote the subdivision graph of $G$, defined as follows.
  To obtain $G'$,
  we split each edge $e = (u,v)$ in $G$
  by introducing
  a new {\em split vertex}  (or \textit{subdivision vertex}) $x_e$ and two \textit{split edges} $(u,x_e)$
  and $(x_e,v)$, both with capacity $c_e$.
  In words, we have $V' = V \union X_E$ where $X_E$ is the set of split vertices of the edges in $E$,
  $E' = \Union_{e = (u,v)\in E} \setof{ (u,x_e), (x_e,v) }$,
  and $c'_{(u,x_e)} = c'_{(v,x_e)} = c_e$ for every edge $e\in E$.

  For a subset of edges $F\subseteq E$, we write $X_F$ to denote the set of split vertices of edges in $E$,
and $c_F$ (or equivalently, $c(F)$) to denote the total capacity of edges in $F$.
\end{definition}

\begin{definition}
    Given $G(\Pcal)$ and another reweighting function $\omegatil$ supported
    on the split edges of the boundary edges of $G(\Pcal)$ in $G'(\Pcal)$.
    Then we define the graph $G'(\Pcaltil = (\Pcal,\omegatil))$ to be the graph
    obtained from $G'(\Pcal)$ by reweighting the split edges of the boundary edges by $\omegatil$.
    That is, for an edge $e$ that is a split edge of a boundary edge, we have
    $c^{G'(\Pcaltil = (\Pcal,\omegatil))}_e = \omegatil(e)\cdot  c^{G'(\Pcal)}_e$.
\end{definition}

We next recall the notion of well-linked edges as used in~\cite{RackeST14}.

\begin{definition}[Well-linked Edges]
  \label{def:well-linked}
  For a capacitated graph $G = (V,E,c)$,
  a set $F\subseteq E$ of edges of total capacity $c_F$ is called {\em $\beta$-well-linked}
  for some $\beta \in (0,1]$ if there is a feasible multicommodity flow in the subdivision graph $G'$
  that sends $(\beta c_e c_f / c_F)$ units of flow from $x_e$ to $x_f$ for all pairs $e,f\in F$ simultaneously.
\end{definition}

Any subset of well-linked edges are also
well-linked.
\begin{proposition}
    \label{prop:subset-well-linked}
    If set $F\subseteq E$ of edges
    is $\beta$-well-linked
    in $G = (V,E,c)$,
    then any subset $F'\subseteq F$ of $F$
    is $\beta$-well-linked.
\end{proposition}
\begin{proof}
    Given a product demand on $F'$, we can first let each edge distribute its demands
    uniformly to all edges in $F$ with each edge getting flow proportional to its capacity.
    This can be done with congestion $1/\beta$ since edges in $F$ are $\beta$-well-linked.
    Then for each demand between two vertices $s,t$ in $F'$, we can compose the routings which
    distributed the flow from $s$ and $t$ to $F$.
    The proposition thus follows.
\end{proof}

\paragraph{\pram\ Primitives.}

We state a few basic \pram\ primitives, whose implementation is discussed
in Appendix~\ref{sec:pramroutine} for completeness.

\begin{proposition}[Prefix Sums \cite{akl1997parallel}]
    There is a \pram\ algorithm that,
    given a list of values $a_1,\ldots,a_n$,
    computes all prefix sums $\sum_{i\leq k} a_i$'s.
    The algorithm has $O(n)$ total work and $O(\log n)$ depth.
\end{proposition}

\begin{proposition}[Subtree Sums]
    There is a \pram\ algorithm that,
    given a tree $T = (V_T,E_T,w_T)$ where $w_T : V_T\to \mathbb{R}$ is
    a vertex weight function,
    a vertex $r\in V_T$,
    computes all subtree sums of the weight function $w_T$ with the $r$ being
    the root of tree.
    The algorithm has total work $O(|E_t|)$ and total depth
    $O(\log |E_T|)$.
\end{proposition}

%% file: size-reductions.tex
\label{sec:size-reducution}

In this section, we detail the steps of Algorithm \ref{alg:congestion-approx} up to the call to \texttt{improve-CA} (which is detailed in Algorithm \ref{alg:improve-ca} and Section \ref{sec:new-rst}).
That is, in this section we detail how to recursively compute a $\alpha'$-congestion approximator for $G$ which is a $O(\log n)$ depth binary hierarchical decomposition tree of $G$ where $\alpha'=O(\polylog n)$.
Section \ref{sec:new-rst} then details how to improve the quality of the congestion approximator to $O(\log^{9}n)$.
\subsection{Reducing the Size of the Graph.}
\subsubsection{Constructing an Ultrasparsifier.}
\label{sec:ultrasparsifier}
To reduce the size, our first step is to construct an ultrasparsifier of the graph.
An ultrasparsifier, informally speaking, consists of a spanning tree plus some additional edges that approximately preserves all graph cuts.
\begin{definition}[Ultrasparsifier]
    \label{def:ultra-sparsify}
    Given a graph $G=(V,E,c)$ with $n$ nodes and $m$ edges, and any parameter $\kappa \geq 1$, a $\kappa$-ultrasparsifier $G_s=(V,E_s,c_s)$ is the union of a spanning tree $T$ of $G$ and a collection of edges $E'$ such that $|E'|=O(m\log^2 n/\kappa)$.
    Moreover, with high probability, the capacity of every cut of $G$ is preserved to within a $\kappa$ factor in $G_s$.
\end{definition}
In Algorithm \ref{alg:congestion-approx}, we use $\kappa = 10\log^{4} n$ to ensure a $\polylog n$ reduction in size compared to the original graph.

\begin{lemma}
    \label{lem:ultra-sparsify}
    There exists a randomized \pram\ algorithm that given an undirected, capacitated graph $G=(V,E,c)$, and any desired quality parameter $\kappa\geq 1$, constructs a $\kappa$-ultrasparsifier of $G$ with $O(\log n)$ depth, and $O(m)$ total work.
\end{lemma}
\begin{proof}
    The algorithm first computes a maximum spanning tree using the result of \cite{pram-mst}, which uses $O(\log n)$ depth and $O(m)$ work.
    Then, scale up the weights on the tree edges by a factor of $\kappa$, and sample each remaining edge with probability $\Theta(\log^2 n/\kappa)$.

    Sampling can easily be done in parallel, and the number of sampled edges in expectation is also immediately as desired.
    So, it remains to show that the final constructed graph preserves all cuts to within a $O(\kappa)$ factor with high probability.

    Let $T$ be the constructed maximum spanning tree of $G$.
    Multiplying the weight of each edge of $T$ by $\kappa$ results in a worst-case cut distortion factor of $\kappa$.
    So, after scaling, we wish to select additional edges in order to approximate cuts (in the graph with newly scaled edges) to within an $O(1)$ factor.
    \cite{FungHNP19} shows that to approximate edges cuts to within a $(1+\varepsilon)$ factor, it is sufficient to sample each edge $e$ with probability at least $Cc_e\log^2 n/(\rho_e \varepsilon^2)$, where $\rho_e$ is the edge-connectivity between the endpoints of $e$, $c_e$ is the capacity of edge $e$, and $C$ is a suitably large constant.
    Here, the connectivity of an edge $e=(u,v)$ is defined as the capacity of the minimum cut separating $u$ from $v$.

    As our algorithm samples off-tree edges (that is, edges in $G$ but not $T$) at rate $\Theta(\log^2 n/\kappa)$, to complete our analysis it suffices to show that the connectivity of any off-tree edge $e=(u,v)$ is at least $c_e\kappa$.
    Since $e$ is off-tree, adding it to $T$ creates a cycle.
    Moreover, $T$ is the maximum weight spanning tree, so $c_e\leq c_{e'}$ for all $e'$ on the cycle created.
    Any cut separating $u$ from $v$ must also involve some tree edge $e'$ from this cycle, so the capacity of any cut (after scaling) is at least $c_{e'}\kappa$.
    Thus, the connectivity of $e$ is also at least $c_{e'}\kappa \geq c_e\kappa$.
\end{proof}

\subsubsection{Constructing a \texorpdfstring{$j$}{j}-tree.}
\label{sec:jtree}
After constructing a $O(\polylog(n))$-ultrasparsifier, we now have a graph $G_s$ which is the union of a spanning tree $T$ and collection of off-tree edges $E'$.
We now transform the ultrasparsifer so that we only need to compute the congestion approximator for a smaller subgraph.
Namely, we convert the ultrasparsifer into a $j$-tree.
A $j$-tree, introduced in \cite{madry10}, is a graph consisting of a forest with $j$ trees (called the \emph{envelope}) connected together by an arbitrary graph (called the \emph{core}) over $j$ vertices, one from each tree.
\begin{definition}[$j$-tree \cite{madry10}]
    \label{def:jtree}
    A weighted graph $G=(V,E,c)$ is a $j$-tree if it is a union of a \textit{core} graph $H$ which is an arbitrary capacitated graph supported on at most $j$ vertices of $G$, and a forest on $V$ such that each connected component in the forest has exactly one vertex in $H$.
    We call this forest the \textit{envelope} of the $j$-tree.
\end{definition}

Lemma 5.8 of \cite{madry10} shows how to sequentially convert any ultrasparsifer to an $O(j)$-tree, where $j$ is the number of nodes incident on off-tree edges in the ultrasparsifer.
In this section, we show how to adapt this sequential process to the \pram\ setting.

The algorithm of \cite{madry10} for converting an ultrasparsifier to a $j$-tree first recursively removes degree 1 vertices.
Let $F$ be the subgraph consisting of these deleted nodes and edges; $F$ is thus a forest, and all removed edges come from the spanning tree $T$ of the ultrasparsifier.
Unfortunately, this process is inherently sequential, so we instead show that an alternative process, which can easily be parallelized and identifies this forest of removed nodes.

\begin{lemma}
    \label{lem:deg-remove}
    Let $G_s$ be a $O(\polylog(n))$-ultrasparsifier with spanning tree $T$ and off-tree edges $E'$.
    Suppose $T$ is rooted at an arbitrary node incident on some off-tree edge.
    Then, a node $u$ is removed by recursive deletion of degree 1 vertices if and only if the subtree of $T$ rooted at $u$ contains no nodes incident on off-tree edges.
    Moreover, there is a \pram\ algorithm with $O(\log n)$ depth and $O(m)$ work which removes all such nodes.
\end{lemma}
\begin{proof}
    Let $r$ be the root of $T$, and let $F$ be the set of nodes removed by recursively deleting degree 1 vertices (so $F$ induces a forest).
    Define $X$ as the set of nodes incident on some off-tree edge.
    If the subtree of $u$ contains no nodes in $X$, then it immediately follows that $u$ (and all of its descendants) must be in $F$ because they induce a tree.

    Now, suppose some node in the subtree of $u$ is incident on an off-tree edge.
    Let $p$ be the parent of $u$, let $v$ be a child of $u$ such that the subtree rooted at $v$ also contains a node in $X$, and let $H_u$ be the graph formed by removing $u$ from $G_s$.
    As $r\in X$, both $p$ and $v$ are connected to $X$ in $H_u$.
    If $p,v$ are in the same connected component of $H_u$, then $u$ is in a cycle, and so it will never be removed by the recursive deletion procedure.

    Otherwise, $p,v$ are connected in $H_u$ to distinct elements of $X$; $p$ is connected to $r$ and $v$ is connected to some $x\in X$ such that $x\neq r$.
    As $p$ and $v$ are both neighbors of $u$, there exists in $G_s$ a $r\to x$ path $P$ such that $u\in P$.
    Consider the rounds of the recursive deletion procedure, where in each round the procedure removes all degree 1 vertices currently present in the graph.
    If before round $i$, no nodes of $P$ have been deleted, then after round $i$, none of $P$ has been removed as well: the interior points have 2 neighbors in the path, thus having degree at least 2, and the endpoints are never removed as they are incident on off-tree edges.
    The entire path is trivially present before the procedure begins, and so no nodes in $P$ are ever removed.
    Thus, as $u\in P$, $u\not\in F$, as desired.

    The algorithm to find $F$ is as follows: first, root $T$ at an arbitrary node incident on some off-tree edge.
    Construct node weights by setting $w_u=1$ if $u$ is incident on some off-tree edge, and $w_u=0$ otherwise.
    Then, run the subtree sum algorithm of Theorem \ref{thm:subtree}, and output all nodes whose subtree sum is 0.
    Identifying which nodes are adjacent to off-tree edges requires at most $O(1)$ depth and $O(m)$ work, and the subtree sum algorithm has depth $O(\log n)$ and total work $O(n)$, giving the desired runtime bounds.
\end{proof}

Let $G'$ be the graph after recursively deleting degree 1 vertices, or, equivalently by Lemma \ref{lem:deg-remove}, removing all nodes whose subtree contains no nodes incident on off-tree edges.
The next step is to find paths in $G'$ involving entirely degree 2 vertices except for the end points $u,v$ which have degree at least 3, and ``move'' capacity from the lowest weight edge into an edge between the endpoints of the path.
In Lemma 5.8, \cite{madry10} proves that after reconnecting the vertices removed by recusively deleting degree 1 vertices, this results in $(3j-2)$-tree, where $j$ is the number of nodes incident on off-tree edges in $G_s$.

\begin{algorithm}{$\TPC(G')$} \\
    \label{alg:j-tree-paths}\nonl\textbf{Input:} Graph $G'$ with no degree 1 vertices, and $j$ nodes incident on off-tree edges. \\
    \nonl\textbf{Output:} A $O(j)$-tree $G''$.\\
    \textbf{Procedure:}
    \begin{algorithmic}[1]
        \State Initialize $W\gets \{u \mid \deg(u)=2\}$, $\mathcal{P}=\emptyset, \mathcal{C}=\emptyset$, and $G''=G'$.
        \State Construct $G_d$ by removing $V \setminus W$ (i.e.~nodes of at least degree 3) from $G'$. 
        \State Run connectivity to identify connected components $C_1,\ldots C_q$ of $G_d$. \label{itm:tpc-connectivity}
        \For{each component $C_i$}
                \State Find $U_1\subseteq C_i$, the set of degree 1 vertices in $C_i$.
                \If{$|U_1|=0$}
                \State Add $C_i$ to $\mathcal{C}$.
                \Else
                \State Let $U_1=\{u_1,u_2\}$, and let $v_1,v_2 \in V \setminus W$ be the additional neighbors in $G'$ of $u_1,u_2$.
                \EndIf
                \If{$v_1=v_2$}
                \State Add the cycle formed by $C_i$ and $v_1$ to $\mathcal{C}$.
                \Else
                \State Add the path from $v_1\to v_2$ through $C_i$ to $\mathcal{P}$.
                \EndIf
        \EndFor
        \For{each $S \in \mathcal{P}\cup \mathcal{C}$}
              \State Find $e_{\min}$, the edge with minimum capacity in $S$.
              \State For all other $e\in S$, update capacity $w(e)=w(e)+w(e_{\min})$ in $G''$.
              \State If $S\in \mathcal{P}$ with endpoints $(v_1,v_2)$ and $(v_1,v_2)\not\in E(G'')$, create edge $(v_1,v_2)$ with capacity $w(e_{\min})$.\;
             \State Delete $e_{\min}$ from $G''$.
        \EndFor
        \State \Return{$G''$}.
    \end{algorithmic}
\end{algorithm}

\begin{theorem}[Parallel Version of Lemma 5.8 of \cite{madry10}]
    \label{thm:parallel-jtree}
    Algorithm~\ref{alg:j-tree-paths} uses $O(\log n)$ depth and $\ot{m}$ total work to construct a $O(j)$-tree, where $j$ is the number of nodes incident on an off-tree edge of $G_s$.
\end{theorem}
\begin{proof}
    For runtime, in Line \ref{itm:tpc-connectivity} we may use the connectivity algorithm shown in \cite{shiloach1980log}, which has $O(\log n)$ depth and $\ot{m}$ work.
    Finding the additional neighbors can be identified in $O(1)$ depth and $O(n)$ work, along with identifying whether they induce a path or a cycle, and a $O(\log n)$ depth algorithm can correctly update the capacities.

    For correctness, all constructed paths have endpoints of degree 3.
    After removing all nodes of degree at least 3, the remaining nodes all have degree 2 in $G'$, as $G'$ has no degree 1 vertices by Lemma \ref{lem:deg-remove}.
    So, every degree 2 vertex and every edge between degree 2 vertices is in exactly one component, and thus every edge is in some path or cycle and all paths and cycles are edge disjoint.

    The result then follows from the proof Lemma 5.8 in \cite{madry10}.
\end{proof}

Once we have a $O(j)$-tree $J$, to identify the core on which we wish to construct a congestion approximator, we first find a spanning tree of $J$.
We then invoke the procedure of Lemma \ref{lem:deg-remove} with respect to this spanning tree to remove any nodes not present in the core.
Then by Lemma 5.8 in \cite{madry10}, the constructed $j$-tree preserves all cut values to within a $\polylog(n)$ factor.

Finally, we then recursively run Algorithm~\ref{alg:congestion-approx} on the core of the $j$-tree, and attach the trees in the envelope (where envelope is defined as in Definition \ref{def:jtree}).

%% file: rst-on-trees.tex
\subsection{Hierarchical Decomposition on Trees.}
\label{sec:rst-on-trees}
After recursing on the core of the $j$-tree, we have a congestion approximator $R'$ for $G$.
However, the depth of $R'$ may be very large, as there is no bound on the depth of trees which make up the envelope of the $j$-tree.
Several parts of our algorithm, such as the procedure to reduce the size of the congestion approximator (Section \ref{sec:ca-contraction}), rely on $R'$ having $O(\log n)$ depth.
As such, before running the procedure to improve the quality of the congestion approximation of $R'$, we emulate the hierarchical decomposition procedure of \cite{RackeST14} \textit{on the tree} $R'$.
That is, we take as input the tree $R'$, and output a hierarchical decomposition tree with depth $O(\log n)$ which is an $O(\log^{4} n)$\footnote{As this loss of $O(\log^{4}n)$ suffices for our purposes, we do not attempt to improve it, and instead rely on the analysis of \cite{RackeST14} in a near-black-box fashion.} congestion approximator for $R'$.
Moreover, to bound the work of the contraction of congestion approximators in Section \ref{sec:ca-contraction}, we require the output congestion approximator to be a binary tree.

The goal of this section is to prove the following theorem:
\begin{theorem}[Hierarchical Decomposition on Trees]
    \label{thm:rst-on-trees}
    Let $T$ be a tree on $n$ nodes.
    There is a $O(\log^3 n)$ depth, $\ot{n}$ work algorithm to compute an $O(\log^{4}n)$-congestion approximator $R$ for $T$ which is a hierarchical decomposition with depth $O(\log n)$.
    Moreover, in the same depth and work, we can ensure that $R$ is a binary tree.
\end{theorem}

The algorithm makes use of a \textit{tree separator node}, which is any node $q$ of a tree $T$ such that removing $q$ results in a forest of components of size at most $|T|/2$.
\begin{definition}[Tree Separator Node]
    \label{def:tree-sep}
    A node $q$ in a tree $T$ is called a tree separator node of $T$ if the forest induced by $T \setminus \{q\}$ consists of trees with at most $|T|/2$ nodes.
\end{definition}
Lemma \ref{lem:tree-sep} gives a $O(\log |T|)$ depth, $O(|T|)$ work \pram\ algorithm to find a tree separator node for any tree $T$.

With this definition, we now describe the algorithm for computing a hierarchical decomposition on trees.
Each level in the hierarchical decomposition algorithm of \cite{RackeST14} consists of two partitioning steps.
Given a set $P$ of nodes, the first determines an $\Omega(1/\log^2 n)$-well-linked (see Definition \ref{def:well-linked}) set of edges within the graph induced by $P$ and for the second, it suffices to find an exact min-cut between these well-linked edges and the edges leaving $P$ (see Lemma 3.9 of~\cite{RackeST14} for more details)\footnote{We reuse the names ``Partition A'' and ``Partition B'' used in Section \ref{sec:new-rst}, to highlight the similar goals of the procedures \PA~and \PB, however we are not invoking or directly implementing these procedures.}.

\begin{definition}[Partition A]
    Let $P$ be a subset of nodes of a graph $G=(V,E,c)$, and let $G'[P]$ be the subdivision graph (see Definition \ref{def:subdivision}) of the subgraph of $G$ induced by $P$.
    Then, Partition A is a partitioning of $P$ into clusters  $P_1,\ldots, P_w$ such that the set of inter-cluster edges in $G'[P]$, namely $\{(u,v) \mid u\in P_i, v\in P_j, i\neq j\}$, is $\Omega(1/\log^2 n)$-well-linked.
    Moreover, we have $|P_i|\leq (1/2)|V(G)|$ for all $i$.
\end{definition}
For the definition of Partition B, we state it in terms of an exact min-cut, which is stronger than what \cite{RackeST14} or Section \ref{sec:new-rst} uses, since when the graph is a tree, we are able to compute exact min-cut efficiently in low depth (see Section~\ref{sec:min-cut-tree} for more details).

\begin{definition}[Min-Cut Partition B]
    \label{def:min-cut-partb}
    Let $P$ be a subset of nodes of a graph $G=(V,E,c)$, let $B_P$ be set of edges with one endpoint in $P$ and one endpoint in $V \setminus P$.
    Construct a graph $H$ as follows.
    Start with $G'[P]$, and for each edge $e$ in $B_P$, add a node $x_e$ to $H$.
    Finally, for each $e=(u,v)$ in $B_P$ such that $v\in P$, add an edge $(x_e,v)$ with capacity $c(e)/\log n$.
    Then, with $W$ the set of nodes of $H$ which corresponds to the set of subdivision nodes of the inter-cluster edges from Partition A, Partition B is a partitioning $(X, P \setminus X)$ such that $(X,P \setminus X)$ is a min-cut between $B_P$ and $W$ in the graph $H$.
\end{definition}
The analysis of \cite{RackeST14} and Section \ref{sec:new-rst} shows that to get a $\polylog(n)$ depth, $\ot{n}$ work \pram algorithm to compute a hierarchical decomposition for a tree, it suffices to implement Partition A and Partition B in $\polylog(k)$ depth and $\ot{k}$ work on a set of $k$ nodes.
We now give a pair of lemmas showing that we can indeed do this.
\begin{lemma}[Computing Partition A on Trees]
    \label{lem:part-a-trees}
    There is a $O(\log k)$ depth, $\ot(k)$ work \pram\ algorithm to compute Partition A on a set of $k$ nodes which induce a tree.
\end{lemma}
\begin{proof}
    Let $T$ be the tree induced by the vertex set we wish to partition.
    Then, compute a tree separator node $c$ of $T$, and set the partitions to be the connected components of $T \setminus \{c\}$ along with $\{c\}$.
    The depth and work are then $O(\log k)$ and $\ot(k)$, using the algorithms for finding a tree separator node and connected components.

    The only edges cut are those incident on the center $c$.
    The set of edges incident on the center are 1-well-linked, since they share an endpoint, meeting the requirements for Partition A.
\end{proof}
\begin{lemma}[Computing Partition B on Trees]
    \label{lem:part-b-trees}
    Given a tree $T$, there is a $O(\log^2 k)$ depth, $\ot{k}$ work \pram\ algorithm to compute Partition B on the subgraph of $T$ induced by a set $K$ of $k$ nodes.
\end{lemma}
\begin{proof}
    First, note that the subgraph induced by $K$ is a forest, as $T$ is a tree.
    Then, with $P=K$, define $B_P$, $W$, and $H$ as in Definition \ref{def:min-cut-partb}.
    Note that as the subgraph induced by $K$ in $T$ is a forest, $H$ is also a forest.
    Add to $H$ a super-source $s$ connected to each node of $B_P$ with capacity $\infty$ and a super-sink $t$ connect to each node of $W$ with capacity $\infty$.
    Then, the result follows from computing a exact $s$-$t$ min-cut on $H$ (using Lemma \ref{lem:min-cut-tree})\footnote{Lemma \ref{lem:min-cut-tree} is written for a tree with the addition of a source $s$ and sink $t$, rather than a forest; as such, we first add a dummy node $y$ that connects to one node of every component of the forest with a capacity 0 edge, which does not change the min-cut value and allows us to use Lemma \ref{lem:min-cut-tree}.}.
\end{proof}

These lemmas allow us to compute hierarchical decompositions on trees.
\begin{lemma}
    \label{lem:tree-hd}
    Given a tree $T$ on $n$ nodes, there is an $O(\log^3 n)$ depth, $\ot{n}$ work \pram algorithm to compute an $O(\log^{4}n)$-congestion approximator $R$ for $T$ which is a hierarchical decomposition with depth $O(\log n)$.
\end{lemma}
\begin{proof}
    We follow the algorithm of \cite{RackeST14} to construct a congestion approximator for $T$.
    As each level of the constructed congestion approximator in \cite{RackeST14} corresponds to a partitioning of the edges, computing Partition A and B for all sets at a given level can be done in $O(\log^2 n)$ depth and $\ot{n}$ work.
    The use of tree separator nodes leads to a constant factor size reduction from each Partition A step, and so there are at most $O(\log n)$ levels, leading to a $O(\log^3 n)$ depth, $\ot{n}$ work algorithm.
    Finally, the analysis of \cite{RackeST14} shows that the constructed hierarchical decomposition is a $O(\log^{4}n)$-congestion approximator for the original tree, and that the tree has $O(\log n)$ depth.
\end{proof}

It thus remains shows how to convert the congestion approximator $R$ into a binary tree; we describe the procedure in Algorithm \ref{alg:rst-binary}.
In Algorithm \ref{alg:rst-binary}, we use the term \textit{first non-binary node} to refer to any node $u\in R$ such that none of the ancestors of $u$ have more than two children, but $u$ has at least three children.
Since $R$ has depth $O(\log n)$, a breadth-first traversal can find all first non-binary nodes in $O(\log n)$ depth and $O(n)$ work.

Finally, for any node $u$ of a hierarchical decomposition tree $R$, we use the notation $R_u$ to refer to the subtree of $R$ rooted at $u$, and $P_u$ the set represented by node $u$.

\begin{algorithm}{$\convb(R)$}\\
    \label{alg:rst-binary}\nonl\textbf{Input:} Hierarchical decomposition tree $R=(V_R,E_R,c_R)$ which is a $\alpha$-congestion approximator for some tree $T$. \\
    \nonl\textbf{Output:} Binary tree $R'$ which is a $\alpha$-congestion approximator for $T$.\\
    \textbf{Procedure:}
    \begin{algorithmic}[1]
        \State Compute the set $\mathcal{B}$ of all first non-binary nodes in $R$
        \State \nonl \algcomment{$R_u$ is the subtree of $R$ rooted at $u$}
        \State Let $\mathcal{F}=\bigcup_{u\in \mathcal{B}}R_{u} \setminus \mathcal{B}$ be the forest of descendants of all first non-binary nodes
        \State Initialize $R' \gets R \setminus \mathcal{F}$
        \label{line:rst-binary-for}\For{each first non-binary node $u\in \mathcal{B}$}
        \State \nonl \algcomment{$P_{v_i}$ is the set represented by $v_i$ in $R$, and $q\geq 3$ as $u\in \mathcal{B}$} 
            \State Let $v_1,\ldots ,v_q$ be the children of $u$ such that $|P_{v_1}|\geq |P_{v_2}| \geq \ldots |P_{v_q}|$ \label{line:childern-sort}
            \State Set $s$ such that $s$ is the smallest index such that $|P_{v_1} \cup \ldots \cup P_{v_s}| \geq |P_u|/4$ \label{line:set-s}
            \State Add to $R'$ new nodes $z_1,z_2$ as children of $u$ with edges of capacity $\infty$. Moreover, in $R'$, $z_1$ represents $P_{v_1}\cup \ldots \cup P_{v_s}$ and $z_2$ represents $P_{v_{s+1}}\cup \ldots \cup P_{v_q}$ \label{line:add-nodes}
            \For{$1\leq i \leq s$}
                \State Add $R_{v_i}$ to $R'$ by adding edge $(z_1,v_i)$ with capacity $c_R((u,v_i))$
            \EndFor
            \For{$s+1 \leq i \leq q$}
                \State Add $R_{v_i}$ to $R'$ by adding edge $(z_2,v_i)$ with capacity $c_R((u,v_i))$
            \EndFor
            \State If $z_1$ (or $z_2$) has exactly one child $v$, remove $z_1$ (or $z_2$) and connect $v$ to $ u$ with an edge of capacity $c_R((u,v))$.
        \EndFor
        \State If $R'$ is binary, return $R'$. Otherwise, return $\convb(R')$. \label{line:end}
    \end{algorithmic}
\end{algorithm}

To prove the correctness of Algorithm \ref{alg:rst-binary}, we first prove the following helper lemma.
\begin{lemma}
    \label{lem:grandparent}
    Let $R'=\convb(R)$, and consider $x$ a node of $R'$.
    Let $y$ be the parent of $x$, and let $z$ be the parent of $y$.
    Then, $|P'_x|\leq (3/4)|P'_z|$, where for any $w\in R'$, $P'_w$ is the set represented by $w$ in $R'$.
\end{lemma}
\begin{proof}
    Every node in $R$ is present in $R'$, as each call to $\convb$ adds nodes but does not remove any.
    As such, $V(R) \subseteq V(R')$.
    We again use the notation $P_w$ to denote the set represented by a node $w$ in $R$, and $P'_w = P_w$ for all $w\in R$.
    Call any node in $R' \setminus R$ an intermediary node.
    First, consider a node $x$ in $R'$ whose parent $y$ is not an intermediary node; that is, $y\in R$.
    If $x\in R$ as well, then by Lemma \ref{lem:part-a-trees}, $|P_x|\leq (1/2)|P_y|$.
    So, suppose $x$ is an intermediary node, and thus $P'_x$ is a union of sets $P_{w_1},\ldots ,P_{w_q}$, where $w_1,\ldots ,w_q$ are the children of $y$ in $R$.
    As in Algorithm \ref{alg:rst-binary}, set $s$ to be the smallest index such that $|P_{w_1}\cup \ldots \cup P_{w_s}|\geq (1/4)|P_{y}|$, and let $S_1 = P_{w_1} \cup \ldots \cup P_{w_s}$ and $S_2 = P_{w_{s+1}} \cup \ldots \cup P_{w_q}$.
    From the fact that $|P_{w_i}| \leq (1/2)|P_{y}|$ (which is again from Lemma \ref{lem:part-a-trees}) and the setting of $s$, we have that $(1/4)|P_{y}|\leq |S_1| \leq (3/4)|P_{y}|$.
    Since $S_1\cup S_2 = P_{y}$, the same inequality holds for $|S_2|$.
    Thus, as either $P'_x=S_1$ or $P'_x=S_2$, we have that $|P'_x|\leq(3/4)|P'_y|$.
    Since $R'$ is a hierarchical decomposition, if $z$ is the parent of $y$, $P'_y \subseteq P'_{z}$, and thus all nodes which are the child of a non-intermediary node satisfy the lemma.

    Now, consider a node $x\in R'$ whose parent $y$ is an intermediary node.
    Let $z$ be the parent of $y$.
    As $R'$ is a hierarchical decomposition by construction, there exists a set $S'\subseteq V(R)$ such that $P'_{z}=\bigcup_{w\in S'}P_w$ and a set $S\subset S'$ such that $P'_y=\bigcup_{w\in S}P_w$.
    If for all $w\in S$, $|P_w|\leq (1/2)|P'_{z}|$, then by the same argument used in the case where $y$ was non-intermediary, $|P'_x|\leq |P'_y|\leq (3/4)|P'_{z}|$.
    Suppose, for contradiction, there exists a $w\in S$ such that $|P_w| > (1/2)|P_{z}|$.
    Since Algorithm \ref{alg:rst-binary} sorts the nodes in Line \ref{line:childern-sort} in decreasing order of size of the sets they represent, if $S$ contains an element $w$ such that $|P_w| > (1/2)|P_{z}|$, $w$ must be the only element of $S$.
    As such, since $S=\{w\}$, $y$ is not an intermediary node, which is a contradiction.
\end{proof}

\begin{lemma}
    \label{lem:rst-binary}
    Given an $\alpha$-congestion approximator $R$ from Lemma \ref{lem:tree-hd} for a tree $T$ with $n$ nodes, Algorithm \ref{alg:rst-binary} is a $O(\log^2 n)$ depth, $\ot{n}$ work \pram~algorithm that constructs a binary tree $R'$ such that $R'$ is a $\alpha$-congestion approximator for $T$.
    Moreover, $R'$ has depth $O(\log n)$.
\end{lemma}
\begin{proof}
    Let $R'=\convb(R)$ be the output of Algorithm \ref{alg:rst-binary} on $R$.
    Lemma \ref{lem:grandparent} immediately implies that $R'$ has depth $O(\log n)$, and Line \ref{line:end} guarantees that it is binary, so it remains to show that $R'$ is a $\alpha$-congestion approximator for $T$ and that Algorithm \ref{alg:rst-binary} has $O(\log^2 n)$ depth and $\ot{n}$ total work.
    Note that for any edge $(u,v)$ in $R$, there is a path from $u$ to $v$ in $R'$ with the same minimum capacity as $(u,v)$ in $R$; namely, all edges on the path from $u$ to $v$ in $R'$ have capacity either $\infty$ or $c_R((u,v))$.
    Let $P_{uv}$ be the path for $(u,v)$ in $R'$, and let $P_{xy}$ be the path for some other edge $(x,y)\in E(R)$ in $R'$.
    By the setting of edge capacities in Algorithm \ref{alg:rst-binary}, all edges along both $P_{uv}$ and $P_{xy}$, if there are any, must have capacity $\infty$.
    As such, any routing in $R'$ can be converted to a routing in $R$ with the same congestion, and vice-versa, so $R'$ is also a $\alpha$-congestion approximator for $T$.

    Let $R_0=R,R_1,\ldots ,R_L$ be the sequence of trees such that for all $i<L$, $\convb(R_i)$ makes a recursive call $\convb(R_{i+1})$ in Line \ref{line:end}, and the call $\convb(R_L)$ returns $R'$ without further recursion.
    Each call to $\convb$~only increases the number of nodes and depth of the tree: no nodes are removed, and each addition of a node $z_i$ in Line \ref{line:add-nodes} can only increase the depth.
    So, it follows that for any $i\in [L]$, the depth of $R_i$ is no more than the depth of $R'$; by Lemma \ref{lem:grandparent}, we have that the depth of $R_i$ is then $O(\log n)$.
    Moreover, by construction, each $R_i$ is a hierarchical decomposition of $T$, so each level $j$ of $R_i$ represents a partitioning of $V(T)$.
    Thus, as there are $O(\log n)$ levels, each node in $T$ can appear in the sets of at most $O(\log n)$ nodes of $R_i$, each at a different level of $R_i$.
    Since there are $n$ nodes in $T$, it follows that there are $O(n\log n)$ nodes in $R_i$.

    Each call $\convb(R_i)$ before the recursive call in Line \ref{line:end} can be implemented in $O(\log |R_i|)$ depth and $\ot{|R_i|}$ work.
    Since $|R_i|=O(n\log n)$, each call can be implemented in $O(\log n)$ depth and $\ot{n}$ work.
    To bound the total depth and work, it thus suffices to bound the total number of recursive calls to \convb~made during a call of $\convb(R)$.
    Define $B(R_i)$ to be the lowest depth (i.e.~distance from the root) of any first non-binary node in $R_i$.
    In each call $\convb(R_i)$, after the \texttt{for} loop of Line \ref{line:rst-binary-for}, all first non-binary nodes have at most two children, and are thus no longer first non-binary nodes.
    In addition, the \texttt{for} loop can only reduce the number of children any node has.
    As such, it follows that $B(R_{i+1}) \geq B(R_i)+1$ for all $i<L$.
    Then, as $R_L$ has depth $O(\log n)$, we must have $B(R_i) = O(\log n)$ for all $i$, and thus $L = O(\log n)$ as well.
    This results in a complete depth of $O(\log^2 n)$ and total work of $\ot{n}$.
\end{proof}

The proof of Theorem \ref{thm:rst-on-trees} then follows from Lemma \ref{lem:tree-hd} and Lemma \ref{lem:rst-binary}.

%% file: newRST14.tex
\label{sec:new-rst}

The goal of this section and the next (Section~\ref{sec:oracle})
is to give a full presentation and
analysis of our Algorithm~\ref{alg:improve-ca},
whose goal is to boost
the approximation quality of a given congestion approximator.
Specifically, given any
congestion approximator of a graph $G$
of arbitrary $\polylog(n)$ distortion,
we will compute an $O(\qual)$-congestion approximator of $G$
in $O(m\polylog(n))$ work and $O(\polylog(n))$ depth.

Our presentation of Algorithm~\ref{alg:improve-ca} will be given as two parts.
First in this section, we show, assuming being able to solve
approximate maximum flows on contracted subgraphs,
how we can compute a congestion approximator of $O(\qual)$ distortion.
Then in the next section (Section~\ref{sec:oracle}), we show
how we can directly extract congestion approximators for contracted subgraphs
from a given congestion approximator of the entire graph, allowing us to then run
Sherman's algorithm~\cite{sherman2013nearly} to compute $(1-\eps)$-approximate
maximum flows on these contracted subgraphs.

We now present in this section a variant of the framework in~\cite{RackeST14} that allows
us to \textit{boost} the approximation of a given congestion approximator
in a manner that is both work-efficient and parallelizable.
The key novelty in our new framework is to avoid running (approximate) maximum flows
on subgraphs of $G$, which are obtained by removing vertices/edges from $G$.
This is because running maximum flows on such subgraphs requires computing congestion approximators
for them using additional recursions, which would have blown up the depth of our algorithm
to at least $n^{o(1)}$.
Rather, we run approximate maximum flows on {\em contracted subgraphs} of $G$,
that are obtained from $G$ by
contracting subsets
of vertices into single nodes.
As we will show in Section~\ref{sec:oracle},
we can extract congestion approximators for contracted subgraphs
{\em directly} from a given congestion approximator of the entire graph,
\textit{without} having to recurse on the subgraphs.

Throughout this section, we assume that
we have a \pram\ algorithm $\Acal$,
whose implementation we describe in the next section
(Section~\ref{sec:oracle}),
that can compute a $(1-\eps)$-approximate maximum flow
on any given contracted subgraph of $G$
with work near-linear in the number of edges of the subgraph and depth $O(\polylog(n))$.
We separate the implementation of $\Acal$ from the presentation of the new framework
here in an effort to make the latter cleaner and more intelligible.

\begin{remark}
Note that the specific implementation of algorithm $\Acal$
will depend on the execution of our framework in this section,
and in particular the execution of our framework and that of $\Acal$ should
alternate with each other.
This is because
the implementation of $\Acal$ involves extracting congestion approximators
for the contracted subgraphs from the given congestion approximator of the entire graph.
However,
only after we have run $\Acal$ on a current cluster $S$
do we know the sub-clusters on which we want to run $\Acal$ subsequently.
The composition of the two algorithms is given in Algorithm~\ref{alg:improve-ca}.
\end{remark}

We state the performance of our assumed \pram\ algorithm $\Acal$ below,
and describe how to implement it in Section~\ref{sec:oracle}. Note that
in our algorithm, we always set
$\zeta$ to be at least $1/\polylog(n)$
so $\Acal$ has near-linear total work
and $\polylog(n)$ depth.

\begin{proposition}[Performance of $\Acal$]
  \label{prop:oracle}
  Let $\CS{G'}{\Pcal = (S,\omega)}$ be the subdivison of
  a contracted subgraph of $G$ with reweighting function $\omega$
  such that the range of $\omega$ is
  within $[\zeta,1]$ for some $\zeta \in (0,1]$.
  Then
  given an arbitrary graph $\CS{G'_{st}}{\Pcal}$ obtained from $\CS{G'}{\Pcal}$
  with maximum $s$-$t$ flow value $F^*$,
  $\Acal$ computes with high probability
  \begin{enumerate}
    \item A feasible $s$-$t$ flow of value at least $(1 - \eps) F^*$
      in $\CS{G'_{st}}{\Pcal}$. %
    \item An $s$-$t$ cut $(T,\bar{T})$ in $\CS{G'_{st}}{\Pcal}$
      with capacity at most $(1 + \eps) F^*$. %
  \end{enumerate}
  $\Acal$ has total work $O(|E(\CS{G'_{st}}{\Pcal})|\zeta^{-2} \eps^{-3}\polylog(n))$ and depth $O(\zeta^{-2} \eps^{-3}\polylog(n))$.
\end{proposition}

\paragraph{Plan for the Rest of the Section.}
We will obtain a hierarchical decomposition tree of the graph that serves as our
congestion approximator. In each step of the decomposition,
we use the maximum flow algorithm $\Acal$ to perform two partitioning steps
on a current cluster $S$, and use the resulting partitions to obtain a
two-level
decomposition tree. Recursively applying this decomposition step to each sub-cluster obtained
ends up giving
us a hierarchical decomposition tree of the graph.
As will be guaranteed by our partitioning steps,
in each decomposition step we reduce the size of the cluster by a constant factor.
As a result, we get a tree of $O(\log n)$ depth.

At a high level, the goals of the two partitioning steps
are similar to those of~\cite{RackeST14}.
Specifically, the first partitioning step is to find a subset of edges
that (i) are well-linked in the sense that
we can route product demands between them with low congestion, and (ii)
separate the current cluster $S$ into sub-clusters whose sizes are a constant factor
smaller.
The second partitioning step is then to find a bottleneck cut that separates
the inter-cluster edges found in step one and the
\textit{boundary edges} that go from $S$ to $V(G)\setminus S$.
The difference between our partitioning steps
and those of~\cite{RackeST14} is
that we obtain these partitions in contracted subgraphs rather than
vertex-induced subgraphs, which
avoids additional recursions on these subgraphs when solving
maximum flows on them,
but on the other hand
requires extra care so as to prevent
the edges from getting over-congested, since
after all an edge could implicitly appear in polynomially many
contracted subgraphs as one inside the contracted vertices.

In the rest of this section,
we will first
present the two partitioning steps needed,
and then show how to use them
to obtain a hierarchical decomposition tree.
In the following, we will interchangeably use the terms \textit{partitioning} and
\textit{clustering}, and the terms \textit{partitions} and \textit{clusters}.
For a given edge subset $F$, the partition \textit{induced by} $F$ is the partition of the vertices correponding to the connected components of the graph after the removal of $F$.

\subsection{The First Partitioning Step.}
\label{sec:parta}

Given a contracted subgraph $\CS{G}{\Pcal}$
with at least two uncontracted vertices,
our first step is to partition the vertices $V(\CS{G}{\Pcal})$
into sets $Z_1,Z_2$
such that the edges between $Z_1,Z_2$ are well-linked in
$\CS{G}{\Pcal}$,
and $Z_1,Z_2$ are size balanced.
To this end, we first use the parallel implementation of the cut-matching game in \cite{RackeST14} to get a balanced partition with \emph{almost} all of the inter-cluster edges well-linked. To simplify our presentation,
we say a partition $Z_1,\ldots,Z_z$ of a set $Z$ is $\gamma$-balanced
for some $\gamma \in (0,1]$ if $\card{Z_i}\leq \gamma \card{Z}$ for all $i$.

By plugging $\Acal$ into the cut matching game of \cite{RackeST14} (their Lemma 3.1), whose parallel version is described in Appendix~\ref{sec:pcmg},
we have the following lemma.
\begin{lemma}\label{lem:parta}
 There exists a \pram algorithm $\PAp$ that given $G(\Pcal)$ for $\Pcal = (S,\omega)$ with $X=V(G)\setminus S$ and the range of $\omega$ within $[\zeta,1]$, and a set of edges $F \subseteq E$ that induces a $3/4$-balanced partition of $V(G(\Pcal))$,
 with high probability computes
 a set of new edges $\Fnew$ such that $\Fnew$ also induces a $3/4$-balanced partition of $V(G(\Pcal))$, and
 \begin{enumerate}
 \item either $c^{G(\Pcal)}(\Fnew)\leq \frac{7}{8}c^{G(\Pcal)}(F)$;
 \item or $\Fnew=A\cup R$ with $A,R$ disjoint, such that $c^{G(\Pcal)}(A)\leq c^{G(\Pcal)}(F)$, $c^{G(\Pcal)}(R)\leq \frac{2}{\log{n}}\cdot c^{G(\Pcal)}(A)$, and edges in $A$ are $\beta$-well-linked in $G(\Pcal)$ for $\beta=\Omega(1/\log^2{n})$.
 \end{enumerate}
 The algorithm $\PAp$ has $O(|E(\CS{G}{\Pcal})|\zeta^{-2} \polylog(n))$ total work and $O(\zeta^{-2} \polylog(n))$ depth.
\end{lemma}

We also need the following lemma,
whose proof is deferred to
Section~\ref{sec:partb}.

\begin{lemma}\label{lem:partaa}
    There is a \pram algorithm $\PApp$ that given $G(\Pcal)$ for $\Pcal = (S,\omega)$ with $X=V(G)\setminus S$ and the range of $\omega$ within $[\zeta,1]$,
    a set of edges $F = A\union B$ inducing a $3/4$-balanced partition of
    $V(G(\Pcal))$ where
    $A,B$ are disjoint with $c^{G(\Pcal)}(B)\leq 2c^{G(\Pcal)}(A)/\log n$,
    \begin{enumerate}
        \item either finds an edge set $\Fnew$ with $c^{G(\Pcal)}(\Fnew)\leq \frac{3}{4} c^{G(\Pcal)}(F)$ that induces a $3/4$-balanced partition;
        \item or finds an edge set $C$ disjoint from $A$ such that
        $\Fnew := A\union C$ induces a $3/4$-balanced partition, and
        there exists a multicommodity flow in $G'(\Pcal)$ with congestion $O(\log n)$
        from $X_C$ to $X_A$ such that (i) each $x_e \in X_C$ sends $c^{G(\Pcal)}_{e}$
        units of flow, and (ii) each $x_f \in X_A$ receives $O(\log n)\cdot c^{G(\Pcal)}_{f}$ units
        of flow.
    \end{enumerate}
    The algorithm $\PApp$ has $O(|E(\CS{G}{\Pcal})|\zeta^{-2} \polylog(n))$ total work and $O(\zeta^{-2} \polylog(n))$ depth.
\end{lemma}

We now prove the main lemma about our first partitioning step.

\begin{lemma}\label{lem:partafix}
 There is a \pram algorithm $\PA$ that
  given $G(\Pcal)$ for $\Pcal = (S,\omega)$
  with $X = V(G)\setminus S$ and the range of $\omega$ within $[\zeta,1]$,
  with high probability partitions the uncontracted vertices into $Z_1,Z_2$ such that
  \begin{enumerate}
    \item $Z_1\union Z_2 = V(\CS{G}{\Pcal}) \setminus \setof{u_{X}}$.
    \item $|Z_i| \leq \frac{7}{8} \sizeof{ V(\CS{G}{\Pcal}) }$ for each $i = 1,2$.
    \item The set of inter-cluster edges
      $
        F: = \setof{ (u,v) | u\in Z_1, v\in Z_2}
      $
      is $\beta$-well-linked in $\CS{G}{\Pcal}$ for $\beta = \Omega(1/\log^{3} n)$.
  \end{enumerate}
  The algorithm $\PA$ has $O(|E(\CS{G}{\Pcal})|\zeta^{-2} \polylog(n))$ total work and $O(\zeta^{-2} \polylog(n))$ depth.
\end{lemma}
\begin{proof}
  By composing Lemmas~\ref{lem:parta} and~\ref{lem:partaa} the same way
  as \cite{RackeST14} compose their Lemmas 3.1 and 3.2,
  we get in desired total work and depth
  a partition $Z'_1,\ldots,Z'_z$ of $V(\CS{G}{P})$ such that
  \begin{enumerate}
    \item $Z'_1\union\ldots\union Z'_z = V(\CS{G}{\Pcal})$. 
    \item $|Z'_i| \leq \frac{3}{4} \sizeof{ V(\CS{G}{\Pcal}) }$ for each $i\in [z]$.
    \item The set of inter-cluster edges
      $
        F' = \setof{ (u,v) | u\in Z'_i, v\in Z'_j,i\neq j}
      $
      is $\beta'$-well-linked in $\CS{G}{\Pcal}$ for $\beta' = \Omega(1/\log^{3} n)$.
  \end{enumerate}
  We can then obtain $Z_1,Z_2$ by the following process. First,
  we remove the supernode $u_{X}$ from $Z'_1,\ldots,Z'_{z}$ to obtain a partition
  $Z''_1,\ldots,Z''_{z}$ of the uncontracted vertices.
  Now the inter-cluster edges between $Z''_1,\ldots,Z''_z$ are still
  $\Omega(1/\log^3 n)$-well-linked in $\CS{G}{\Pcal}$ since they are a subset
  of the inter-cluster edges between $Z'_1,\ldots,Z'_z$.
  Then to get a bi-partition, we merge the subsets $Z''_1,\ldots,Z''_z$ as follows.
  Let $i$ be the largest integer s.t. 
  $\sizeof{ Z_1'' \union Z_2'' \union\ldots \union Z_i''}\leq \frac{7}{8} \sizeof{V(\CS{G}{\Pcal}}$.
  Then we merge the $Z''_1,\ldots Z''_i$ into one partition $Z_1$
  and the remaining $Z''_i$'s into another partition $Z_2$
  to obtain our bi-partition $Z_1,Z_2$.
  Then because $Z''_1,\ldots,Z''_z$ is a partition of the uncontracted vertices,
  so is $Z_1,Z_2$.
  Since each $Z''_i$ has size at most $\frac{3}{4} \sizeof{ V(\CS{G}{\Pcal}) }$,
  by the definition of $i$, $Z_1$ has size between
  $[\frac{1}{8} \sizeof{ V(\CS{G}{\Pcal}) }, \frac{7}{8} \sizeof{ V(\CS{G}{\Pcal}) }]$.
  Therefore, $Z_2$ has size at most $\frac{7}{8} \sizeof{ V(\CS{G}{\Pcal}) }$ since $Z_1,Z_2$ is a partition
  of the uncontracted vertices in $V(\CS{G}{\Pcal})$.
  Finally, inter-cluster edges
  (call them $F$)
  from $Z_1$ to $Z_2$ are still $\Omega(1/\log^3 n)$-well-linked
  in $\CS{G}{\Pcal}$ since they are a subset of the
  inter-cluster edges between $Z''_1,\ldots,Z''_z$.

  Note that the partitioning into $Z_1,Z_2$
  can be done in parallel in $O(\log n)$ depth by
  computing a prefix sum and a binary search.
  The total work and depth then follows from the performance of the first partitioning step of~\cite{RackeST14}.
\end{proof}

\subsection{The Second Partitioning Step.}
\label{sec:partb}

Let $B$ denote the {\em boundary edges} in $\CS{G}{\Pcal}$, 
namely those that go from the contracted vertex $u_{X}$ to the uncontracted vertices.
In other words,
$B := E(u_{X}, V(G)\setminus X)$.

Our second step is to find a cut in $\CS{G}{\Pcal}$ separating
the boundary edges $B$ from the inter-cluster edges $F$ that we identified in the
first partitioning step.
Here, we want the property that there is a low-congestion routing from the cut edges we find to the boundary edges $B$,
as well as from the cut edges to the inter-cluster edges $F$,
such that each cut edge sends out flow equal to
its capacity.
Specifically,
we prove the following lemma.
Recall that
for a graph $H$ and an edge $e\in H$,
we write $c_e^{H}$ to denote the capacity of edge $e$ in $H$.

\begin{lemma}\label{lem:partb}
  There is a \pram algorithm $\PB$ that
  given $G(\Pcal)$ for $\Pcal = (S,\omega)$ with $X = V(G)\setminus S$
  and the range of $\omega$ being within $[\zeta,1]$ with $\zeta \in (0,1]$,
  two disjoint edge subsets $B,F\subset E(G(\Pcal))$,
  and a parameter $\psi \in (0,1]$,
  with high probability returns a subset of edges
  $Y$ (potentially intersecting both $B$ and $F$) in $\CS{G}{\Pcal}$ such that
  \begin{enumerate}
    \item $X_Y$ separates $X_B$ from $X_{F}$ in $\CS{G'}{\Pcal}$, that is, in $\CS{G'}{\Pcal}$
      every path between a vertex in $X_B$ and a vertex in $X_{F}$
      must contain a vertex in $X_Y$. Moreover, in $G(\Pcal)$,
      the total capacity of $Y$ is at most twice the capacity of $B$ and at most twice the capacity of $F$.
    \item There is a multicommodity flow in $\CS{G'}{\Pcal}$
      from $X_Y$ to $X_B$ such that
      \begin{enumerate}
          \item The congestion on edges incident on $u_X$ in $G'(\Pcal)$ is
          $O(\psi \log n)$, while
          the congestion on other edges in $G'(\Pcal)$ is $O(\log n)$.
          \item Each $x_y\in X_Y$ sends $c^{\CS{G}{\Pcal}}_y$ units of flow while each
      $x_b\in X_B$ receives at most $O(\log n) \cdot c^{\CS{G}{\Pcal}}_b$ units of flow.
      \end{enumerate}
     \item Similarly, there is a multicommodity flow in $\CS{G'}{\Pcal}$
      from $X_Y$ to $X_{F}$ such that
      \begin{enumerate}
          \item The congestion on edges incident on $u_X$ in $G'(\Pcal)$ is
          $O(\psi \log n)$, while
          the congestion on other edges in $G'(\Pcal)$ is $O(\log n)$.
          \item Each $x_y\in X_Y$ sends $c^{\CS{G}{\Pcal}}_y$ units of flow,
          while each $x_f\in X_{F}$ receives at most $O(\log n) \cdot c^{\CS{G}{\Pcal}}_f$ units of flow.
      \end{enumerate} 
  \end{enumerate}
  The algorithm $\PB$ has $O(|E(\CS{G}{\Pcal})|\zeta^{-2} \psi^{-2} \polylog(n))$ total work and $O(\zeta^{-2} \psi^{-2} \polylog(n))$ depth.
\end{lemma}

\begin{proof}
  We will do an iterative refinement process to find the desired set of cut edges $Y$.
  Initially, we start with $Y$ being the smaller (in capacity) of $B$ and $F$, which has the desired
  property that $X_Y$ separates $X_B$ from $X_F$ in $G'(\Pcal)$.
  We will maintain this property, while ``refine'' the set $Y$. In particular, we will classify the edges in $Y$ into
  \textit{good} edges $\Yg$ and \textit{bad} edges $\Yb$, and try to reduce the total
  capacity of the bad edges $\Yb$.
  Initially, $\Yg = \emptyset$ and $\Yb = Y$.

  In each iteration, we do the following refinement step.
  Define a reweighting function $\omegatil$ on the split edges
  (in $G'(\Pcal)$)
  of the boundary edges of $G(\Pcal)$ such that
  \begin{align*}
      \omegatil(f) =
      \begin{cases}
          \psi  & \text{$f$ incident on $u_X$} \\
          1 & \text{otherwise}.
      \end{cases}
  \end{align*}
  and look at the graph
  $G'(\tilde{\Pcal} = (\Pcal, \omegatil))$.
  Let $\eps =  \log^{-1} n$.

  Let $T_1 := B$ and $T_2 := F$.
  For $j \in \setof{1,2}$,  
  we consider the graph $G'_{s_jt_j}(\Pcaltil)$ where
  we connect $s_j$ to each $x_y\in X_{\Yb}$ with capacity
  $c^{G(\Pcal)}_y$ and connect each $x_e \in X_{T_j}$ to $t_j$ with capacity
  $c^{G(\Pcal)}_e$.
  We run $\Acal$ (as defined in Proposition \ref{prop:oracle}) on $G'_{s_jt_j}(\Pcaltil)$ to find a
  $(1-\eps)$-approximate maximum $s_j$-$t_j$ flow $f'_j$ and
  a $(1+\eps)$-approximate $s_j$-$t_j$ minimum cut with edges $Y'_j$.

  We say an edge $y\in \Yb$ \textit{has become good}, if
  for both $j=1,2$,
  the edge $(s_j,x_y)$ in $G'_{s_jt_j}(\Pcaltil)$
  carries (in the direction from $s_j$ to $x_y$)
  at least $c^{G(\Pcal)}_y / 4$ units of flow in $f'_j$.
  
  We then do a case analysis as follows:
  \begin{enumerate}[label=\textbf{Case \arabic*},leftmargin=55pt]
      \item If for both $j=1,2$, we have
      \begin{align*}
          c^{G'_{s_jt_j}(\Pcaltil)}(Y'_j) \geq \frac{9}{10}
          c^{G(\Pcal)}(\Yb),
      \end{align*}
      then move the edges in $\Yb$ that have become good to $\Yg$.
      \label{casea}
      \item Else, let $j^*\in \setof{1,2}$ be such that
      \begin{align*}
          c^{G'_{s_{j^*}t_{j^*}}(\Pcaltil)}(Y'_{j^*}) < \frac{9}{10}
          c^{G(\Pcal)}(\Yb).
      \end{align*}
      Then let $\Yb \gets Y_{j^*}$, where $Y_{j^*}$ is constructed as follows:
      \begin{enumerate}
          \item Let $Y''_{j^*}$ be $Y'_{j^*}$ with edges incident on $u_X$ removed.
          \item Include any edge $e\in E(G(\Pcal))$ in $Y_{j^*}$ if $x_e$ is incident on at least one edge in $Y''_{j^*}$.
      \end{enumerate}
      \label{caseb}
  \end{enumerate}

  \begin{claim}
      In~\ref{casea}, the edges $y \in \Yb$'s that
      have become good contribute
      at least $1/5$ of the total capacity of $\Yb$.
  \end{claim}
  \begin{proof}
      For $j\in\setof{1,2}$,
      consider the flow $f'_j$ in $G'_{s_jt_j}(\Pcaltil)$.
      The flow value is at least $\frac{4}{5} c^{G(\Pcal)}(\Yb)$ given the conditions
      of \ref{casea} and that $f'_j$ is a $(1-\eps)$-approximate maximum flow.
      Therefore at least $3/5$ (in capacity) of $y\in\Yb$'s
      (call them $\Yg^{(j)}$)
      satisfies that the edge $(s_j,x_y)$ in $G'_{s_jt_j}(\Pcaltil)$
      carries (in the direction from $s_j$ to $x_y$)
      at least $c^{G(\Pcaltil)}_y / 4$ units of flow in $f'_j$,
      since otherwise the total flow value would be
      \begin{align*}
          \kh{ \frac{3}{5} + \frac{2}{5}\cdot \frac{1}{4} } c^{G(\Pcal)}(\Yb)
          < \frac{4}{5} c^{G(\Pcal)}(\Yb),
      \end{align*}
      a contradiction.
      The claim then follows by taking the intersection of $\Yg^{(1)}$ and $\Yg^{(2)}$.
  \end{proof}

  \begin{claim}
      In \ref{caseb}, we have
      \begin{align*}
          c^{G(\Pcal)}(Y_{j^*}) \leq c^{G'_{s_{j^*}t_{j^*}}(\Pcaltil)}(Y'_{j^*}).
      \end{align*}
      Moreover, $X_{Y_{j^*} \union \Yg}$ separates $X_B$ from $X_F$.
  \end{claim}
  \begin{proof}
      For the capacity condition, notice that we can simply charge the capacity of every edge
      in $e\in Y_{j^*}$ to the capacity of an edge in $Y''_{j^*}$ that is incident on $x_e$.
      Let $\Yg'$ denote the set of all split edges of edges in $\Yg$ in $G'(\Pcal)$.
      Define edge set $Z_{j^*}$ by including edges $e\in E(G(\Pcal))$ for which
      $x_e$ is incident on at least on edge in $Y'_{j^*}$.
      Since $Y'_{j^*}$ is an $s_j$-$t_j$ cut in $G_{s_jt_j}'(\Pcaltil)$,
      $X_{Z_{j^*}}$ separates $X_{\Yb}$ from $X_{T_{j^*}}$.
      Since $X_{\Yb}\union X_{\Yg}$ separates $X_B$ and $X_F$,
      $X_{Z_{j^*}} \union X_{\Yg}$ also separates $X_B$ and $X_F$.
      Now notice that the only difference between $Y_{j^*}$ and $Z_{j^*}$ is that
      the former does not have the edges $e$ for which the only incident edge of $e$ in $Y'_{j^*}$ is $(u_X,x_e)$.
      However, these edges are only useful for moving between $X_B$ through $u_X$, and thus does not affect whether
      or not $X_B$ is separated from $X_F$. Hence $X_{Y_{j^*} \union \Yg}$ also separates $X_B$ from $X_F$.
  \end{proof}

  Therefore, if we repeat the above refinement process,
  each time we shrink the capacity of $\Yb$ by a constant factor.
  So after $O(\log n)$ iterations,
  we are done finding a desired set of cut edges $Y$ that can route to both
  $B$ and $F$ with the desired congestion. Moreover,
  in each iteration,
  we can determine in parallel which edges
  in $\Yb$ have become good
  by simply examining our flow solution.
  The lemma thus follows.
\end{proof}

We now prove Lemma \ref{lem:partaa}.

\begin{proof}(Lemma \ref{lem:partaa})
    The proof is essentially the same as Lemma 3.2 of \cite{RackeST14}, with their second partitioning step replaced by ours.
    Using Lemma~\ref{lem:partb} with $\psi = 1$, we can find an edge set $C$ such that $X_C$ separates $X_A$ and $X_B$ in $G'(\Pcal)$ and
    there exists a desired multicommodity flow routing from $X_C$ to $X_A$ as guaranteed by Lemma~\ref{lem:partb}.
    Moreover, the total capacity of $C$ is at most twice of the total capacity of $B$.
    Using the fact that $X_C$ separates $X_A$ and $X_B$ in $G'(\Pcal)$,
    and that $A\union B$ induces a $3/4$-balanced partition,
    due to Claim 1 in the Proof of Lemma 3.2 in \cite{RackeST14}, either $A\union C$ or $B\union C$ induces
    a $3/4$-balanced partition. If it is $A\union C$, then we have achieved the second case of the lemma.
    If it is $B\union C$, we have achieved the first case of the lemma.
\end{proof}

\subsection{Recursive Construction of Congestion Approximators.}
\label{sec:rc-ca}

We now show how to recursively construct a congestion approximator for $G$
using our two partitioning steps above.
Note that imperatively,
we use a lower reweighting factor for the boundary edges in $\PA$ than in $\PB$
($\log^{-12} n$ vs. $\log^{-4} n$) to avoid a blowup in the congestion when doing
the routing ``fixing'' phase, which is needed because we route on
contracted subgraphs rather than vertex-induced subgraphs.

\begin{algorithm}{$\RCA(\CS{G}{\Pcal})$}\\
    \label{alg:congestion-sapprox-details}\textbf{Input:} Reweighted contracted subgraph $\CS{G}{\Pcal}$, where $\Pcal = (S, 1)$. \\
    \nonl \textbf{Output:} A hierarchical tree decomposition tree $\TT$ of $G$.\\
    \textbf{Procedure:}
    \begin{algorithmic}[1]
        \If{$\CS{G}{\Pcal}$ only has one uncontracted vertex, plus the supernode $u_{X}$}
        \State Return the single uncontracted vertex as our congestion approximator and abort.
        \EndIf
        \State Run $\PA$ on $\CS{G}{\Pcal_A := (S,\log^{-12} n)}$
        to get a partition $Z_1,Z_2$ of
          the uncontracted vertices $V(\CS{G}{\Pcal})\setminus \setof{u_X}$,
          with $F$ being the inter-cluster edges between $Z_1,Z_2$.
        \State Run $\PB$ on $\CS{G}{\Pcal_B := (S,\log^{-4} n)}$ with $\psi = \log^{-6} n$ to obtain a set of cut edges $Y$ separating boundary edges $B$
        from inter-cluster edges $F$.
        \State\nonl \algcomment{Recall that $B$ is the set of edges incident to $u_X$ in $G(\Pcal_B)$.}
      \State Let $L_1$ denote the vertices in $V(\CS{G}{\Pcal})$ that cannot reach any edge in $B$ after the removal of $Y$,
      and let $L_2$ denote the other vertices.
      \State \nonl \algcomment{$L_{2}$ are the vertices that can reach $B$ after the removal of $Y$}
      \State Let $\Zcal:= \setof{ \Zcal_1 := Z_1\intersect L_1, \Zcal_2 := Z_2\intersect L_1, \Zcal_3 := Z_1\intersect L_2,
          \Zcal_4 := Z_2\intersect L_2}$ be
          the partition obtained by taking the intersection of the partitions returned by $\PA$ and $\PB$.
        \For{$\Zcal_{i} \in \Zcal$}
        \State Recursively run $\RCA$ on $\CS{G}{\Pcal_{i}}$ to get a tree $\Tcal_{i}$, where $\Pcal_i = (\{S_i:= \Zcal_i\}, 1)$
        \EndFor
        \State Create a new tree node $r$ as the root of our tree, and two other nodes
          $v_{L_1}, v_{L_2}$ as $r$'s children corresponding to vertex subsets $v_{L_1},v_{L_2}$ respectively.
          We also create nodes $v_1,v_2,v_3,v_4$ corresponding to $\Zcal_1,\Zcal_2,\Zcal_3,\Zcal_4$ respectively,
          and let $v_1,v_2$ be the children of $L_1$ and let $v_3,v_4$ be the children of $L_2$.\label{line:combine1}
        \State For each $v_i$, let $v_i$ be the root of the tree $\Tcal_{i}$ that
          we computed using recursive calls to $\RCA$. In other words,
          the children of $L_1$ are the roots of $\Tcal_1$ and $\Tcal_2$,
          and the children of $L_2$ are the roots of $\Tcal_3$ and $\Tcal_4$.
          \label{line:combine2}
        \State For each tree node corresponding to a subset of vertices $S\subset V(G)$,
        weight the tree edge that connects this tree node
        to its parent by the total capacity of the cut $(S,V(G)\setminus S)$ in $G$.
        \State \Return{The tree rooted at $r$ constructed above.}
    \end{algorithmic}
\end{algorithm}
\bigbreak
Similar to~\cite{RackeST14}, our congestion approximator will also be
a hierarchical decomposition tree
of the graph.
For a tree node corresponding to a subset of vertices $S\subset V$, the tree edge that connects this tree node
to its parent will have capacity equal to the total capacity of edges leaving $S$ in $G$
(i.e. the capacity of the cut $(S,V(G)\setminus S)$ in $G$).

We give the pseudocode of our construction of a congestion approximator of $G$ below.
As we highlighted at the beginning of this section,
this algorithm implements our Algorithm~\ref{alg:improve-ca} modulo
being able to solve approximate maximum flows on contracted subgraphs,
which we will show how to do in the next section (Section~\ref{sec:oracle}).
During the execution of our algorithm,
we do not reweight the graph that we recurse on, but only do reweighting
when running the two partitioning steps.
We slightly abuse notations by writing a number $\gamma$ to denote
a constant reweighting function that evaluates to $\gamma$ on every boundary edge.
Initially, we call the algorithm on the entire graph with $X = \emptyset$.

\begin{theorem}\label{thm:RCA}
  The algorithm $\RCA(G)$ has $O(m\, \polylog(n))$ total work and $O(\polylog(n))$ depth.
  Moreover, the output tree $\Tcal = \RCA(G)$ is an $O(\log^9 n)$-congestion approximator of $G$ with high probability.
\end{theorem}

\begin{proof}
  The total work and depth follows easily from the performance of the two
  partitioning steps above.
  We thus focus on proving that the returned tree $\Tcal$ is an $O(\qual)$-congestion approximator.
  Specifically, we show that
  \begin{enumerate}
    \item  Any multicommodity flow demands that can be routed in $G$ with congestion $1$
      can also be routed on the tree with congestion at most $1$. \label{item:ca1}
    \item Any multicommodity flow demands that can be routed on the tree with congestion $1$
      can also be routed in $G$ with congestion $O(\log^9 n)$. \label{item:ca2}
  \end{enumerate}
  Here~\ref{item:ca1} is clear, since the tree cut induced by
  each tree edge corresponds to a cut in the original graph with exactly the same
  capacity, and thus the set of these tree cuts are a subset of the cuts in the original graph.
  We then prove~\ref{item:ca2}.
  Consider that there is a demand $D_{st}$ between each vertex pair $s,t\in V(G)$ such that
  these demands $D_{st}$'s can be routed simultaneously on the tree with congestion
  at most $1$.
  We then describe a routing scheme to show that they can also be simultaneously routed in
  $G$ with congestion $O(\log^9 n)$.
  Our routing scheme will be
  invoking an essentially same routing routine in~\cite{RackeST14},
  plus an additional routing \textit{fixing} phase,
  which is necessary due to the fact that we work in contracted subgraphs rather than vertex-induced subgraphs.

  To illustrate a routing between $s$ and $t$ of flow value $D_{st}$, we let each of $s,t$ send out a message of size
  $D_{st}$.
  We will thus use the terms {\em flow} and {\em message},
  as well as {\em message passing} and {\em flow routing}, interchangeably.
  We will move these messages up along the hierarchical decomposition tree by repeatedly
  moving the messages from the boundary edges of the current cluster to the boundary edges of the parent cluster.
  If at some point, to some edge, we have routed the same amount of flow from
  $s$ and $t$,
  then we can construct a routing from $s$ to $t$ of this flow amount by reversing
  the message trail of $t$; whenever this happens in our routing process,
  we discard these paired-up messages and never consider them in subsequent routing steps.
  Crucially, we move these messages for different $s,t$ pairs \textit{simultaneously}.

  We will specify this message passing process inductively.
  Note that in each recursive call of the tree construction,
  we always do two steps of partitioning and obtain a two-level tree before recursing on the leaves of the two-level tree corresponding to smaller clusters.
  Let us call the two levels in each single recursive call a {\em level block}.
  We will route the messages level block by level block.
  For each level block with root being $S=V(G)\setminus X$,
  we specify a routing in the graph $\CS{G'}{\Pcal = (S,1)}$,
  the subdivision of the contracted subgraph.
  When we say we route some flow to an edge in $G(\Pcal)$ (the contracted subgraph
  without subdivision),
  we mean we route the flow to the split vertex of that edge.
  We will maintain the following invariant:
  \begin{quote}
    \textbf{Invariant:}
    \textit{After we are done with the routing for a level block with root being
    $S = V(G)\setminus X$, all messages originating from
    vertices in $S$
    either have been discarded, or
    reside on the boundary edges
    (i.e. the ones that go between $S$ and $u_{X}$)}.
    \hfill $(\star)$
  \end{quote}

  We will also do the routing ``fixing'' inductively.
  Specifically, in each inductive step,
  even before we specify the routing,
  we first fix the routing for each smaller cluster $\Zcal_i$ that
  we specified in the smaller contracted subgraph $\CS{G'}{\Pcal_i = (\Zcal_i, 1)}$,
  so that they become routings in the bigger contracted subgraph
  $\CS{G'}{\Pcal = (S,1)}$.
  Only after the fixing is done for each $\Zcal_i$ do we proceed with the routing
  of the messages in $\CS{G'}{\Pcal}$.

  As a result, our inductive step consists of two phases,
  namely a \textit{fixing phase}, and a \textit{routing phase},
  where the routing phase will be essentially the same as in~\cite{RackeST14}.
  We now fully describe our inductive routing scheme below, as well
  as analyze the congestion caused. Notice that, throughout the proof,
  we always analyze the congestion of edges with respect to their
  \textit{original} capacities in $G$ (without considering the reweighting factors
  that we use in contracted subgraphs).
  We will \hl{highlight all the congestion} we calculate along the way
  of presenting our routing scheme.
  Since we route in subdivision graphs, we will need to distinguish, for a boundary edge
  $e$, the congestion on its split edge incident on $u_X$, and the congestion on its other split edge incident on $S$.
  We will call the former \hl{outer congestion}, and the latter \hl{inner congestion}.
  
  Note that for our goal of proving the obtained tree is a congestion approximator,
  it suffices to prove the existence of such a low-congestion routing. We will
  then plug in the tree into Sherman's algorithm to compute approximate maximum flows.

  \paragraph{Base Case.} Initially, as the base case of the inductive routing process,
  for each demand $D_{st}$, $s$ and $t$ will each distribute the total amount of
  flow $D_{st}$ uniformly
  to their outgoing edges - a.k.a. the boundary edges of these singleton clusters -
  with each edge getting flow proportional to its capacity.
  Thus we have established the invariant $(\star)$.
  
  This routing causes congestion at most $2$ on the edges of $G$, 
  since a node cannot send or receive
  a total flow amount greater than its weighted degree
  given that the demands are routable with congestion $1$,
  and an edge is incident on two vertices.

  \paragraph{Inductive Step: Fixing Phase.}
    For $i = 1,2,3,4$,
    given a routing in $\CS{G'}{\Pcal_i}$, we aim to obtain another routing
    in the bigger contracted subgraph $\CS{G'}{\Pcal}$ that route the same demands.
    Recall that $S$ denotes the set of nodes in consideration, i.e., $S=V\setminus X=\Zcal_1\cup\Zcal_2\cup\Zcal_3\cup\Zcal_4$.
    Suppose first we take the exact same given routing in $\CS{G'}{\Pcal_i}$
    and simply expand (namely, undo the contractions of)
    the vertices in $S\setminus \Zcal_i$,
    obtaining a routing in $\CS{G'}{\Pcal}$.
    Due to the expanding we have done,
    we may have created extra deficits and excesses on the vertices in
    $(S\setminus \Zcal_i)\union\setof{u_X}$, which we now have to fix.
    To this end, it suffices to give for each demand
    a routing with the deficits/excesses on $(S\setminus \Zcal_i)\union\setof{u_X}$ switched.
    These routings should be simultaneously achievable with low congestion.

    Let $\alpha_i \geq 0$ be the maximum outer congestion on the boundary edges of $G(\Pcal_i = (\Zcal_i,1))$,
    and let $\alpha := \max\setof{\alpha_1,\alpha_2,\alpha_3,\alpha_4}$.
    Let $\gamma_i \geq 0$ be the maximum inner congestion on the boundary edges of $G(\Pcal_i = (\Zcal_i,1))$,
    and let $\gamma := \max\setof{\gamma_1,\gamma_2,\gamma_3,\gamma_4}$.
    \begin{claim}\label{claim:fixing}
        We can fix the routings in $G(\Pcal_i)$'s so that the become routings in $G(\Pcal)$ with the same demand while
        causing outer congestion of boundary edges \hl{$O(\alpha \log^{-3} n)$},
        inner congestion of boundary edges \hl{$O(\alpha \log n)$}, and congestion on edges inside $S$
        \hl{$O(\alpha \log^8 n)$}.
    \end{claim}
    \begin{proof}
    For $\Zcal_3, \Zcal_4$, we fix the extra deficits/excesses by a routing from $Y\setminus B$ to $B$, followed by a mixing
    to uniform (with each edge getting flow proportional to its capacity) over $B$.
    The routing causes outer congestion of boundary edges \hl{$O(\alpha \log^{-9} n)$},
    inner congestion of boundary edges \hl{$O(\alpha \log^{-3} n)$},
    and congestion of edges inside $S$ \hl{$O(\alpha \log n)$}.
    The mixing causes congestion \hl{$O(\alpha \log^{-3} n)$}. The last
    congestion is because each edge in $B$ receives \hl{ $O(\alpha \log^{-3} n)$ } times its capacity units of flow,

    For $\Zcal_1,\Zcal_2$, we fix the extra deficits/excesses by a routing from $Y$ to $F$, followed by a mixing
    to uniform (with each edge getting flow proportional to its capacity) over $F$.
    The routing causes outer congestion of boundary edges \hl{ $O(\alpha \log^{-5} n)$ },
    inner congestion of boundary edges \hl{ $O(\alpha \log n)$ }, and congestion of edges inside $S$
    \hl{ $ O(\alpha \log^ 5n)$}. The mixing causes congestion on boundary edges \hl{$O(\alpha \log^{-4} n)$},
    and congestion on edges inside $S$ \hl{$O(\alpha \log^8 n)$}.
    \end{proof}

  \paragraph{Inductive Step: Routing Phase.}
  This phase is essentially the same as in~\cite{RackeST14}, except that
  we do the routing in the contracted subgraph $\CS{G'}{\Pcal}$ rather than in the vertex
  induced subgraph, and therefore we also have to analyze the congestion
  on the boundary edges $B$.
  
  For the routing phase,
  we divide the messages originating in $S$
  that haven't yet been discarded (thus still need to be rounted)
  into following
  types:
  \begin{enumerate}[label=\textbf{Type \arabic*},leftmargin=55pt]
      \item The messages corresponding to demands $D_{st}$'s for which
      both $s,t$ lie in $L = \Zcal_1\union \Zcal_2$ (vertices
      not reachable from $B$ after removing $Y$).
    \label{type1}
      \item The messages corresponding to demands $D_{st}$'s
      for which one of $s,t$ lies in $L$ and the other lies outside of $L$.
      \label{type2}
      \item The messages corresponding to demands $D_{st}$'s
      for which both $s,t$ lie outside $L$ (but at least one of $s,t$ lies
      in $R = S\setminus L$, as otherwise they are not considered in this level block).
      \label{type3}
  \end{enumerate}

  Recall that, by the invariant $(\star)$ that we maintain,
  all three types of messages reside on the boundary edges
  of the smaller clusters $\Zcal_i$'s.
  Note that, by the construction of the $\Zcal$'s,
  the boundary edges of clusters $\Zcal_1,\Zcal_2$ are subsets of
  $F\cup Y$, while the boundary edges of $\Zcal_3,\Zcal_4$ are subsets of $Y\cup B$.
  The goal of our inductive routing step is as follows:
  \begin{enumerate}[label=\textbf{Goal \arabic*},leftmargin=55pt]
      \item For~\ref{type1} messages, we show that
      we can simultaneously pair up the messages from $s$ and the ones from $t$.
      \label{goal1}
      \item For~\ref{type2} and~\ref{type3} messages, we show how to
      route them to the boundary edges $B$ of the bigger cluster $S$.
      \label{goal2}
  \end{enumerate}
  Here~\ref{goal1} means that we can discard all~\ref{type1} messages
  afterwards, since by doing so we have already found a
  simultaneous routing of the corresponding demands $D_{st}$'s;
  whereas~\ref{goal2} means that
  after our routing,
  \ref{type2} and \ref{type3} messages always reside on the boundary edges of $S$.
  These together imply that we have
  maintained our invariant $(\star)$.
  
  We now describe how we achieve these goals, as well as analyze the congestion
  caused. Throughout, let $\beta_i \geq 0$ be the maximum ratio of the amount
  of flow received to the (original) edge capacity over the boundary edges of $\Zcal_i$,
  when we finish
  routing in $\CS{G}{\Pcal_i}$.
  Then let $\beta = \max\setof{\beta_1,\beta_2,\beta_3,\beta_4}$.

  \paragraph{Inductive Step: Routing \ref{type1} Messages.}
  By our invariant $(\star)$, \ref{type1} messages reside on the boundary edges
  of $\Zcal_1\union \Zcal_2$, which are $Y\union F$.
  We first route \ref{type1} messages that reside on $Y\setminus F$ to $F$.
  By the guarantee of $\PB$, this can be done with
  outer congestion of boundary edges \hl{ $O(\beta \log^{-5} n)$ },
  inner congestion of boundary edges \hl{ $O(\beta \log n)$ },
  and congestion of edges inside $S$ \hl{ $O(\beta \log^5 n)$ }.
  
  After the routing,
  all \ref{type1} messages reside on $F$,
  and the flow that each edge in $F$ carries is
  at most \hl{ $2\beta + O(\beta \log^5 n) = O(\beta \log^5 n)$ } of its capacity.
  We then simultaneously mix each of \ref{type1} messages uniformly over $F$,
  in the sense that each edge in $F$ gets an amount proportional to its capacity.
  After the mixing, we have successfully paired up all \ref{type1} messages
  and thus can discard them all. %
  
  This mixing causes
  congestion on boundary edges \hl{ $O(\beta \log^{-4} n)$ },
  and congestion on edges inside $S$ \hl{ $O(\beta \log^8 n)$ }.
  Combined this with the congestion we get in the $Y\setminus F\to F$ routing,
  the added outer and inner congestion of $B$ in $\CS{G}{\Pcal}$
  is at most \hl{$O(\beta \log^{-4} n)$} and \hl{$O(\beta \log n)$}, whereas the added congestion on edges inside $S$ in
  {$\CS{G}{\Pcal}$} is at most \hl{$O(\beta \log^8 n)$}.

  \paragraph{Inductive Step: Routing \ref{type2}, \ref{type3} Messages.}
  By our invariant $(\star)$, both \ref{type2} and \ref{type3} messages reside
  on the edges in $Y\union B\union F$.
  We first route the messages that reside on $F\setminus Y$ to $Y$.
  This routing consists of first mixing the message uniformly on $F$ and
  then reversing the routing from $Y$ to
  a uniform distribution over $F$, whose existence is guaranteed by $\PB$.
  After the routing, edges in $F\setminus Y$ carry no flow on them,
  and each edge in $Y$ carries flow that is at most
  \hl{$2\beta + 1$} of its capacity in $\CS{G}{\Pcal}$.

  By a similar analysis as in the routing of \ref{type1} messages,
  the routing can be done with
  outer congestion of boundary edges \hl{ $O(\beta \log^{-5} n)$ },
  inner congestion of boundary edges \hl{ $O(\beta \log n)$ },
  and congestion of edges inside $S$ \hl{ $O(\beta \log^5 n)$ }.

  We then route message on $Y\setminus B$
  to $B$.
  The routing causes outer congestion of boundary edges \hl{$O(\beta \log^{-9} n)$},
  inner congestion of boundary edges \hl{$O(\beta \log^{-3} n)$},
  and congestion of edges inside $S$ \hl{$O(\beta \log n)$}.
  After the routing, each edge in $B$ carries flow that is at most
  \hl{$\beta + (2\beta + 1) O(\log^{-3} n) = (1 + O(1/\log^{3} n)) \beta$}
  of its capacity in $\CS{G}{\Pcal}$.

  \paragraph{Total Congestion Analysis.}
  We first obtain an upper bound on $\beta$, the
  total amount of flow carried by a boundary edge divided
  by its capacity in $G$.
  Initially, we have $\beta = O(1)$ in our base case.
  Then in each inductive step, $\beta$ grows by $1 + O(1/\log^{3} n)$,
  as we have analyzed above when routing \ref{type2}, \ref{type3} messages,
  which is the only place where we route flows to $B$.
  Since the depth of the tree
  is $O(\log n)$, we have $\beta = O(1)$ throughout.

  We next obtain an upper bound on $\alpha$,
  the total outer congestion of any boundary edge,
  and $\gamma$, the total inner congestion of any boundary edge,
  In the fixing phase, the total outer congestion and inner congestion added on a boundary edge is
  $O(\alpha_i / \log^3 n)$ and $O(\alpha \log n)$ respectively.
  In the routing phase, the total outer and inner congestion added on a boundary edge is
  $O(\beta \log^{-4} n)$ and $O(\beta \log n)$.
  Therefore, by a simple induction and the fact that
  the tree has $O(\log n)$ depth, the total outer and inner congestion
  accumulated over descendant routing steps
  on a boundary edge is bounded by $O(1)$ and $O(\log^2 n)$, respectively.
  
  Finally, in the routing phase and the fixing phase, the total congestion added on the edges inside $S$
  is bounded by $O(\beta \log^8 n) + O(\alpha \log^8 n) = O(\log^8 n)$.
  Since each edge appears $O(\log n)$ times as inside $S$, the total congestion
  accumulated is $O(\log^9 n)$.
  This finishes the proof of the theorem.
\end{proof}

%% file: oracle.tex
We finally describe the implementation details of 
the approximate maximum flow oracle $\Acal$ in Section~\ref{sec:new-rst} whose
performance guarantees are summarized in Proposition~\ref{prop:oracle}. We would like to remind the reader that this oracle, at its heart, utilizes Sherman's framework to compute approximate maximum flows which requires as input $\polylog n$-congestion approximators for the flow instance upon which it is invoked. While we defer the \pram implementation details of Sherman's algorithm to Appendix~\ref{sec:shermans}, we discuss in this section, three important routines for constructing congestion approximators required for the contracted subgraphs generated by our new framework for constructing high-quality congestion approximators. These are $(i)$ (partially) compressing the global congestion approximator to obtain one for the contracted subgraphs, $(ii)$ obtaining a near-linear work, low-depth implementation of Sherman's algorithm when given as input these (partially) compressed congestion approximators, and $(iii)$ constructing congestion approximators for contracted subgraphs with an arbitrarily attached super-source and super-sink. We would crucially like to remind the reader that at all points in our overall approach, we build and maintain congestion approximators for \emph{subdivision graphs} and \emph{contracted subdivision graphs}, as these are precisely the flow instances upon which the max-flow oracle is invoked.           

\subsection{Compressing Congestion Approximators.}
\input{ca-contraction.tex}

%% file: ca-contraction.tex
\label{sec:ca-contraction}

We start by describing the \texttt{ca-contraction} subroutine of Algorithm \ref{alg:improve-ca} that given a contracted subgraph and a congestion approximator for a much larger graph, compresses it to have size proportional to that of the contracted subgraph.
Specifically, we are given as input a $O(\polylog n)$-congestion approximator $R$ for some larger graph $G=(V,E,c)$ along with a subset $S\subset V$ of $k$ vertices that have the following special property: all edges that leave the set $S$ are well-linked in $G$. We prove such a property for the graphs we encounter:
   \begin{claim}\label{claim:well-linked-part}
        In any $G(\Pcal = (S,1))$ with $X = V(G)\setminus S$ and $S$ being a cluster corresponding to a leaf of a level block
        of the tree constructed by $\RCA$,
        the boundary edges incident on $u_X$ are $\Omega(1/\log^9 n)$-well-linked in $G$.
    \end{claim}
    \begin{proof}
        Note that in $G(\Pcal)$, these edges are $1$-well-linked.
        We then consider the multicommodity flow routing between them in $G(\Pcal)$.
        We now fix this routing using our fixing step with performance guaranteed by Claim~\ref{claim:fixing} in a bottom-up
        manner until we have converted the routing into a valid one in $G$ with the same demands.
        Then by Claim~\ref{claim:fixing} the total congestion is $O(\log^9 n)$, implying the claim.
    \end{proof}

Our goal is to output a $O(\polylog n)$-congestion approximator of size $O(k\log n)$ for the contracted subgraph $G(S)$ where $V \setminus S$ is compressed into a single node\footnote{with some edges properly reweighted; see Section \ref{sec:rc-ca}. Since this does not affect the algorithm or analysis of this section, we assume this reweighting has been done prior to the call to \texttt{ca-contraction}.}, and the aforementioned well-linkedness property will be crucial in achieving it. This compression step is critical in order to achieve near-linear work; while the original congestion approximator $R$ of $G$ is also a $O(\polylog n)$-congestion approximator for the contracted graph $G(S)$, its size may be very large relative to that of $G(S)$, and naively using it for computing approximate max-flows in $G(S)$ would substantially blow up the total work. This subroutine precisely addresses this issue by reducing the size of the congestion approximator without degrading its quality substantially.

If we were simply to contract into a \textit{single node} all the nodes in $R$ which represent only subsets of $V \setminus S$, the resulting graph would not be a tree, and so we would not be able to use it for computing max-flows with Sherman's algorithm.
As such, we must use an alternative approach to shrink the size of $R$ that preserves the tree structure.
We accomplish this with the following algorithm, which uses all three properties guaranteed by the transformation in Section \ref{sec:rst-on-trees}: $R$ is a hierarchical decomposition and also a binary tree of depth $O(\log n)$.
As $R$ is a hierarchical decomposition, for all $v\in S$, there is a leaf of $R$ corresponding to the set containing only $v$.
Assign a node weight of 1 to these leaves, and assign a weight of 0 to all other nodes in $R$.
Then, run the subtree sum algorithm of Theorem \ref{thm:subtree} and contract all subtrees whose subtree sum is 0.
As before, for a graph $G$ and set $S \subseteq V(G)$, let $G(S)$ be $G$ with $V(G) \setminus S$ contracted into a single node.
We use the fact that $R$ is binary and has $O(\log n)$ depth to bound the size of resulting tree.

\vspace{5pt}

\begin{algorithm}{$\texttt{ca-contraction}(R,S)$}\\
    \label{alg:ca-contraction}\nonl \textbf{Input:} $\alpha$-congestion approximator $R$ (for a graph $G=(V,E,c)$), which is a hierarchical decomposition and a binary tree of depth $O(\log n)$; a subset $S\subset V$ of $|S|=k$ uncontracted vertices. \\
    \nonl \textbf{Output:} Tree $R'$ with $O(k\log n)$ nodes that is an $(\alpha\cdot\polylog n)$-congestion approximator for $G(S)$. \\
    \textbf{Procedure:}
    \begin{algorithmic}[1]
        \State Assign a node weight of 1 to each leaf $u$ of $R$ such that $P_u=\{v\}$ for some $v\in S$, where $P_u\subseteq V$ is the partition of vertices the node $u$ corresponds to in the hierarchical decomposition.
        \State Assign a node weight of 0 to all other nodes of $R$.
        \State Compute the subtree sums with respect to these weights using the algorithm of Theorem \ref{thm:subtree}.
        \State From top-down, contract each subtree whose sum is $0$ into a single supernode.
    \end{algorithmic}
\end{algorithm}

It is important to note that this algorithm does \textit{not} require all leaves of the input $R$ to correspond to single vertices, as is the case in a standard hierarchical decomposition tree.
In Algorithm \ref{alg:improve-ca}, we contract congestion approximators that themselves have been contracted from a previous tree, and so not all leaves may correspond to single vertices.

\begin{lemma}
    \label{lem:ca-contraction}
    The $\texttt{ca-contraction}(R,S)$ subroutine has depth $O(\log n)$ and $O(|R|\log n)$ total work, and outputs a tree $R'$ with $O(k\log n)$ nodes.
\end{lemma}
\begin{proof}
    By Theorem \ref{thm:subtree} and the fact that $R$ has $O(\log n)$ depth, the algorithm has $O(\log n)$ depth and $O(|R|\log n)$ total work.
    To bound the size, first note by the fact that $R$ is a hierarchical decomposition, if $u_1,\ldots ,u_q$ are the nodes of $R$ at some level $i$, the $P_{u_1},\ldots ,P_{u_q}$ are a partitioning of $V$.
    So, at each level of $R$, there can be at most $k$ nodes $u$ such that $P_u\cap S \neq \emptyset$, and so at most $k$ nodes can remain uncontracted at each level after contraction.
    Moreover, since $R$ is a binary tree, every uncontracted node can have at most two supernodes as children, and, by definition, all supernodes are leaves. 
    Since $R$ has depth $O(\log n)$, it thus follows that the size of the resulting tree $R'$ is $O(k\log n)$, as desired.
\end{proof}

It thus remains to show that the contracted tree $R'$ can still be used to route flow with low congestion in the graph with $V \setminus S$ contracted into a single node.
$R'$ is constructed by contracting some subtrees of $R$; call the root of these contracted subtrees (along with any remaining leaves corresponding to single nodes from $V \setminus S$) $u^{*}_1,\ldots ,u^{*}_q$ and let $T_i=P_{u^{*}_i}$, ordering arbitrarily.
So, $T_1\cup\ldots \cup T_q = V \setminus S$.
Let $\bar{G}(S)$ be the graph with each $T_i$ contracted to a single node, but these contracted nodes are \textit{not} further contracted.
Since $R'$ is constructed by contracting the nodes corresponding to each $T_i$ in $R$, $R'$ is a $\alpha$-congestion approximator for $\bar{G}(S)$ (where $\alpha$ is the congestion achieved by $R$ for routing on $G$).
Our goal is thus to show that $R'$ can be used as an $(\alpha\cdot\polylog n)$-congestion approximator for $G(S)$.

Let $x$ be the contracted node (i.e.~the node formed by contracting $V \setminus S$ in $G$) in $G(S)$, and suppose we are given a demand vector $b$ on $G(S)$.
Furthermore, for each $i\in [q]$, let $x_i$ be the node in $\bar{G}(S)$ formed by contracting $T_i$.
If we were able to efficiently split the demand $b_x$ into demands for $x_1,\ldots ,x_q$ without inducing much additional congestion, then we would be able to use $R'$ as a congestion approximator for $G(S)$.
Namely, we first convert the demand on $G(S)$ into the corresponding demand on $\bar{G}(S)$, and then use $R'$ to route the flow on $\bar{G}(S)$ which is also a routing on $G(S)$ (by replacing any $x_i$ with $x$).
So, it remains to show that we may indeed split the demand $b_x$ into demands $b'_{x_1},\ldots ,b'_{x_q}$ such that $b_x=\sum_{i\in [q]}b'_{x_i}$ and $b'$ can be routed on $\bar{G}(S)$ with low congestion.
This follows from the fact that we are guaranteed the set of edges leaving $S$ in $G$ are $\Omega(1/\log^9n)$-well-linked (Claim \ref{claim:well-linked-part}), and is formalized in the following lemma.
Importantly, since $R'$ remains an $\alpha$-congestion approximator for $\bar{G}(S)$, for any $Q \subseteq S$, $\texttt{ca-contraction}(R',Q)$ is also an $\alpha$-congestion approximator for $\bar{G}(Q)$\footnote{We slighly abuse notation to refer to $\bar{G}(Q)$ as the graph with the same components contracted as in the tree output of $\texttt{ca-contraction}(R',Q)$}.
As such, the error does not accumulate when repeatedly contracting a congestion approximator, as we do in Algorithm \ref{alg:improve-ca}.

\begin{lemma}
    \label{lem:split-demands-gu}
    Let $b$ be a demand vector on $G(S)$ which can be satisfied with congestion 1.
    Let $W_x=\sum_{(u,x)\in E}c((u,x))$ be the sum of capacities of edges incident on $x$ in $G(S)$, and define $W_{x_i}$ similarly for each $x_i$ in $\bar{G}(S)$.
    Then, with $b'$ the demand vector on $\bar{G}(S)$ such $b'_{u}=b_{u}$ for all $u\in S$ and $b'_{x_i}=W_{x_i}b_{x}/W_x$, $b'$ can be satisfied with congestion $O(\log^9n)$.
\end{lemma}
\begin{proof}
    Consider some flow $f$ in $G(S)$ which satisfies $b$ with congestion 1.
    Let $S=\{(u,x)\mid u\in S\}$ be the set of edges leaving $S$ in $G(S)$, and similarly let $S_i=\{(u,x_i)\mid u\in S\}$ be the set of edges incident on $x_i$ in $\bar{G}(S)$.
    Note that for each $(u,x)\in S$, there exists a corresponding edge $(u,x_i)$ (for some $x_i$) in $\bar{G}(S)$ with the same capacity, by the construction of $x$ and $x_1,\ldots,x_q$.
    So, we may convert $f$ into a flow on $\bar{G}(S)$; call this $f'$.
    Define the flow vector $b^{*}$ such that $b^{*}_{x_i}$ is the net incoming flow to $x_i$ induced by $f'$, and $b^{*}_{u}=b_{u}$ for all $u\in S$.
    $f'$ has congestion 1, by assumption on $b$, and satisfies the demand vector $b^{*}$, so it follows that $b^{*}$ can be satisfied in $\bar{G}(S)$ with congestion 1 as well.
    As $S$ is $\Omega(1/\log^9 n)$-well-linked for each $S$ on which we call \texttt{ca-contraction} in Algorithm \ref{alg:improve-ca} (as shown in Claim \ref{claim:well-linked-part}), each $S_i$ is as well by Proposition \ref{prop:subset-well-linked}.
    Thus, by definition of well-linked and the fact that $b$ and $b^{*}$ differ only on the $x_i$, it follows that $b'$ can be satisfied on $\bar{G}(S)$ with congestion $O(\log^9n)$.
\end{proof}
There is one last point to check: that this splitting procedure can be implemented in $O(\log n)$ depth and $O(|R|)$ work.
This is not as simple as it may initially seem: since there could be potentially $\Theta(k\log n)$ contracted nodes $x_i$ in $\bar{G}(S)$, naively summing the weight on all edges could result in $\Omega(k^2)$ work.
Fortunately, given access to the contracted graph $G(S)$ (which are computed in Algorithm \ref{alg:improve-ca}), we can use subtree sums to implement the demand splitting efficiently.
Iterate through these edges, and for each $(u,v)\in B$, where $B$ is the boundary edges of $G(S)$, such that $u\in S$ and $v\in V \setminus S$, assign a node weight of $c((u,v))$ to leaf of $R$ which corresponds to $v$.
Assigning these weights can be implemented in work $O(|B|)$; this suffices to prove that Algorithm \ref{alg:improve-ca} requires only $\ot{m}$ total work and does not affect the runtime of \texttt{ca-contraction}.
Once the node weights have been assigned, compute the subtree sums for each node in $R$, which can be done in $O(\log n)$ depth and $O(|R|)$ work by Theorem \ref{thm:subtree}.
The sum $W_{x_i}$ used in Lemma \ref{lem:split-demands-gu} is then exactly the subtree sum of the node corresponding to the set $T_i$, allowing us to correctly distribute the demands.

\subsection{Implementing Sherman's Algorithm on the Contracted Subgraph.}

Next, we discuss how we achieve a linear-work, low-depth implementation of Sherman's algorithm on the contracted subgraph $G(S)$, where $S\subset V$ is the set of $|S|=k$ uncontracted vertices, with all other vertices $V\setminus S$ being contracted into a single super-vertex $x$. We encourage the reader to familiarize themselves with the vanilla implementation of Sherman's algorithm outlined in Appendix~\ref{sec:shermans} to obtain a better understanding of the discussion that follows.

Recall from the preceding section, that we are given access to an $\alpha'$-congestion approximator $R'$ for a slightly larger graph $\bar{G}(S)$, where the vertices $V\setminus S$ have been partitioned and contracted into multiple super-vertices $x_1,\ldots,x_q$. In order to use this congestion approximator $R'$ for our desired contracted graph $G(S)$, we need to translate demands on vertices in $G(S)$ into demands on vertices in $\bar{G}(S)$. The optimization problem within the $\texttt{AlmostRoute}$ subroutine of \cite{sherman2013nearly} therefore becomes a minimization problem over the new congestion potential 
\[\phi(f) = \lm(C^{-1}f) + \lm(2\alpha' R'P(Bf-b)),\]
where $B$ is the vertex-edge incidence matrix for $G(S)$, and $P$ is a linear operator that projects any demands $b$ supported over vertices of $G(S)$ to demands $b'$ supported over vertices of $\bar{G}(S)$ as described in Lemma~\ref{lem:split-demands-gu}. Due to the fact that $R'$ is a $O(k\log n)$ size, $O(\log n)$ depth tree, evaluating this potential is easy; its computation remains unchanged from the vanilla case described in Appendix~\ref{sec:shermans}. The only major challenge here is efficiently computing the derivatives of this new potential, specifically, the derivative of the second term
\[\phi_2(f) := \lm(2\alpha' R'P(Bf-b)).\]

  First, observe that the operation $P\cdot B$ effectively constructs a new vertex-edge incidence matrix\footnote{Note that this new vertex-edge incidence matrix $B'$ may not be the same as the actual incidence matrix of $\bar{G}(S)$. However, this is how it effectively appears to Sherman's algorithm when invoked with $R'$ as its input.} $B'$ for $\bar{G}(S)$ from the incidence matrix $B$ in the following way: for every edge $e=(u_1,u_2)$ in $G(S)$ with only uncompressed vertices $u_1,u_2\in S$ as its endpoints, this operation replicates the edge exactly in $B'$, i.e. $B'_{v,e} = B_{v,e} \in \{-1,1\}$, for $v\in \{u_1,u_2\}$ and $0$ otherwise.
  However, for any edge $e=(u,x)$ (or $(x,u)$) with one of its endpoints being the contracted vertex $x$, this operation splits the edge $e$ into fractional copies $e_1=(u,x_1),\ldots,e_q=(u,x_q)$ with copy $e_i$ having fractional value $\rho_{x_i} := W_{x_i}/W_x$ as defined in Lemma~\ref{lem:split-demands-gu}. i.e. for each $i\in [q]$, $B'_{x_i,e_i} = \rho_{x_i}\cdot B_{x,e}$, $B'_{u,e_i} = \rho_{x_i}\cdot B_{u,e}$, and $0$ otherwise. Note that we never consider edges going between two partially contracted vertices $x_i,x_j$, as they are absent in $G(S)$. While we do not explicitly compute this new incidence matrix $B'$ as doing so naively might exceed our linear (in size of $G(S)$) work requirement, it will serve as a useful intermediate object for analyzing the gradients.

  Now let $\mathcal{I}'$ be the set of all cuts considered by our congestion approximator $R'$, and for any cut $i = (S_i,\overline{S_i})\in \mathcal{I}'$, let $y_i = 2\alpha'[R'P(Bf - b)]_i$ be the congestion induced by the residual demands across cut $i$. We have that for any edge $e\in G(S)$, the partial derivative
\[\frac{\partial \phi_2(f)}{\partial f_e} = \sum_{i\in \mathcal{I}'} \frac{\partial \phi_2(f)}{\partial y_i}\cdot \frac{\partial y_i}{\partial f_e} = \sum_{i\in I'} \frac{\exp(y_i)-\exp(-y_i)}{\exp(\phi_2(f))}\cdot \frac{2\alpha' B'_{S_i,e}}{c(S_i,\overline{S_i})},\]
where $c(S_i,\overline{S_i})$ is the capacity of cut $i=(S_i,\overline{S_i})$ in $\bar{G}(S)$ considered in our congestion approximator $R'$, and (with some abuse of notation) $B'_{S_i,e}$ represents the total ``fraction'' of the edge $e$ crossing the cut $(S_i,\overline{S_i})$; for edges $e=(u_1,u_2)$ with only uncompressed vertices as its endpoints, this quantity is $B'_{S_i,e} = \sum_{v\in S_i} B'_{v,e} = \sum_{v\in S_i\cap S} B_{v,e}$, and for edges $e=(u,x)$ (or $(x,u)$) with one of its endpoints being the compressed vertex, this quantity is $B'_{S_i,e} = \sum_{j\in [q]} B'_{S_i,e_j} = \sum_{j\in [q]}\sum_{v\in S_i} B'_{v,e_j} = \sum_{v\in S_i\cap S} B_{v,e} + \sum_{v\in S_i\setminus S} \rho_v\cdot B_{x,e}$, where the set $\{e_j\}_{j\in [q]}$ correspond to the fractional copies of edge $e$ constructed by the operation $P\cdot B$. 

In order to efficiently compute this gradient, we shall again exploit the fact that $R'$ is represented by a rooted hierarchical decomposition tree $T'$ of size $O(k\log n)$ and depth $O(\log n)$. To do so, we use the same node-potential trick as in Appendix \ref{sec:shermans}: for any internal node $j$ in $T'$ (which in turn corresponds to a cut $(S_j,\overline{S_j})$), we define the node potential $\pi_j$ as
\[\pi_j := \sum_{i\in T'_{j,r}} \frac{\exp(y_i)-\exp(-y_i)}{\exp(\phi_2(f))}\cdot \frac{2\alpha' }{c(S_i,\overline{S_i})},\]
where $T'_{j,r}$ denotes the path in $T'$ from node $j$ to the root $r$ of $T'$. Now observe that, following an identical calculation as in Appendix ~\ref{sec:shermans}, the gradient of $\phi_2(f)$ w.r.t. the flow $f_e$ on any edge $e=(u_1,u_2)$ with only uncompressed vertices as its end-points remains unchanged from the vanilla case.
\[\frac{\partial \phi_2(f)}{\partial f_e} = \sum_{i\in T'_{u_1,u_2}} \frac{\exp(y_i)-\exp(-y_i)}{\exp(\phi_2(f))}\cdot \frac{2\alpha B'_{S_i,e}}{c(S_i,\overline{S_i})} = \pi_{u_2}-\pi_{u_1},\]
where $T'_{u_1,u_2}$ denotes the unique path between $u_1,u_2$ in the tree $T'$. For any edge $e=(u,x)$ (or $(x,u)$) with one of its end points being the contracted vertex $x$ that is in turn partitioned into fractional edges $e_1=(u,x_1),\ldots, e_q=(u,x_q)$ in $B'$, observe that this gradient
\begin{align*}\frac{\partial \phi_2(f)}{\partial f_e} &= \sum_{j\in [q]}\sum_{i\in T'_{u,x_j}} \frac{\exp(y_i)-\exp(-y_i)}{\exp(\phi_2(f))}\cdot \frac{2\alpha B'_{S_i,e_j}}{c(S_i,\overline{S_i})}\\
&= \sum_{j\in [q]} \rho_{x_j}(\pi_{x_j} - \pi_{u}) =  \left(\sum_{j\in [q]} \rho_{x_j}\pi_{x_j}\right) - \pi_u,
\end{align*}
where the final equality follows from the fact that $\sum_{j\in[q]} \rho_{x_j}=1$. We can easily compute all these node potentials $\pi_v$ (through a prefix sum over an Eulerian tour of the congestion approximator tree $T'$ starting at its root) and precompute the quantity $\sum_{j\in[q]} \rho_{x_j}\pi_{x_j}$ in the \pram model with $O(|R'|)=O(k\log n)$ work and $O(\log n)$ depth. Since this is all we need, namely, be able to efficiently evaluate congestions and compute gradients over the supplied congestion approximator, we have an efficient implementation of Sherman's algorithm on the contracted subgraph $\bar{G}(S)$. 

\subsection{Computing a Congestion Approximator for \texorpdfstring{$G\cup \{s,t\}$}{G+st}.}
\label{sec:gst}
The final requirement is the following: in the cut-matching game of Section \ref{sec:pcmg}, we need to compute $(1-1/\polylog n)$-approximate max flow on $G$ with the addition of a source $s$ and sink $t$ which are arbitrarily connected to $G$.
As such, we must convert our $\alpha$-congestion approximator $R$ for $G$ into a $O(\alpha~\polylog n)$-congestion approximator $R'$ for $G\cup \{s,t\}$.
In this section, we specify how this is achieved.

\begin{lemma}
    \label{lem:ca-gst}
    Let $R$ be a $\alpha$-congestion approximator for $G$ which is a hierarchical decomposition, and let $s,t$ be two additional vertices connected arbitrarily (and with arbitrary capacities) to $G$.
    Then, there is a $\polylog k$ depth, $\ot{k}$ work \pram~algorithm which computes a $O(\alpha~\polylog n)$-congestion approximator $R'$ for $G\cup \{s,t\}$, where $k=|R|$.
\end{lemma}

Note that $R\cup \{s,t\}$, with $s$,$t$ connected to the leaves of $R$ corresponding to their neighbors, can be used to route any feasible flow on $G \cup \{s,t\}$ with congestion $\alpha$, since $R$ is an $\alpha$-congestion approximator for $G$.
However, $R\cup \{s,t\}$ is not a tree, and thus cannot be readily plugged into Sherman's algorithm
to compute maximum flows in $G\union \setof{s,t}$.
It thus remains to obtain a congestion approximator (that is a tree) for $G \cup \{s,t\}$, which boils down to computing a hierarchical decomposition (as in \cite{RackeST14}) for $R \cup \{s,t\}$.
The algorithm is quite similar to the procedure of Section \ref{sec:rst-on-trees}, but with a few key modifications to account for the nodes $s$ and $t$.

Recall from Section \ref{sec:rst-on-trees} that it suffices to implement the two partitioning steps Partition A and Partition B; we use the same definitions as in Section \ref{sec:rst-on-trees}.

\begin{lemma}
    Partition A\footnote{If $s\in Q$ or $t\in Q$, the partitions are not balanced, which does not meet the exact definition of Partition A.  However, there are only be 2 such paritionings, and so this can only increase the depth of the final tree by 2.} can be implemented in $O(\log |Q|)$ depth and $\ot{|Q|}$ work on any subset $Q$ of nodes of $R \cup \{s,t\}$.
\end{lemma}
\begin{proof}
    We use the following procedure:
    \begin{enumerate}
        \item If $s\in Q$, then output the partition $(\{s\}, Q \setminus \{s\})$
        \item If $t\in Q$ and $s\not\in Q$, then output the partition $(\{t\}, Q \setminus \{t\})$
        \item If $s\not\in Q$ and $t\not\in Q$, then $Q$ induces a subtree of $R$, and we run the procedure of Lemma \ref{lem:part-a-trees}
    \end{enumerate}
    The depth and work are immediate from Lemma \ref{lem:part-a-trees}.
    It remains to show that the edges between the partition that is output are 1-well-linked.
    If $s\in Q$ or $t\in Q$, then all the edges between the outputted partitions are incident on $s$ or $t$ and are thus 1-well-linked.
    If $s\not \in Q$ and $t\not\in Q$, then the edges between the outputted partitions are 1-well-linked by Lemma \ref{lem:part-a-trees}.
\end{proof}

For Partition B, we only need to apply it on partitions without $s$ or $t$, or on partitionings where all boundary edges are incident on either $s$ or $t$, by the construction of Partition A.
As such, Partition B reduces to finding the (exact) min-cut on a tree with a source and sink added, exactly as in Section \ref{sec:rst-on-trees}.
So, we may use the algorithm of Appendix \ref{sec:min-cut-tree} to compute Partition B and the analysis follows from Lemma \ref{lem:part-b-trees}.

The proof of Lemma \ref{lem:ca-gst} is then essentially identical to the proof of Lemma \ref{lem:tree-hd}.

%% file: shortcut.tex
\label{sec:shortcutting}
\newcommand{\din}{\mathtt{deg}_{\mathrm{{in}}}}
\newcommand{\dout}{\mathtt{deg}_{\mathrm{{out}}}}
\newcommand{\Nin}{N_{\mathrm{{in}}}}
\newcommand{\Nout}{N_{\mathrm{{out}}}}
\newcommand{\iin}{\mathrm{in}}
\newcommand{\out}{\mathrm{out}}
\newcommand{\rem}{\mathrm{rem}}

We describe our \pram flow decomposition subroutine in this section, and we will begin by specifying the relevant notation.  
Consider an $s$-$t$ flow
$f$ specified by a weighted, directed graph $H = (V,E,f)$ containing a source vertex $s\in V$, and sink vertex $t\in V$, with the flow on any edge given by 
$f: E \to \mathbb{R}^{+}$. Note that this flow network $H$ is restricted to only the subset of edges that carry \emph{positive} flow, and will be iteratively updated by our algorithm as we make progress towards our flow-decomposition objective. Moreover, while the initial graph $H$ specifying the flow $f$ does not contain any parallel edges, such edges will inevitably end up being created in our flow-decomposition process. Therefore in this section, we will more generally deal with \emph{multigraphs}, whose edges are assumed to be uniquely indexed. Let $|f| := \sum_{(s,v)\in E} f_{(s,v)} - \sum_{(v,s)\in E} f_{(v,s)}$ be the value of the $s$-$t$ flow.
Lastly, we use $S = \Nout^H(s)$ to denote the out-neighbors of $s$ corresponding to the ``source-side'' vertices, and $T = \Nin^H(t)$ to denote the in-neighbors of $t$ corresponding to the ``sink-side'' vertices. %

Our goal is to determine how much flow in $f$ is routed between each pair $x\in S,y\in T$. We do so by computing \emph{pairwise} demands\footnote{For the objective of computing pairwise demands, we assume that the source and sink-side vertices are non-overlapping, i.e. $S \intersect T = \emptyset$, since this will always be the case in our application which is the implementation of the cut-matching game. The parallel flow-decomposition result however, is more generally applicable.} $d: S\times T \to \mathbb{R}^{+}$ such that $\norm{d}_1 = |f|$ and $d$ can be routed in $H$ exactly with $f$ being the edge-capacity constraints, and as in the parallel setting, this objective more easily admits a small work and low-depth implementation.
We will build a DAG data structure that implicitly encodes the necessary information.

\begin{definition}[Flow Decomposition DAG]
\label{def:flow-decomp-dag}
  An $\ell$-layered directed graph $\Dcal$ is a flow decomposition DAG of an $s$-$t$ flow $f$ specified by a weighted, directed graph $H$ iff
  \begin{enumerate}
    \item Each node of $\Dcal$ corresponds to a directed edge (not necessarily in $H$)
      between two vertices $u,v$ in $V$ associated with a flow value $h_{(u,v)}\in \mathbb{R}^+$.
      We use $(u,v,h)$ to represent this node in $\Dcal$, dropping the subscript $(u,v)$ in the flow value $h_{(u,v)}$ when it is unambiguous for ease of exposition. Note that there can be multiple nodes of the form $(u,v,h_{(u,v)})$ at the same layer of $\Dcal$ due to the existence of parallel $(u,v)$ edges, each with potentially different flow values $h_{(u,v)}$. We assume that all nodes in $\Dcal$ are uniquely indexed to avoid ambiguity.  
    \item The nodes $(u,v,h)$ at the lowest layer $1$ do not have any predecessors.
      In particular, each node $(u,v,h)$ in layer $1$ corresponds to a directed edge $(u,v)\in E$ in $H$,
      and has the same flow value $h_{(u,v)}=f_{(u,v)}$ as in $H$.
    \item Each node $(u,v,h)$ in any other layer $l\in \{2,\ldots,\ell\}$ has at most two predecessors in the previous layer $l-1$. If $(u,v,h)$ has two predecessors, then they have the form $(u,w,g)$, $(w,v,g')$ corresponding to a length-two path between $u,v$; this means that $(u,v,h)$ is obtained by ``merging'' (a part of the) flows $(u,w,g)$ and $(w,v,g')$. If $(u,v,h)$ has only one predecessor, it must have the form $(u,v,g)$. This means that $(u,v,g)$ had some residual flow $h\leq g$ that was not merged with any other node. 
    \item The nodes at each layer $l\in \{1,\ldots,\ell-1\}$ satisfy flow conservation with the succeeding layer $l+1$, i.e. for each node $(u,v,h)$ at any layer $l$ having successors $\Pi(u,v,h)$ at layer $l+1$,
      we have $\sum_{(x,y,g)\in \Pi(u,v,h)} g_{(x,y)} = h_{(u,v)}$.
    \item The nodes at the top-most layer $\ell$ are only of the form $(s,t,h)$, and have total flow value equal to $|f|$, i.e. these nodes all correspond to parallel $(s,t)$ edges, which together account for all of the $s$-$t$ flow $f$.   
  \end{enumerate}
\end{definition}

The size of this data-structure is measured in terms of the total number of nodes and edges it contains, where the edges denote successor-predecessor relationships between nodes across consecutive layers. If the size and the number of layers of a flow decomposition DAG are both low, we can efficiently perform various \pram computations with low work and depth therein.
Of particular importance to our approximate max-flow algorithm, for a $\eta$-size $\ell$-layer flow decomposition DAG, we can compute the second vertex (which is a neighbor of $s$) and penultimate vertex (which is a neighbor of $t$) of every flow path in layer $\ell$ simultaneously with $O(\eta)$ work and $O(\ell+\log \eta)$ depth using a simple algorithm; we show this in Lemma \ref{lem:matching-dag}.

We now show that a small-size and low-depth flow decomposition DAG can indeed be found efficiently by slightly relaxing property 5 in Definition \ref{def:flow-decomp-dag}; specifically, given a parameter $\delta\in (0,1)$, we have that the nodes of the form $(s,t,\cdot)$ (i.e. parallel $(s,t)$ edges) in layer $\ell$ together account for a $(1-\delta)$ fraction of the $s$-$t$ flow value $|f|$ (at a small cost to its size and depth). This is sufficient for all our applications, since we are only concerned with \emph{approximate} max flows. The precise guarantees are given in the following main theorem of this section.

\begin{theorem}[Parallel Flow Decomposition]
  \label{lem:short-cut}
  There exists a \pram algorithm $\PFD$ that
  given any parameter $\delta \in (0,1)$ and a polynomially bounded $s$-$t$ flow specified by a weighted, directed graph $H=(V,E,f)$ with flow value $|f|$,
  finds with high probability a flow decomposition DAG $\Dcal$ of $\ell = O(\log(n/\delta))$ layers
  such that at the topmost layer $\ell$, the total value of the flow captured by nodes $(s,t,\cdot)$ representing (parallel) $(s,t)$ edges is at least $(1-\delta)|f|$.
  The algorithm $\PFD$ has total work $O(m~\polylog(n) \log (n/\delta))$ and the total depth
  is $O(\polylog(n) \log(n/\delta))$.
  The total number of nodes, and edges representing successor-predecessor relationships between nodes across consecutive layers  in $\Dcal$ are both bounded by $O(m \log (n/\delta))$.
\end{theorem}

The idea behind our proof of Theorem \ref{lem:short-cut} can be illustrated by the following relatively intuitive process\footnote{This is only an illustration of our idea, see Section~\ref{lem:bmatchig} and the proof of Proposition \ref{proposition:bmatching} for what we actually do.}: repeatedly ``shortcut'' the flow graph $H$
by replacing a length-two flow path $u\to w\to v$ with a single edge $(u,v)$ having maximal flow value $h_{(u,v)} = \min\{g_{(u,w)},g_{(w,v)}\}$, and a residual edge $(x,y)\in \{(u,w),(w,v)\}$ having flow value $h_{(x,y)} = |g_{(u,w)}-g_{(w,v)}|$ iff there is any non-zero leftover flow not accounted for by this shortcut edge $(u,v)$. Consequently in the flow-decomposition DAG, the nodes $(u,w,g_{(u,w)})$ and $(w,v,g_{(w,v)})$ become the predecessors of this ``shortcut edge'' $(u,v,h_{(u,v)})$, and the node $(x,y,g_{(x,y)})$ becomes the predecessor of the residual edge $(x,y,h_{(x,y)})$ if there is any leftover flow.  
In order to achieve low depth, our objective is to find a collection of length-two flow paths that together account for a large fraction of the (total $\ell_1$ norm of the) flow that does not directly go from $s$ to $t$ which we can then shortcut
in parallel. We show that we can find such a collection of flow paths efficiently with the following technical lemma.

\begin{lemma}
  \label{lem:bmatchig}
  Let $f$ be an $s$-$t$ flow of value $|f|$ specified by a weighted, directed multigraph
  $H = (V,E,f)$ containing source vertex $s\in V$, sink vertex $t\in V$ with no edges directly connecting $s$ to $t$ and with no self-loops, and the flow on any edge being given by $f: E \to \mathbb{R}^{+}$. Then we can find in
  $O(m \, \polylog(n))$ work and $O(\polylog(n))$ depth, a collection
  of ``shortcut paths'' represented as
  tuples $\mathcal{P}:=\setof{(e^{(1)}_i, e^{(2)}_i, h_i)}$ where $e^{(1)}_i\neq e^{(2)}_i\in E$, $h_i\in \mathbb{R}^{+}$, along with a collection of ``residual edges'' given by $\mathcal{R} := \setof{(e_i,r_i)}$ where $e_i\in E$ are disjoint, and $r_i\in \mathbb{R}^{+}$ such that
  \begin{enumerate}
    \item (Length-two paths) For each $i\in [|\mathcal{P}|]$, $e^{(1)}_i, e^{(2)}_i$ form a path of length two.
    \item (Flow constraints) For each edge $e\in E$,
      the total flow accounted for by the shortcut paths involving this edge, which is given by the summation of $h_i$ over the tuples in $\mathcal{P}$ in which $e$ appears along with its residual capacity $r_i$ if any (if $e$ is present in $\mathcal{R}$), is exactly $ f_e$:
      \begin{align*}
        \sum\nolimits_{i\in [|\mathcal{P}|]: e\in \setof{e^{(1)}_i,\ e^{(2)}_i}} h_i + 
        \sum\nolimits_{i \in[|\mathcal{R}|]: e=e_{i}} r_{i} = f_e.
      \end{align*}
    \item (Large $\ell_1$-norm) With probability at least $1/15$, $\sum_{i\in [|\mathcal{P}|] } h_i \geq  \norm{ f }_1/16$.
    \item (Non-increasing flow-support) %
    $|\mathcal{P}| + |\mathcal{R}| \leq |E|$.
  \end{enumerate}
\end{lemma}

\begin{proof}(Theorem~\ref{lem:short-cut}) Given a flow $f$ specified by a weighted directed graph $H=H_1$, we begin by creating the lowest layer $1$ of our flow decomposition DAG $\Dcal$ from $H_1$ as specified in property 2 of Definition \ref{def:flow-decomp-dag}. It is easy to see that this can be achieved with $O(m)$ total work and $O(1)$ depth. We also create a set $\mathcal{S}$, initially empty, to track tuples of the form $(s,t,h_{(s,t)})$ (i.e. direct $(s,t)$ edges) which correspond to ``fully processed'' $s$-$t$ flow paths. If $H_1$ contains such a direct edge, we delete it from $H_1$ and add the tuple $(s,t,f_{(s,t)})$ to $\mathcal{S}$. The subsequent proof (and algorithmic procedure) then follows from a repeated application of Lemma~\ref{lem:bmatchig}; the $l>1$-th iteration, given as input an $s$-$t$ flow specified by a weighted, directed \emph{multigraph} $H_{l-1}$, we have:
  \begin{enumerate}
    \item Invoke Lemma~\ref{lem:bmatchig} on $H_{l-1}$ to find the desired collection of shortcut paths 
      $\mathcal{P}_l=\setof{(e^{(1)}_i, e^{(2)}_i, h_i)}$, and residual edges $\mathcal{R}_l = \setof{(e_i,r_i)}$. Set $\mathcal{S}_l$ to be initially empty. 
    \item Construct a new flow graph $H_l$, initially empty, as follows: for each tuple $(e^{(1)}_i,e^{(2)}_i,h_i) \in \mathcal{P}_l$, add a directed edge $(u_i,v_i)$ with flow value $f_{(u_i,v_i)} = h_i$, where $u_i,v_i$ are the endpoints\footnote{Since we are only guaranteed an approximate max-flow $f$, it may contain circulations that we may discover in this process, i.e. the two endpoints $u_i,v_i$ are identical. In this case, we can simply delete all the self-loops.}
      of the length-two path $(e^{(1)}_i, e^{(2)}_i)$. For each residual edge $(e_i,r_i)\in \mathcal{R}_l$, add a directed edge $e_i$ with flow value $r_i$. Note that this may lead to the creation of parallel edges. Delete any edge $(s,t)$ in $H_l$ (may be multiple), as these correspond to ``fully processed'' flow paths that will be tracked separately in $\mathcal{S}_l$.
      Also delete any self-loops (edges with both endpoints being the same vertex) from $H_l$.%
    \item Construct layer $l$ of the flow-decomposition DAG $\mathcal{D}$ as follows: for each tuple $\{e^{(1)}_i,e^{(2)}_i,h_i\} \in \mathcal{P}_l$, add to the $l^{th}$ layer of $\Dcal$ a node $(u_i,v_i,h_i)$, where $u_i,v_i$ are the endpoints\footnote{If $u_i = v_i$, then skip this tuple without adding any nodes/edges.}
      of the length-two path $(e^{(1)}_i, e^{(2)}_i)$. We set the predecessors of $(u_i,v_i,h_i)$ to be the nodes $(u^{(1)}_i,v^{(1)}_i,f_{(u^{(1)}_i,v^{(1)}_i)})$ and $(u^{(2)}_i,v^{(2)}_i,f_{(u^{(2)}_i,v^{(2)}_i)})$ from layer $l-1$, where $ e^{(j)}_i = (u^{(j)}_i,v^{(j)}_i)$ for $j\in\{1,2\}$. For each residual edge $(e_i,r_i)\in \mathcal{R}_l$, add to the $l^{th}$ layer of $\Dcal$, a node $(u_i,v_i,r_i)$ where $e_i=(u_i,v_i)$. We set the predecessor of $(u_i,v_i,r_i)$ to be the node $(u_i,v_i,f_{(u_i,v_i)})$ from layer $l-1$. For each newly created node of the form $(s,t,h)$ at layer $l$, add this node to $\mathcal{S}$. For each node $(s,t,h_i)\in \mathcal{S}_{l-1}$, add to the $l^{th}$ layer of $\Dcal$, a new node $(s,t,h_i)$, and set its predecessor to be the corresponding $(s,t,h_i)$ node from layer $l-1$. Finally, create $\mathcal{S}_l$ by adding all newly created $(s,t,\cdot)$ tuples in $\mathcal{S}$ to $\mathcal{S}_{l-1}$. 
  \end{enumerate}
  
  Observe that in each iteration, the updated flow specified by the multigraph $H_l$ (prior to deleting $(s,t)$ edges and self-loops if any) can trivially be routed in the preceding graph $H_{l-1}$ by simply ``undoing'' the shortcutting that produced $H_l$. Moreover, after accounting for all the $(s,t)$ edges specified by tuples in $\mathcal{S}_l$, the total flow value leaving $s$ is preserved across all iterations in our procedure. Therefore, the ``routability'' property desired of the flow decomposition procedure is trivially satisfied by our aforementioned process.
  
  We shall now prove the work and depth guarantees of this algorithm, along with the near linear-size and polylogarithmic depth of the resulting flow-decomposition DAG $\Dcal$. Since in every iteration, the $\ell_1$-norm of the flow $f$ reduces by $\sum_{i\in \mathcal{P}} h_i$, which by property (3) of Lemma~\ref{lem:bmatchig} is at least a constant-fraction of the $\ell_1$-norm of the flow $f$ at the start of the iteration with constant probability, we can deduce that
  after $\ell = \Theta(\log\frac{n \norm{f}_1}{\delta |f|}) = \Theta(\log(n/\delta))$ iterations of the above process, with polynomially high probability (in $n$, follows by a straightforward Chernoff bound), the edges that remain in the flow graph $H_{\ell}$ (i.e. that do not directly go from $s$ to $t$) contribute at most a $\delta$ fraction
  of the total (initial) amount of flow $|f|$ leaving $s$. Since in all our applications, we are only concerned with finding an approximate max-flow, we can safely ignore this residual flow for small enough $\delta$. The second equality in the above bound follows by observing that $|f|\geq \min_{e\in E} w_e$, and $\norm{f}_1\leq \sum_{e\in E} w_e \leq m\cdot \max_{e\in E} w_e$, and the aspect ratio $\max_{e\in E} w_e/ \min_{e\in E} w_e$ is assumed to be polynomially bounded. 

  In each iteration $l$, our procedure involves obtaining the relevant collection of tuples $\mathcal{P}_l,\mathcal{R}_l$ given an input flow multigraph $H_{l-1}$, which in turn are used to add a new layer $l$ to the flow-decomposition DAG $\Dcal$, and to update the flow multigraph $H_{l-1}\rightarrow H_l$. The total number of edges $|E_l|$ in the updated flow multigraph $H_l$ is at most  $|\mathcal{P}_l| + |\mathcal{R}_l| - |\mathcal{S}_l\setminus \mathcal{S}_{l-1}|$, which by property (4) of Lemma \ref{lem:bmatchig} is at most $|E_{l-1}|$, the total number of edges in the flow multigraph $H_{l-1}$ from the previous iteration $l-1$. Moreover, the total number of nodes added to layer $l$ in $\Dcal$ is at most $|\mathcal{P}_l| + |\mathcal{R}_l| + |\mathcal{S}_{l-1}| = |\mathcal{P}_l| + |\mathcal{R}_l| - |\mathcal{S}_l\setminus \mathcal{S}_{l-1}| + |\mathcal{S}_l\setminus \mathcal{S}_{l-1}| + |\mathcal{S}_{l-1}| \geq |E_l| + |S_l|$. However, by property (4) of Lemma \ref{lem:bmatchig}, we also have that $|\mathcal{P}_l| + |\mathcal{R}_l| \leq |E_{l-1}|$, due to which we have that $|\mathcal{P}_l| + |\mathcal{R}_l| + |\mathcal{S}_{l-1}| \leq |E_{l-1}| + |\mathcal{S}_{l-1}|$. Therefore, we can infer that the quantity $|E_{l'}| + |\mathcal{S}_{l'}|$ is non-increasing across $l'$, due to which we can conclude that $|E_{l'}| + |\mathcal{S}_{l'}| \leq |E_1| + |\mathcal{S}_1| = |E|$. The above bounds, and the fact that the number of edges $|E|$ in the initial flow graph $H$ provided as input to our algorithm is at most $m$ combined with the computational guarantees of Lemma \ref{lem:bmatchig} gives us that any iteration of our algorithm (finding tuples, updating $H_{l-1}\rightarrow H_l$, creating a new layer in $\Dcal$) can be implemented in $O(m \, \polylog{n})$ work and $O(\polylog{n})$ depth. Moreover, each new layer created in the flow decomposition DAG has at most $m$ nodes, and at most $2m$ edges between it and the preceding layer (since every node has at most two predecessors from the previous layer). Since there are a total of $\Theta(\log\frac{n \norm{f}_1}{\delta |f|}) = O(\log (n/\delta))$ iterations of our algorithm for polynomially bounded flows, we have that the total work and depth of our algorithm is $O(m \, \polylog(n)\log(1/\delta))$ and $O(\polylog(n)\log(1/\delta))$, respectively, and the size (number of nodes,edges) and depth of the flow decomposition DAG $\Dcal$ produced is at most $O(m\log (n/\delta))$, and $O(\log (n/\delta))$, respectively as claimed.   
\end{proof}

\subsection{Proof of Lemma~\ref{lem:bmatchig}.}

One can formulate the problem as a (uncapacitated) $b$-matching problem. Specifically,
we create a $b$-matching instance $H_b$ from the flow multigraph $H=(V,E,f)$ (with no edge directly connecting $s$ to $t$) as follows.
For each edge $e\in E$, we create a vertex $u_e$ with demand $f_e$.
Then for every pair of edges $e^{(1)},e^{(2)}$ that form a length-two path by sharing a vertex $v\in V$,
we connect their corresponding vertices $u_{e^{(1)}},u_{e^{(2)}}$ with an edge of infinite capacity.
It suffices to find a large $b$-matching whose size is a constant fraction of 
$\norm{ f}_1$, the total $\ell_1$-norm of the flow $f$.

First, we shall prove such a $b$-matching exists.
To this end, let us consider a bipartite $b$-matching instance $H'_b$, constructed as follows.
For each edge $e\in E$, we create two vertex copies $u_{e_{\iin}}$ and $u_{e_{\out}}$ both with demand
$f_e$.
Then for every pair of edges $(e^{(1)},e^{(2)})$ that form a length-two path by sharing a vertex $v\in V$,
we add an edge between $u_{e^{(1)}_{\iin}}$ and  $u_{e^{(2)}_{\out}}$ with infinite capacity.
We claim that there is a $b$-matching of size at least $\norm{f}_1/2$, by the following simple construction.
For each vertex $v\in V$ such that $v\neq s$ and $v\neq t$, we use a perfect $b$-matching between the vertices corresponding to the ``in'' copies of the incoming edges of $v$
and the vertices corresponding to the ``out'' copies of the outgoing edges of $v$, whose existence is guaranteed by the flow conservation property. Then both vertices corresponding to every edge $e\in E$ is fully matched except for the vertices corresponding to the ``out'' copies of edges leaving the source $s$, and the vertices corresponding to the ``in'' copies of edges entering the sink $t$. By a simple charging argument that assigns the $b$-matching value for every vertex $v\in V$ to the ``in'' copies of the vertices corresponding to the incoming edges into $v$, we can bound the total demand of these aforementioned unmatched edges by at most $\norm{f}_1/2$, which gives us our claim.
Now consider for each edge $e\in E$, keeping either $u_{e_{\iin}}$ or $u_{e_{\out}}$ uniformly at random and discarding the other vertex, letting the resulting subsampled graph be $H_b$.
Then in expectation, observe that the maximum $b$-matching in the resulting subsampled graph is at least
$\norm{f}_1/8$, since each matched edge is kept with probability $1/4$ (i.e. if its two end points, one of which is a vertex corresponding to the ``in'' copy of an incoming edge, and the other is a vertex corresponding to the ``out'' copy of an outgoing edge are both sampled in the subsampled graph $H_b$). Therefore, by a standard Markov argument, we have that with probability at least $1/15$, the matching size in the subsampled graph is at least $\norm{f}_1/16$.

We can in fact algorithmically compute such a matching efficiently in light of the above construction.
As described above, we can subsample the vertices corresponding to the edge-copies in the graph $H'_b$ and find the maximum
$b$-matching in the resulting graph by locally finding, for each vertex $v\in V \setminus \{s,t\}$,
the maximum $b$ matching between the (subsampled) vertices corresponding to ``in'' copies of incoming edges into $v$, and (subsampled) vertices corresponding to ``out'' copies of outgoing edges from $v$ in $H$. This is correct because after we keep either $u_{e_{\iin}}$ or $u_{e_{\out}}$ and discard the other
for each $e$ in $H$, the graph becomes a union of disjoint connected components,
where each component consists of the ``in'' copies of the incoming edges and the ``out'' copies
of the outgoing edges of a single vertex. It then remains to show that we can efficiently find the maximum $b$-matching
locally for every component, along with the non-increasing flow-support property for which it suffices to prove the following proposition.

\begin{proposition}\label{proposition:bmatching}
  Given two sets of elements $A=\{a_1,\ldots,a_{n_{\iin}}\}$ and $B=\{b_1,\ldots,b_{n_{\out}}\}$, along with a positive function $f:A\cup B\rightarrow \mathbb{R}^+$
  we can find in $O( (n_{\iin} + n_{\out})\, \polylog(n_{\iin} + n_{\out}) )$ total work and $O(\log(n_{\iin}+n_{\out}))$ depth, collections
  of tuples $\mathcal{P} :=\{(x_i,y_i,h_{i})\}$ where $x_{i}\in A,y_{i}\in B,h_{i}\in\mathbb{R}^{+}$, and $\mathcal{R} := \{(z_i,r_i)\}$ where $z_i\in A\cup B$ are disjoint and $r_i\in\mathbb{R}^+$ such that
  \begin{enumerate}
    \item (Flow constraints) For each $a\in A$ (and each $b\in B$), we have 
      \begin{align*}
        &\sum_{i\in [|\mathcal{P}|]: x_{i} = a} h_{i} +
        \sum_{i\in [|\mathcal{R}|]: a=z_i} r_{i}
        = f_a, \quad \text{and}\\
        &\sum_{j \in [|\mathcal{P}|]: y_{j} = b} h_{j}  +
        \sum_{j\in [|\mathcal{R}|]: b=z_j} r_j
        = f_b.
      \end{align*}
    \item (Non-increasing support) $|\mathcal{P}|+|\mathcal{R}| \leq n_{\iin} + n_{\out}$
    \item (Maximality) $\sum_{i\in [|\mathcal{P}|]} h_{i} = \min\setof{ \sum_{a\in A} f_a, \sum_{b\in B} f_b }$.
  \end{enumerate}
\end{proposition}
\begin{proof}
  The proof is standard and it appeared in earlier work like \cite{andoni2020parallel}, though we describe it explicitly for completeness.
  We begin by computing the prefix sum of the values in both $A$ and $B$; with some abuse of notation, let $A_{k} = \sum_{i=1}^{k}f_{a_{i}}$ and $B_{k} = \sum_{i=1}^{k} f_{b_{i}}$. By a standard \pram algorithm (\cite{akl1997parallel}), this procedure takes $O(n_{\iin}+n_{\out})$ work and $O(\max\{\log n_{\iin},\log n_{\out}\})=O(\log(n_{\iin}+n_{\out}))$ depth. 
  Next, we rank (sort) all the values in $\{A_{i}\}_{i\in [n_{\iin}]}\cup \{B_{j}\}_{j\in [n_{\out}]}$, which takes $O((n_{\iin}+n_{\out})\, \log(n_{\iin}+n_{\out}))$ work and $O(\log(n_{\iin} + n_{\out}))$ depth; let $C_{k}$ be the $k$-th ranked value in the resulting sorted prefix sum values, with $c_k$ denoting the element $a_i$ or $b_j$ depending on whether $C_k = A_i$ or $C_k = B_j$. If there are ties (i.e. the total value $A_i$ of the first $i$ elements of $A$ is exactly equal to the total value $B_j$ of the first $j$ elements of $B$ for some $i,j$), then set $c_k = a_i$, and drop the $C_{k'}$ term corresponding to $B_j$ after setting $b_j$ to be the successor of $c_k$, which we define next.    
  For each element $c_k$, find its successor ${c}_{\text{succ-}k}$ to be the element from the other array corresponding to the smallest prefix sum that is at least as large as $C_k$, i.e., if $c_{k} = a_i$ for some $i$ (i.e. $C_k=A_i$), then the successor $c_{\text{succ-}k}=b_j$ where $j = \argmin_{\ell: B_{\ell}\geq C_k=A_i} B_{\ell}$. If there is no such value (i.e. $C_k=A_i>B_{n_{\out}}$), then we set the successor $c_{\text{succ-}k}$ to be a dummy element $\perp$. The case where $c_k = b_j$ follows symmetrically. We also let $C_{0}=0$ so that the subsequent process is well defined. Finally let $C$ be the resulting processed, sorted array, and let $n_{\text{tot}}$ be its length (i.e. array $C$ contains $C_0,C_1,\ldots,C_{n_{\text{tot}}-1}$). This procedure can also be done in $O((n_{\iin}+n_{\out})\cdot \log(n_{\iin} + n_{\out}))$ work and $O(\log(n_{\iin} + n_{\out}))$ depth by the standard doubling trick. Now starting with both $\mathcal{P},\mathcal{R}$ being initially empty, do the following: for each $\ell \in \{1,\ldots, n_{\text{tot}}-1\}$ in parallel: if $c_\ell = a_i$, and $c_{\text{succ-}\ell} \neq \perp$, then add the tuple $(c_\ell,c_{\text{succ-}\ell},C_\ell - C_{\ell-1})$ to $\mathcal{P}$; if $c_\ell = b_j$, and $c_{\text{succ-}\ell} \neq \perp$, then add the tuple $(c_{\text{succ-}\ell},c_\ell,C_\ell - C_{\ell-1})$ to $\mathcal{P}$; otherwise, add the tuple $(c_{\ell},C_{\ell}-C_{\ell-1})$ to $\mathcal{R}$. This entire process requires just $O(n_{\iin}+n_{\out})$ work and $O(1)$ depth. This proves our computational guarantees.

  We shall now prove the flow constraint, non-increasing support, and maximality properties outlined in the proposition statement. To prove the flow constraint property, consider any fixed element $b_j$ for some $j$. By nature of our algorithm, $b_j$ appears in tuples due to one of two reasons: either (a) it was the successor $b_j = c_{\text{succ-}k'}$ for some elements $c_{k'}$ (the number of such elements is $\geq 0$), or (b) when the element $C_k=B_j$ was processed by itself (exactly once if $B_j$ was not tied with some $A_i$, in which case it was combined with its successor $c_{\text{succ-}k}$ to form a tuple and added to set $\mathcal{P}$ if $c_{\text{succ-}k}\neq \perp$, and to set $\mathcal{R}$ otherwise. If tied with some $A_i$, then this does not occur as there is no $k:C_k=B_j$). In the former, observe that $b_j$ will be the successor of all elements $c_{k'}=a_{i}$ with value $C_{k'}=A_i$ where $\,B_{j-1}<C_{k'}=A_i\leq B_{j}$. If we sum the $h_{k'}$ values corresponding to all such $k'$, including the final $h_k$ value (which occurs only if there is a $k:\, C_k=B_j$), we get $\sum_{k': B_{j-1}<C_{k'}\leq B_{j}} h_{k'} + \boldsymbol{1}(\exists\,k:C_k=B_j)h_k = \sum_{k': B_{j-1}<C_{k'}\leq B_{j}}  (C_{k'}-C_{k'-1}) + \boldsymbol{1}(\exists\,k:C_k=B_j) (C_k-C_{k-1}) = B_{j} - B_{j-1} = f_{b_j}$, where the final equality follows by telescoping summation. The argument where $c_k=a_i$ follows symmetrically. The non increasing support property trivially follows by observing that the total number of tuples added to $\mathcal{P},\mathcal{R}$ together is exactly $n_{\text{tot}}-1\leq n_{\iin}+n_{\out}$ by definition of $C$. Lastly, to prove maximality, observe that the largest term $C_{n_{\text{tot}}-1}$ must be achieved at either $A_{n_{\iin}}$ or $B_{n_{\out}}$. Let us assume that it is $B_{n_{\out}}$, in which case it must be the case that every $C_{k}$ term corresponding to some $A_i$ value will have a successor in $B$, in which case the total capacity of $A_{n_{\iin}} = \sum_{i=1}^{n_{\iin}} f_{a_i}$ must be accounted for by the tuples in $\mathcal{P}$, since no $a_i$ element will end up in the residual set $\mathcal{R}$. The argument for the other case where the largest term $C_{n_{\text{tot}}-1}$ is achieved at $A_{n_{\iin}}$ follows symmetrically.    
\end{proof}

\subsection{Computing Fractional Matching.}
It remains to show how, for each neighbor $u$ of $s$, we may determine how much flow through $u$ is routed through each neighbor of $t$.
This allows us to use the flow to compute a (fractional) matching between the neighbors of $s$ and the neighbors of $t$, which we use in the cut matching game (see Section \ref{sec:pcmg}).
Note also that in our use, the flow $f$ that we want to decompose
never contains any edges connecting $s$ directly to $t$.
The following lemma shows that this can be done using a flow-decomposition DAG; combined with Theorem \ref{lem:short-cut}, this lemma shows that this can be done in logarithmic depth and near-linear work.

\begin{lemma}
    \label{lem:matching-dag}
    Given a graph $H$, and vertices $s,t\in V(H)$, let $S$ and $T$ be the set of neighbors of $s$ and $t$ in $H$, respectively.
    Given a flow $f$ without edges connecting $s$ directly to $t$ and a flow-decomposition DAG $\Dcal$ of size $\eta$ and depth $\ell$, let $P$ be the set of flow paths
    corresponding to the nodes of the form $(s,t,\cdot)$ at
    the final layer $\ell$.
    There exists a $O(\ell + \log \eta)$ depth, $O(\eta)$ work \pram algorithm which computes $r_{x,y}$ for each $x\in S$ and $y\in T$ such that
    a total of $r_{x,y}$ units of flow is routed by the flow paths in $P$
    whose second vertex is $x$ and penultimate vertex is $y$.
\end{lemma}
\newcommand{\se}{\texttt{search}}
\begin{proof}
    For each node $u$ of the form $(s,t,\cdot)$ at layer $\ell$
    of the flow-decomposition DAG $\Dcal$, our goal is to compute $\se(u)$, which is the set $\{x,y\}$ such that $x\in S$ is the second node on the path represented by $u$ and $y\in T$ is the penultimate node.
    To aid in presentation, for each DAG node $v$ of the form $v=(s,a,\cdot)$ where $a\neq t$, we define $\se(v)$ to be only the second node $x\in S$ along the path represented by $v$; we analogously define $\se(v)$ to be only the penultimate node on the path when $v=(a,t,\cdot)$.
    Note that we are only interested in $\se(v)$ of nodes $u=(v_1,v_2,\cdot)$ of $\Dcal$ such that $v_1=s$ or $v_2=t$.

    Starting from each node $u$ of the form $(s,t,\cdot)$ at level $\ell$, in parallel we recursively compute $\se(u)$ as follows:
    \begin{itemize}
        \item If $u$ has exactly one predecessor $p$, return $\se(p)$.
        \item If $u$ is at level 1 in $\Dcal$, then either $u=(s,x,\cdot)$ for some $x\in S$ or $u=(y,t,\cdot)$ for some $y\in T$
        (recall that the initial given flow $f$ does not contain edges directly connecting $s$ to $t$); return $x$ (or $y$, respectively).
        \item Otherwise, $u$ has two predecessors $u_1$ and $u_2$.
            \begin{itemize}
                \item If $u=(s,t,\cdot)$, then $u_1=(s,a,\cdot)$ and $u_2=(a,t,\cdot)$, for some $a\in V(H)$. Return $\{\se(u_1),\se(u_2)\}$.
                \item Otherwise, $u=(v_1,v_2,\cdot)$, where either $v_1=s$ or $v_2=t$ (but not both), and $u_1=(v_1,a,\cdot)$ and $u_2=(a,v_2,\cdot)$ for some $a\in V(H)$.
                    If $v_1=s$, return $\se(u_1)$, and if $v_2=t$, return $\se(u_2)$.
            \end{itemize}
    \end{itemize}
    The correctness follows by induction on the level of $u$.
    If $u$ is at level 1, $\se(u)$ is trivially correct, and if $u$ has exactly one predecessor $p$, then $\se(u)=\se(p)$, which is correct by induction.
    Otherwise, suppose $u$ has two predecessors $u_1$ and $u_2$.
    If $u=(s,t,\cdot)$, then by the inductive hypothesis, $\se(u_1)$ is the second node on the path represented by $u_1$ and $\se(u_2)$ is the penultimate node on the path represented by $u_2$; since the path represented by $u$ is the union of the paths represented by $u_1$ and $u_2$, it then follows that $\se(u)$ is also correct.
    The correctness of $\se(u)$ when $u=(v_1,v_2,\cdot)$, with $v_1=s$ or $v_2=t$, follows similarly.

    The depth of the algorithm is bounded by the depth of $\Dcal$, which is $\ell$.
    Similarly, for work, the work to compute any \textit{one} $\se(u)$ is at most $O(\ell)$, as $\Dcal$ has $\ell$ layers.
    So, the total work is $O(\zeta \ell)=O(\eta)$, where $\zeta$ is the number of nodes at level $\ell$, and $\zeta\ell=O(\eta)$ by property 4 of Lemma \ref{lem:bmatchig}.
    The desired values $r_{x,y}$ can then be computed by, for each $u$ at level $\ell$, adding the flow value of $u$ to $r_{\se(u)}$
    (which is initially set to $0$), which takes
    $O(\eta)$ work
    $O(\log \eta)$ depth by parallel summation.
\end{proof}

%% file: applications-sparsest-cut.tex
\subsection{Sparsest Cut and Balanced Min-cut.}
\label{sec:sparse-cut}

The sparsest cut problem is a classic problem in graph theory that informally asks to partition a given graph while removing as little edge-mass as possible. More precisely, given an undirected, weighted graph $G=(V,E,c)$ the objective is to find a cut $(X, V\setminus X)$ of minimum \emph{sparsity}, which is formally defined as 
\begin{align*}
\phi(X) = \frac{c(E(X, V\setminus X))}{\min\setof{\card{X},\card{V\setminus X}}},
\end{align*}
where $c(E(X,V\setminus X))$ is the total weight of edges going across the cut $(X,V\setminus X)$.
A closely related problem, the $\beta$-balanced minimum cut asks for a cut $(X, V\setminus X)$ such that the smaller side of the partition has at least $\beta n$ vertices for a given parameter $\beta>0$, and the total edge weight $c(E(X, V\setminus X))$ is minimized among all such partitions.
The $\beta$-balanced sparsest cut is similarly defined as a cut whose $i).$ smaller side has at least $\beta n$ vertices and $ii).$ sparsity is minimized among all such partitions.

All the aforementioned problems are known to be NP-hard, and the best-known polynomial-time algorithms achieve $O(\sqrt{\log{n}})$-approximation \cite{AroraRV09} to their corresponding objective, albeit at the expense of a large (sequential) polynomial running time. For the balance constrained variants of these cut problems, bicriteria approximations that allow multiplicative factors on both the balance parameter as well as the cut size are also commonly studied.
This notion can be formally defined as follows.
\begin{definition}[Bicriteria approximation for balanced cut problems]
\label{def:pseudo-approx-balance}
Let $\beta'<\beta\leq \frac{1}{2}$ be real numbers, and let $(X^{*}, V\setminus X^{*})$ be an optimal $\beta$-balanced min-cut  (resp. $\beta$-balanced sparsest cut of $G$). We say that a cut $(X,V\setminus X)$ is an $(\alpha, \beta')$-bicriteria approximation of the $\beta$-balanced min-cut (resp. $\beta$-balanced sparsest cut) if
\begin{enumerate}
\item $(X,V\setminus X)$ is a $\beta'$-balanced cut, and;
\item Approximation guarantees:
\begin{enumerate}
\item For $\beta$-balanced sparsest cut, $\phi(X)\leq \alpha \cdot \phi(X^{*})$ .
\item For $\beta$-balanced min-cut, $c(E(X,V\setminus X))\leq \alpha \cdot c(E(X^{*}, V\setminus X^{*}))$ .
\end{enumerate}
\end{enumerate}
\end{definition}

While algorithms for sparsest and balanced-min cuts have been developed for the distributed $\mathsf{CONGEST}$ model of computation \cite{KuhnM15,ChangS19}, no parallel \pram algorithms with nearly-linear work and polylogarithmic depth are known. 
We resolve this state-of-the-affair by designing the first \pram $(\polylog{n}, \polylog{n})$-bicriteria approximation for both problems using our parallel approximate max-flow algorithm. Formally,
\begin{theorem}
\label{lem:sparse-and-balance-cut}
There is a randomized \pram\ algorithm that given an undirected weighted graph $G=(V,E,c)$, computes with high probability a cut $(X, V\setminus X)$ that achieves 
\begin{itemize}
\item An $O(\log^3 n)$-approximation for sparsest cut;
\item An $\kh{O\kh{\log^3{n}}, O\kh{\frac{\beta}{\log^2{n}}}}$-bicriteria approximation for $\beta$-balanced sparsest cut;
\item An $\kh{O\kh{\frac{\log^3{n}}{\beta}}, O\kh{\frac{\beta}{\log^2{n}}}}$-bicriteria approximation for $\beta$-balanced min-cut.
\end{itemize}
This algorithm has $O(m\cdot \polylog {n})$ work and $O(\polylog{n})$ depth.
\end{theorem}

Our algorithm builds upon the cut-matching game framework developed by \cite{KhandekarRV06} that effectively reduces the computation of all of these aforementioned problems to polylogarithmically many single-commodity max-flow computations (as well as a flow-decomposition of the corresponding max-flow solutions). As a consequence, this framework provides algorithms for computing $\polylog n$ approximations to all these cut problems with work and depth that matching that of the max-flow (and flow-decomposition) oracle utilized in their implementations up to a $\polylog{n}$ factor. While the original idea of \cite{KhandekarRV06} required an \emph{exact} max-flow oracle in its implementation, \cite{ns17} showed in a fairly straightforward extension that an approximate max-flow oracle also suffices to achieve morally the same result.

Our algorithm, formally described in Algorithm~\ref{alg:cut-matching-sparsest-cut}, is a straightforward extension of this aforementioned result of \cite{ns17} to the \pram setting. By crucially utilizing our parallel approximate max-flow result in Theorem~\ref{rst:max-flow}, the parallel flow-decomposition result in Lemma~\ref{lem:short-cut}, and the near-linear work, logarithmic depth flow-rounding algorithm of \cite{cohen95}, this algorithm achieves guarantees formally described in Lemma~\ref{lem:balance-sparse-cut}, which in turn imply the guarantees claimed in Theorem~\ref{lem:sparse-and-balance-cut}.

\begin{algorithm}{$\texttt{Parallel cut-matching game for sparsest cuts}(G,\alpha,\beta)$}\\
    \label{alg:cut-matching-sparsest-cut}\textbf{Input:} Graph $G=(V,E,c)$ with $n=|V|,m=|E|$, a sparsity parameter $\alpha$, a balance parameter $\beta$. \\
    \nonl \textbf{Output:} \\
    \nonl \textbf{Either Cut Case:} a cut $(X,V\setminus X)$ such that $\phi(X)\leq \alpha$ and $\beta n \leq \card{X}\leq \frac{n}{2}$; \\
    \nonl \textbf{Or Expander Case: } a graph $H$ embeddable in $G$ with congestion at most $O(\log^3 n/\alpha)$ such that every $(\beta \log^2{n})$-balanced cut in $H$ has sparsity at least $\Omega(1)$.\\
    \textbf{Procedure:}
    \begin{algorithmic}[1]
        \For{\normalfont round $t=1,\ldots , c_{1} \log^2{n}$, where $c_{1}$ is a sufficiently large constant}
        \State \nonl \algcomment{The cut player:}
        \State Sample a random $n$-dimensional unit vector $\rr$ orthogonal to $\boldsymbol{1}$.
        \If{$t>1$}
            \State $\uu\gets \MM_{t-1}(\MM_{t}(\ldots (\MM_1\rr)))$, where for any $t'<t$, the $(n\times n)$ (sparse) matrix $\MM_{t'}$ is the probability transition matrix corresponding to perfect matching output by the matching player in round $t'$.
        \Else
            \State $\uu \gets \rr$
        \EndIf
        \State Return cut $X_{t} \gets$ vertices corresponding to the smallest $n/2$ entries in $\uu$.
        \State \nonl \algcomment{Matching player:}
        \State Fix $\eps = 1/10$, and let $c_2 = c_2(\eps)$ be a sufficiently large constant.
        \For{\normalfont sub-round $j=1,\ldots ,c_{2}\log{n}$}
         \State Maintain $Sl^{j}_{t}$ and $Sr^{j}_{t}$: at the beginning, let $Sl_{t}^{1}=X_{t}$; $Sl^{j}_{t}= Sl^{j}_{t}\setminus V(\MM^{j-1}_{t})$, where $V(\MM^{j-1}_{t})$ is the matching computed in the $(j-1)$-th sub-round of the matching player. Update $Sr_{t}^j$ analogously.
         \State \nonl \algcomment{Note that the sizes of $Sl^{j}_{t}$ and $Sr^{j}_{t}$ always remain equal by construction across all sub-rounds.}
        \State Connect a source $s$ to $Sl^{j}_{t}$ and a sink $t$ to $Sr^{j}_{t}$ with edge capacities $1$, and scale all other edge capacities in $G$ by $\alpha^{-1}$.
        \State Compute a $(1-\eps)$-approximate max-flow, and round the flow to be integral using the algorithm in \cite{cohen95}; also obtain the corresponding approximate min-cut $(C^{j}_{t},V\setminus C_t^j)$.
        \State\nonl \algcomment{by the dual variables of Sherman's framework.}
        \If {$c(E(C^{j}_{t}, V\setminus C^{j}_{t}))< \card{Sl^{j}_{t}}-\beta n$}
        \State Output $(C^{j}_{t}, V\setminus C^{j}_{t})$ as the desired $\alpha$-sparse cut and terminate the game.
        \Else
        \State Compute a \emph{partial} matching $\MM^j_t$ between the vertex sets $Sl^j_t,Sr^j_t$ given by the flow decomposition.
        \EndIf\EndFor
        \State Let $Sl^{J+1}_t = Sl^J_t\setminus V(\MM^J_t)$, and $Sr^{J+1}_t = Sr^J_t\setminus V(\MM^J_t)$, where $J=c_2\log n$ is the final sub-round of the matching player.
        \If{$\card{Sl^{J+1}_{t}}=\card{Sr^{J+1}_{t}}<\beta n$}
            \State Compute an arbitrary matching $\MM'_{t}$ between $Sl^{J+1}_{t},Sr^{J+1}_{t}$.
            \State Return the perfect matching $\MM_{t} = \paren{\bigcup~ \MM^j_t} \cup \MM'_t$ to be the union of the partial matchings across all sub-rounds of the matching player.
        \EndIf
        \EndFor
        \State If the matching player outputs a cut in any (sub)round, then it is the desired $\alpha$-sparse cut; otherwise, $H=\bigcup \paren{\MM_{t} \setminus \MM'_{t}}$ is the graph embeddable in $G$ with low congestion.
    \end{algorithmic}
\end{algorithm}

\begin{lemma}[Parallel version of \cite{ns17} Lemma B.18, cf. \cite{KhandekarRV06}]
\label{lem:balance-sparse-cut}
There is a randomized \pram\ algorithm that given an undirected, weighted graph $G=(V,E,c)$, a sparsity parameter $\alpha>0$, and a balance parameter $\beta=O(\frac{1}{\log^2{n}})$, with high probability computes 
\begin{itemize}
\item either an $\alpha$-sparse cut $X$ such that $\beta n \leq \card{X}\leq \frac{n}{2}$;
\item or a graph $H$ embeddable in $G$ with congestion at most $O(\log^3/\alpha)$ such that every $(\beta \log^2{n})$-balanced cut in $H$ has sparsity at least $\Omega(1)$.
\end{itemize}
The algorithm has $O(m\cdot \polylog {n})$ work and $O(\polylog{n})$ depth.
\end{lemma}
\begin{proof}
These aforementioned guarantees are achieved by Algorithm~\ref{alg:cut-matching-sparsest-cut}, and its correctness follows directly from \cite{ns17} with no change to the proofs. Specifically the correctness for an $\alpha$-approximate (unconstrained) sparsest cut follows from Lemma B.16 and Corollary B.21, and for an $\alpha$-approximate, $\beta$-balanced sparsest cut follows from Lemma B.18 and Corollary B.22.

We now show that this algorithm admits a $O(m\cdot \polylog {n})$ work and $O(\polylog{n})$ depth \pram implementation. First, observe that in the $t^{th}$ round of the cut-matching game, the cut player's strategy can implemented in $O(mt)$ work and $O(t\log{n})$ depth as described in Lemma \ref{lem:flow-vector-mixing}. Now consider the matching player's strategy in the $t^{th}$ round of the cut-matching game, which consists of $O(\log n)$ sub-rounds. We claim that each sub-round $j$ of the matching player requires $O(m\cdot \polylog {n})$ work and $O(\polylog{n})$ depth. Observe that maintaining $Sl$ and $Sr$ and scaling edge weights can be done in $O(m)$ work and $O(1)$ depth. Since we pick $\eps=O(1)$, the $(1-\eps)$ max-flow algorithm runs in $O(m\cdot \polylog {n})$ work and $O(\polylog{n})$ depth as in Corollary \ref{rst:max-flow}.
Rounding the flow to an integral solution can be done in $O(m)$ work and $O(\log{n})$ depth by \cite{cohen95}. Finally, since the edge capacities are bounded polynomials, we can use our parallel flow decomposition algorithm in Lemma \ref{lem:short-cut} with $\delta= \frac{\eps}{\norm{c}_{1}}$. By Lemma \ref{lem:short-cut}, this takes $O(m\cdot \polylog {n})$ work and $O(\polylog{n})$ depth, and it preserves flow decomposition for a $(1-\eps)$ approximate max-flow. Since there are at most $O(\log n)$ sub-rounds of the matching player in any (outer) round of our cut-matching game, the cut-player's strategy in the $t^{th}$ round can also be implemented in $O(m\polylog n)$ work and $O(\polylog n)$ depth. Finally, since the number of rounds $t$ of the cut-matching game is bounded by $O(\log^2 n)$, the total work and depth of our algorithm is bounded by $O(m\polylog n)$ and $O(\polylog n)$, respectively.
\end{proof}

\paragraph{The algorithm and analysis of Lemma \ref{lem:sparse-and-balance-cut}.} The algorithm follows from polylogarithmically-many parallel applications of \Cref{lem:balance-sparse-cut}.
Concretely, when setting $\beta'=\beta/(10c\log^2{n})$ (or $\beta'=0$ for the sparsest cut), we can guess the value of $\alpha$ as $({c_{\min}}/{n}) \cdot 2^i$ for integer $i$ and return the smallest guess value that returns a sparse cut.
Any graph $G$ whose eligible cuts have sparsity less than $\alpha/\log^3{n}$ will not be embeddable for $H$, which gives us a cut whose sparsity is at least an $O(\log^3{n})$ approximation.
Finally, note that by the balance of partition, an $\alpha$-approximation of the $\beta$-balanced sparsest is always an $O(\alpha/\beta)$-approximation for the $\beta$-balanced min-cut.

To analyze the efficiency, under the assumption that the ratio between the maximum and the minimum capacities is polynomial, the geometrically-increasing guess can be done in parallel with an $O(\log{n})$ multiplicative factor of work and $O(1)$ depth overhead.
Each call to the algorithm of Lemma \ref{lem:balance-sparse-cut} takes $O(m\cdot \polylog {n})$ work and $O(\polylog{n})$ depth, and so the complete algorithm has total work $O(m\cdot \polylog {n})$ and depth $O(\polylog{n})$.

%% file: applications-HC.tex
\newcommand{\Xsparse}{\ensuremath{(X^{spr}, V\setminus X^{spr})}\xspace}
\newcommand{\Xbalance}[1]{\ensuremath{(X_{#1}^{bal}, V\setminus X_{#1}^{bal})}\xspace}

\subsection{Minimum Cost Hierarchical Clustering.}
Hierarchical clustering is a fundamental data analysis tool used to organize data into a dendogram. Given data represented as an undirected weighted graph $G=(V,E,c)$, where the vertices represent datapoints and (positive) edge weights represent similarities between their corresponding end points, the goal is to build a hierarchy, represented as a rooted tree $\mathcal{T}$; the leaves of this tree correspond to individual datapoints (vertices $V$), and the internal nodes correspond to a \emph{cluster} $S \subseteq V$ consisting of their descendent leaves. Intuitively, this tree can be viewed as clustering the vertices of $G$ at multiple levels of granularity simultaneously, with the clustering becoming increasingly fine-grained at deeper levels. \cite{Dasgupta16} initiated the study of this problem from an \emph{optimization} perspective by proposing the following cost function
for similarity-based hierarchical clustering-
\begin{align}
\label{eq:dasgupta}
\cost_G(\mathcal{T})=\sum_{(u, v)\in E} c_{uv} \cdot |\text{leaves}(\mathcal{T}_{uv})|,
\end{align}
where $c_{uv}$ is the weight (similarity) of the edge $(u,v) \in E$ and $\mathcal{T}_{uv}$ is the subtree of $\mathcal{T}$ rooted at the least common ancestor of $u$ and $v$, and $|\text{leaves}(\mathcal{T}_{uv})|$ is the number of descendent leaves in this subtree $\mathcal{T}_{uv}$. 
Intuitively, this objective imposes a large penalty for separating ``similar'' vertices at higher levels in the tree, thereby placing ``similar'' vertices 
closer together.

This minimization problem was shown to be NP-hard, and moreover, no polynomial time constant factor approximation factor was shown to be possible for this objective assuming the \emph{small-set expansion} hypothesis \cite{CharikarC17}. Nevertheless, this objective is considered a reasonable one; \cite{Dasgupta16} and a follow-up work by \cite{Cohen-AddadKMM18} show that it satisfies several properties desired of a ``good'' hierarchical clustering. As a consequence, this objective has been well studied in the literature (cf. \cite{Dasgupta16,CharikarC17,RoyP16,ChatziafratisNC18,Cohen-AddadKMM18,AssadiCLMW22,AKLPNeurips22}, and references therein), where it was shown to have strong algorithmic connections to the sparsest cut-problem. 
Specifically, \cite{Dasgupta16, CharikarC17, Cohen-AddadKMM18} show that 
recursively partitioning the subgraph induced at each internal node using any $\alpha$-approximate sparsest cut oracle results in a cluster tree that is a 
$O(\alpha)$ approximation for Dasgupta's objective. 

While our results do imply a near-linear work, polylogarithmic depth parallel algorithm for computing polylog-approximate sparsest cuts, this by itself does not suffice to give a low-depth, work-efficient parallel algorithm for (approximately minimum-cost) hierarchical clustering. The reason is precisely that the sparsest cut subroutine may produce highly imbalanced partitions, resulting in a cluster tree with super-logarithmic depth. Due to the dependent nature of the recursively generated subgraphs on which the sparsest cut oracle is invoked, the resulting clustering algorithm would have not just super-logarithmic depth, but also super-linear work. 

Fortunately, the solution is relatively simple, which is to recursively partition the graph using balanced min-cuts (with $\beta = \Omega(1)$) instead, guaranteeing that the resulting hierarchy would have logarithmic depth. When utilizing such a balanced min-cut subroutine, \cite{CharikarC17,AssadiCLMW22} show morally the same guarantees for the resultant hierarchical clustering as the unbalanced case. Specifically, they show that recursively partitioning the subgraph induced at each internal node using any $(\alpha,\beta')$-bicriteria approximation oracle for $\beta$-balanced min-cuts $(0<\beta'\leq \beta\leq 1/2)$ results in a hierarchy that is an $O(\alpha/\beta')$ approximation for Dasgupta's cost function.     

This result in combination with our parallel $O(\beta^{-1}\log^3 n , \beta\log^{-2} n)$-bicriteria approximation algorithm for $\beta$-balanced min-cuts presented in Theorem \ref{lem:sparse-and-balance-cut} directly gives us the first \pram algorithm for minimum-cost hierarchical clustering. Formally,

\begin{theorem}
\label{lem:hc-alg-sp-cut}
There is a randomized \pram\ algorithm that given an undirected weighted graph $G=(V,E,c)$, computes with high probability, a hierarchical clustering tree $\mathcal{T}$ that is a $O(\log^5 n)$-approximation for Dasgupta's objective (\Cref{eq:dasgupta}). This algorithm has $O(m\cdot \polylog {n})$ work and $O(\polylog{n})$ depth.
\end{theorem}

Combined with the simulation result of \cite{KarloffSV10,GoodrichSZ11}, the above result implies the first \emph{fully-scalable} \mpc\ algorithm for this problem; our algorithm computes a $O(\log^5 n)$-approximate minimum cost hierarchical clustering in $O(\polylog n)$ rounds, where each machine has local-memory $O(n^{\delta})$ for any constant $\delta > 0$, and the total memory is $O(m\polylog n)$. Prior to this work, the state-of-the-art \mpc\ algorithm for this problem required $\Omega(n\,\polylog n)$ local (per-machine) memory \cite{AKLPNeurips22}.

%% file: applications-faircuts-GH.tex
\subsection{Fair Cuts and Approximate Gomory-Hu Trees.}

In a recent work, \cite{Li+23} introduced the notion of fair
cuts and showed that it is useful in several applications, one of them being the constructing approximate Gomory-Hu trees. Formally, a fair cut is defined as follows.

\begin{definition}[Fair Cut \cite{Li+23}]
Let $G = (V,E,c)$ be an undirected graph with edge capacities $c \in R^E_{>0}$.
For any two vertices $s$ and $t$, and parameter $\alpha \ge 1$, we say that a cut $(S,T)$
is a $\alpha$-fair $(s,t)$-cut if there exists a feasible $(s,t)$-flow such that $f(u,v) \ge \frac{1}{\alpha}\cdot c(e)$ for every $e \in E(S,T)$.
\end{definition}

Note that an $\alpha$-fair $(s,t)$-cut is also
an $\alpha$-approximate $(s,t)$-min-cut but not vice-versa. We show that the results presented in our paper also extend to this stronger notion of approximate min-cuts, improving upon the algorithmic result of \cite{Li+23}. In particular, \cite{Li+23} give a \pram algorithm for computing a $(1+\eps)$-fair cut with $n^{o(1)}/\poly(\eps)$ depth and $m^{1+o(1)}/\poly(\eps)$ work. Importantly, the computational bottleneck in their approach is the construction and resultant quality of a congestion approximator for the input graph, in the sense that their algorithm has depth and work $\poly(\alpha,\epsilon^{-1},\log n)$, and $m~\poly(\alpha,\epsilon^{-1},\log n)$, respectively, plus the depth and work required to construct such an $\alpha$-congestion approximator. The same bottleneck persists in the application of their fair cut idea for computing isolating cuts and approximate Gomory-Hu trees. In their work, \cite{Li+23} utilize a \emph{boundary-linked expander decomposition} of \cite{ChangS19,goranci2021expander} as their congestion approximator, which requires $n^{o(1)}$ depth and $m^{1+o(1)}$ work to construct, producing a $n^{o(1)}$-congestion approximator which gives them their \pram result\footnote{\cite{Li+23} only show their results for unweighted graphs since
their parallel construction of congestion approximators
are only for unweighted graphs; though they mention
in their paper that they believe
known techniques imply their same results for weighted
graphs. We give an explicit construction even for the weighted case.}. Our new polylogarithmic depth, near-linear work construction of congestion approximators from Theorem \ref{thm:alg1} when used as a blackbox in the algorithm of \cite{Li+23} immediately imply the following improvements.

\begin{theorem}
    There is a randomized \pram\ 
   algorithm that given as input an undirected, weighted graph $G = (V,E,c)$, vertices $s,t \in V$, and desired precision $\epsilon >0$, with high probability, computes
   a $(1+\epsilon)$-fair $(s,t)$-cut in $O(m~\poly(\epsilon^{-1},\log n))$ work and $O(\poly(\epsilon^{-1},\log n))$ depth.
\end{theorem}

\begin{theorem}
    There is a randomized \pram\ 
   algorithm that given as input an undirected, weighted graph $G = (V,E,c)$, and desired precision $\epsilon >0$, computes with high probability
   a $(1+\epsilon)$-approximate Gomory-Hu tree in $O(m~\poly(\epsilon^{-1},\log n))$ work and $O(\poly(\epsilon^{-1},\log n))$ depth.
\end{theorem}

%% file: primitives.tex
\section{\texorpdfstring{\pram}{PRAM} Primitives.}\label{sec:pramroutine}

\subsection{Eulerian Tours and Subtree Sum.}
Consider a tree $T$, and for each undirected edge $(u,v)$ create two anti-parallel arcs $(u,v)$ and $(v,u)$; call this new graph $T'$.
Then, $T'$ admits an Eulerian tour, which can also be viewed as DFS traversal of $T$.
This Eulerian tour can be used to compute a number of properties of $T$ efficiently in parallel, and in this section we describe several of use to us.
Most of these results are folklore and can be found in several textbooks on parallel algorithms, such as \cite{akl1997parallel}.

\begin{theorem}[Eulerian Tour \cite{akl1997parallel}]
    There is a $O(1)$ depth, $O(n)$ total work \pram~algorithm which outputs an ordered list of nodes corresponding to an Eulerian tour of a tree $T$ with $n$ nodes starting at a root $r$.
\end{theorem}
Eulerian tours are usually computed via a linked list representation, where each directed arc points to the next element in the tour.
This can also be converted to a random access ordered array, which is useful in several results below.

It is also known that one can compute all prefix sums of an array in $O(\log n)$ depth.
\begin{theorem}[Prefix Sums \cite{akl1997parallel}]
    Given a list of values $a_1,\ldots ,a_n$, there is an $O(\log n)$ depth, $O(n)$ work \pram~algorithm which outputs a new list $b_1,\ldots ,b_n$ such that $b_k = \sum_{i\leq k}a_i$
\end{theorem}

Using prefix sums and Eulerian tours, we can compute several useful properties of nodes in a tree.
\begin{theorem}[\cite{akl1997parallel}]
    \label{thm:misc-euler}
    Let $T$ be a tree rooted at $r$.
    Then, there is a \pram~algorithm which in $O(\log n)$ depth and $O(n)$ total work computes
    \begin{enumerate}
        \item The first and last time each node is visited by an Eulerian tour starting at $r$
        \item The parent of each node
        \item The depth of each node
    \end{enumerate}
    whenever there are at least $\Omega(n)$ processors.
\end{theorem}
\begin{proof}
    Note that the length of the Eulerian tour array is always $O(n)$, as each original edge is traversed exactly twice and all trees over $n$ nodes have $n-1$ edges.
    To find the first time each node is visited by a tour starting at $r$, assign one processor to each pair of consecutive elements in the list.
    Let $u,v$ be these elements, appearing at indices $i-1,i$.
    The processor assigned to the pair then sets $b[(u,v)]=i$, where $b$ is a new lookup table for the ordering of edges.
    To determine the first time a node $u$ is visited by the tour, it suffices to compute $\min_v b[(v,u)]$, and similarly the last time can be determined by $\max_v b[(u,v)]$.
    These minimums and maximums can then be computed in depth $O(\log n)$ \cite{akl1997parallel}.
    The parent of any node first visited at position $i$ in the list is the $i-1$ element in the Eulerian tour list.

    For the depth each node, we say $(u,v)$ is a ``forward'' arc if $u$ is the parent of $v$, and a ``reverse'' edge if $v$ is the parent of $u$.
    In the ordered list of arcs traversed by the Eulerian tour, assign each forward arc a value of $+1$ and each reverse arc a value of $-1$.
    Then, run prefix sum.
    Any cycle in the tour must have zero sum, since it must use both copies of each undirected edge it visits, and so the prefix sum whenever the tour vists $v$ is the depth of $v$.
    Using the previously computed first time each node is visited, we may in parallel output the depth of every node.
\end{proof}

One of the most useful results of this this section is that we can compute \textit{sumtree sums}.
\begin{theorem}[Subtree Sum]
    \label{thm:subtree}
    Let $T$ be a tree where each node $v$ is associated with a weight $w_v$.
    Then, there is a \pram~algorithm with $O(\log n)$ depth and $O(n)$ total work that computes, for every node $u$, the sum of the $w_v$ in the subtree rooted at $u$.
\end{theorem}
\begin{proof}
    As in the proof of Theorem \ref{thm:misc-euler}, classify each arc in the Eulerian tour of $T$ as either forward or reverse.
    Set the weight of each reverse arc to be 0, and the weight of each forward arc $(u,v)$ to $w_v$.
    Then, run prefix sum on the Eulerian tour list of arcs.
    At the last occurrence of $u$, all arcs in the subtree of $u$ must have already been traversed (since the Eulerian tour uses each arc exactly once, and does not visit $u$ again), so the prefix sums includes the sum of all weights in this subtree.
    Moreover, between the first and last visit to $u$, the tour cannot leave the subtree of $u$; if it did, then one of the arcs between $u$ and its parent must have been used at least twice.
    So, the sum of the subtree is exactly $w_u + p[u^{(-1)}]-p[u^{(1)}]$, where $p$ is the array of prefix sums and $u^{(-1)},u^{(1)}$ are the indices of the last and first occurrences of $u$ in the tour, respectively.
    The prefix sum along with first and last occurrences of each node can be computed in $O(\log n)$ depth and $O(n)$ work, and the final sum requires only $O(1)$ depth and $O(n)$ work.
\end{proof}

For example, subtree sums can be used to find the LCA of two nodes in a tree, which we use in Section \ref{sec:min-cut-tree}.
\begin{theorem}
    \label{thm:lca}
    Let $T$ be a tree rooted at $r$.
    Then, for any $u,v\in T$, there is an $O(\log n)$ depth, $O(n)$ work \pram~algorithm to find the least-common ancestor of $u$ and $v$.
\end{theorem}
\begin{proof}
    The algorithm first sets weights of 0 on all nodes besides $u$ and $v$, and sets weight 1 on $u$ and $v$ before computing subtree sums.
    Using the algorithm of Theorem \ref{thm:misc-euler}, also compute the depth of each node.
    Then, output the lowest-depth (i.e.~furthest from the root) node who has subtree sum 2.

    Computing the subtree sum and depth of each node can be done in $O(\log n)$ work and $O(n)$ work.
    Identifying all nodes with sum 2 can be done in $O(1)$ depth and $O(n)$ work, and taking the max depth of these can be done in $O(\log n)$ depth and $O(n)$ work, so the entire algorithm has the desired runtime.

    For correctness, the algorithm computes the lowest-depth node whose subtree contains both $u$ and $v$, which is exactly the LCA of $u$ and $v$.
\end{proof}

\subsection{Tree Separators.}
As also defined in Section \ref{sec:rst-on-trees}, a tree separator node of a tree is a node whose removal results is a forest with components of at most half the size of the original tree.

\begin{definition}[Tree Separator Node, Definition \ref{def:tree-sep}]
    A node $q$ in a tree $T$ is called a tree separator node of $T$ if the forest induced by $T \setminus \{q\}$ consists of trees with at most $|T|/2$ nodes.
\end{definition}

A very useful folklore results shows that every tree contains at least one tree separator node.
\begin{lemma}[folklore]
    \label{lem:tree-sep-exist}
    For all trees $T$, there exists a node $q\in T$ such that $q$ is a tree separator node of $T$.
\end{lemma}
\begin{proof}
    Suppose for contradiction $T$ does not contain a tree separator node.
    Define $M_u$ for each node $u\in T$ to be the size of the largest component of $T \setminus \{u\}$, and let $u'=\argmin_{u\in T}M_u$.
    By assumption, $M_{u'} > |T|/2$, and let $C'$ be the component of $T \setminus \{u'\}$ with size $M_{u'}$.
    Since $T$ is a tree, there is a unique neighbor of $u'$ in $C$; call this node $v$.
$|C'| > |T|/2$, so $|T \setminus C'|\leq |C'| + 1$ and $|T \setminus (C' \cup \{u'\})|\leq |C'|$.
    Thus, $C' \setminus \{v\}$ must be the largest component in $T \setminus \{v\}$, as the sum of all other components can be at most $|C'|-1$.
    But then we have $M_{v} < M_{u'}$, contradicting that $M_{u'}$ is minimal.
\end{proof}

We now give an algorithm to find a tree separator node with logarithmic depth and linear work, and uses the subtree sum algorithm of Theorem \ref{thm:subtree}.

\begin{lemma}
    \label{lem:tree-sep}
    There is a $O(\log k)$ depth, $O(k)$ work \pram~algorithm to find a tree separator node of any tree with $k$ nodes.
\end{lemma}
\begin{proof}
    Let $T$ be a tree with $k$ nodes, and arbitrarily root $T$.
    The algorithm first assigns every node weight 1, and uses the algorithm of Theorem \ref{thm:subtree} to compute subtree sums.
    Let $s_u$ be the value of this subtree sum at node $u$.
    Then, in parallel, for each $u\in T$, check if $s_v \leq k/2$ for all $v$ children of $u$, and check that $k-s_u\leq k/2$.
    Output any such $u$ as a tree separator node.

    Removing any tree separator node, by definition, results in a forest with components of size at most $k/2$, which is the exact condition the algorithm checks.
    A tree separator node exists for all trees (Lemma \ref{lem:tree-sep-exist}), so the algorithm always outputs a tree separator node.
    The depth of the algorithm comes from the fact that the subtree sums and examining all nodes can be done in $O(\log k)$ depth.
    For work, note that the value of every node $u$ needs to be checked twice: once when determining if $k-s_u\leq k/2$, and once when its (unique) parent checks if $s_u\leq k/2$.
    So, combined with the $O(k)$ work for subtree sums, the total work is $O(k)$.
\end{proof}

%% file: cut-matching.tex
\newcommand{\inner}[2]{\langle #1\,,\,#2 \rangle}
\newcommand{\flowwork}[2]{\ensuremath{T_{#1, #2}}\xspace}
\newcommand{\flowdepth}[2]{\ensuremath{D_{#1, #2}}\xspace}

\section{Parallel Implementation of the Cut-Matching Game of \texorpdfstring{\cite{RackeST14}}{RST14}.}
\label{sec:pcmg}

In this section, we discuss the implementation of $\PAp$ in \Cref{sec:parta}, which in turn is built on top of the cut-matching game in \cite{RackeST14}. In the remainder of this section, note that we always operate over the subdivision graph as defined in \Cref{def:subdivision}, i.e.,  we have $V' = V \union X_E$ where $X_E$ is the set of split vertices of the edges in $E$,
$E' = \Union_{e = (u,v)\in E} \setof{ (u,x_e), (x_e,v) }$, and $c'_{(u,x_e)} = c'_{(v,x_e)} = c_e$ for every edge $e\in E$. Moreover, for ease of exposition, whenever we refer to a set of edges $Y$, we in fact refer to the subdivision vertices $X_Y$ corresponding to set $Y$. It is straightforward to see that constructing the subdivision graph only takes $O(m)$ work and $O(1)$ depth. 

All the algorithms that are presented in this section are identical to that of \cite{RackeST14}; we merely discuss them for the sake of completeness and showing that they admit a $O(m\, \polylog{n})$ work and $O(\polylog{n})$ depth \pram implementation. This entire process requires a blackbox access to only two subroutines: a $(1-\epsilon)$-approximate max-flow algorithm $\mathcal{F}_{\eps}$ that also returns the $(1+\epsilon)$-approximate minimum cut, and a $(1-\delta)$-approximate flow-path decomposition algorithm $\mathcal{D}_\delta$ (whose implementation is given in \Cref{sec:shortcutting}). We shall henceforth refer to the work and depth of these algorithms $\mathcal{A}\in \{\mathcal{F}_\epsilon,\mathcal{D}_\delta\}$ on an $m$-edge graph as $T(\mathcal{A},m)$, and $D(\mathcal{A},m)$. To build intuition, we begin by discussing the implementation specifics for the uncapacitated case in \Cref{subsec:flow-vector-sim} through \Cref{subsec:well-linked-edges}, and extend our result to the general capacitated case in \Cref{subsec:racke-weighted}.

\paragraph{Using a cut-matching game to find well-linked edges.} 

Before discussing \cite{RackeST14}, we shall find it useful to first understand the cut-matching game framework of \cite{KhandekarRV06}. It is a technical tool connecting the problem of approximating sparsest cuts or alternatively, certifying expansion to max flows, and is vital in the aforementioned result of \cite{RackeST14}. Conceptually, it is an alternating game between a cut-player and a matching-player, where the former produces a bisection of the vertices, and the latter responds by producing a perfect matching across this bisection that does not necessarily belong to the underlying graph, but can be embedded in it\footnote{Roughly speaking, we say a matching can be embedded into the graph, if there exist flow paths connecting every pair of matched nodes such that the net flow does not exceed the capacity of any edge.}.
The game ends when either the cut player produces a bisection for which the matching player cannot find a perfect matching that can be embedded in it (i.e. a sparse cut has been found), or the union of the perfect matchings produced thus far form an expander (i.e. an expander can be embedded in the underlying graph, certifying its expansion). The objective of the cut player is to terminate this game as quickly as possible, whereas the matching player seeks to delay this. In their work, \cite{KhandekarRV06} showed that there is a near-linear time strategy for the cut-player that guarantees that this game terminates in $O(\log^2n)$ rounds, regardless of the matching player's strategy, and it is easiest to understand when viewed as the following \emph{multicommodity flow problem}: initially, every vertex starts off with a unit of its own unique commodity (which we can represent as a $n$-dimensional one-hot encoded vector, which we shall henceforth refer to as a \emph{flow vector} held by the vertex), and the goal is to uniformly distribute these commodities across all vertices with low congestion. In each round of the game, every matched pair of vertices (according to the perfect matching produced by the matching player) mix (average) their currently held commodities with each other through the matching edge. Since the perfect matching could be embedded into the underlying graph, this operation is guaranteed to be feasible. Now supposing in some round, every vertex ended up with a (near) uniform spread over all unique commodities, then we are done; it implies that the union of the perfect matchings must induce an expander, which moreover can be embedded into the underlying graph. A good cut strategy therefore must find bisections that mix these commodities quickly, and the key to arguing its existence is a potential function that measures the total distance of each vertex's currently held commodities from uniform. In particular, one can show that there always exists a bipartition that guarantees a multiplicative $(1-\Omega(1/\log n))$ factor reduction in this potential, regardless of the perfect matching produced by the matching-player, guaranteeing that the game terminates in just $O(\log^2 n)$ rounds. Moreover, this bisection can be found efficiently. While this sketches the \emph{analysis}, this procedure would be far too inefficient to implement as it would require explicitly maintaining the vector of currently held commodities for every vertex (which would naively take $O(n^2)$ work). Fortunately, \cite{KhandekarRV06} showed that it suffices to consider a \emph{sketch} that is a \emph{random projection} of these flow vectors in every round and still obtain the same convergence guarantees with high probability. The matching players strategy is essentially a max-flow computation with the sources and sinks being the two sides of the bipartition. If the max flow value is $n/2$, then we can obtain a perfect matching via a flow-path decomposition. Otherwise, a perfect matching cannot be embedded into this cut.

The above cut-matching game provides the basis of the algorithm of \cite{RackeST14} for finding well linked edges; intuitively, a set $F$ of edges is well-linked if we can embed an \emph{expander} between the subdivision vertices of these edges, and the cut-matching game is precisely the technical tool used to verify this. The algorithm of \cite{RackeST14} starts off with a candidate set of edges $F$, and the cut-matching game is used to either certify well-linkedness of this set $F$, or to produce a witness of non-expansion (a highly congested cut), in which case $F$ is updated to a new set $F_{\textnormal{new}}$ (intuitively, to the edges in the most congested cut). However, in the event that $F$ is updated, the entire game would have to restart, which would be too expensive (unless the updated set is a constant factor smaller than the old candidate set, in which case it is okay because the number of restarts would be bounded by $O(\log n)$). The modified game of \cite{RackeST14} circumvents this issue by \emph{reusing} rounds of the old game in the event that the candidate set $F$ does not change by much. This is achieved by \emph{moving} flow vectors from the old candidate set $F$ to the new set $F_{\textnormal{new}}$ along paths of constant congestion (found by the flow decomposition algorithm $\mathcal{D}$). However, not all edges in the new set $F_{\textnormal{new}}$ may end up receiving a flow vector from the old set $F$. Following the notation of \cite{RackeST14}, we shall therefore denote all candidate sets $F:= A\uplus R$, where $A$ refers to the set of \emph{active} edges that have a flow vector, and $R$ refers to the remaining edges in $F$. The exact details of this matching player are discussed in \Cref{subsec:pram-matching-player}. The cut-player is also slightly different from the one in \cite{KhandekarRV06}, where instead of a strict bipartition, the player produces a disjoint edge sets corresponding to the ``source'' side and ``sink'' side satisfying certain necessary properties. The exact details of this cut player are discussed in \Cref{subsec:pram-cut-player}.

\subsection{Computing the projections of flow vectors.}
\label{subsec:flow-vector-sim}

We first discuss how to efficiently compute the \emph{sketch}, i.e. random projection of current state of the flow vectors held by every active edge (which evolves starting with a one-hot encoded vector, mixing and moving in every round depending on the matching player's strategy in that round). We remind the reader that given a candidate set of active edges $A$, there are $|A|$ flow vectors, each in $\mathbb{R}^{m}$, and the naive way to average and move flow vector would take $O(|A|\cdot m) = O(m^2)$ work. However, we note that in the algorithmic process, the flow vectors are only used for cut player to find the sources and sinks, and the procedure is implemented by a \emph{projection} onto a random unit vector which we show can be simulated in nearly-linear work. We give a subroutine for such an implementation.
\begin{lemma}
\label{lem:flow-vector-mixing}
Let $\rr$ be any unit vector of dimension $m$, and let $\MM_{t}=M_{t}M_{t-1}\cdots M_{1}$ be a $m \times m$ matrix such that each $M_{i}$ is supported over at most $m'$ non-zero coordinates,
and let $\{\vv_i\}_{i=1}^{m}$ be $m$ standard basis vectors that represent indicators for each of the $m$ unique commodities. 
Then, there exists a \pram~algorithm that computes $\mu^{(t)}_i = \rr^T \, \MM_{t}\, \vv_i$ for every $i$, in $O(t \log n)$ depth and $O((m+m')t)$ work.
\end{lemma}
\begin{proof}
We perform the multiplication from
left to right:
$
\mu_i^{(t)} =
((\rr^T M_t) M_{t-1}) M_{t-2} \ldots
$, where each time
we multiply a vector with a matrix
of at most $m'$ nonzeros, which
takes $O(\log n)$ depth and $O(m')$ work.
Summing over $t$ iterations gives the desired
result.
\end{proof}

As we will see shortly, the operations on the flow vectors in the $t^{th}$ round of the cut-matching game can all be simulated by matrix multiplication of $\rr^T \, \MM_t\, \vv_i$ as prescribed in \Cref{lem:flow-vector-mixing}. Moreover, in all applications where the above subroutine is invoked, we always have the sparsity property $m'=O(m~\polylog n)$ for each matrix $M_i$ since they correspond to fractional matrices that mix or move flows produced by the flow decomposition algorithm $\mathcal{D}$, and $t=O(\polylog n)$ due to the convergence guarantees of the cut-matching game in \cite{RackeST14}. 
We will refer to these matrices $M_i$'s as mix-or-move matrices.
We remark that the idea was implemented by both \cite{KhandekarRV06} and \cite{RackeST14}, albeit it was less explicitly stated.

\subsection{Parallel implementation of the cut player.}
\label{subsec:pram-cut-player}

We now describe the parallel implementation of 
a single iteration of the cut player that achieves the guarantees in Lemma 3.3 
in \cite{RackeST14}.
Recall that the cut player's strategy
given a set of active edges $A$,
is to find two different subsets $A_S \subseteq A$ and  
$A_T \subseteq A$
whose commodities are not \emph{well-mixed}. This is done in the following way: for every edge $e\in A$, let $\mathbf{f}_e$ be the flow vector currently held by $e$. In round $t$ of the game, this is precisely given by $\mathbf{f}_e = M_tM_{t-1}\ldots M_1 \vv_e$, where $\vv_e$ is the indicator (one-hot encoded vector) of edge $e$'s unique commodity, and each $M_i$ represents the fractional mix-or-move matrix representing the matching player's strategy (i.e. mixing or moving) in round $i$. Let  $\boldsymbol{\mu} := (1/|A|)\sum_{e\in A}\mathbf{f}_e$ represent the average flow vector in that round. Given these quantities (neither explicitly computed), and a random vector $\rr$, for every edge $e\in A$, let $\mu_e = \inner{\mathbf{f}_e}{r}$, and $\bar{\mu} := \inner{\boldsymbol{\mu}}{\rr}$ denote the projection of the flow vector of edge $e$, and the average flow vector onto the random vector, respectively. 
For any $B\subseteq A$,
the potential $P_B$ is defined as
\[P_{B}:=\sum_{f\in B}(\mu_f - \bar{\mu})^2\,.\]
Now given a set of active edges $A$, the cut player precisely seeks to identify a set of source edges $A_S$, a set of sink edges $A_T$, along with a value $\eta$ such that (rf. Lemma 3.3 in \cite{RackeST14})
\begin{enumerate}
    \item $\eta$ separates the sets, i.e. $\max_{e\in A_S} \mu_e\leq \eta\leq \min_{f\in A_T} \mu_f$ or $\min_{e\in A_S} \mu_e \geq \eta \geq \max_{f\in A_T} \mu_f$,
    \item $|A_S| \leq |A|/8$ and $|A_T|\geq |A|/2$
    \item for every source edge $e\in A_S$, $|\mu_e-\eta|^2 \geq (1/9)|\mu_e-\bar{\mu}|^2$,
    \item $\sum_{e\in A_S} |\mu_e-\bar{\mu}|^2 \geq (1/160) \sum_{e\in A} |\mu_e-\bar{\mu}|^2$,
\end{enumerate}

Note that, unlike \cite{KhandekarRV06},
$A_S$ and $A_T$ does not need to be an 
exact bi-partition of $A$. The following procedure describes implementation of the cut player's strategy that meets the above requirements in detail.

\begin{tbox}
\textbf{Subroutine} \texttt{find-sources-and-sinks} \cite{RackeST14}

\smallskip
	
\textbf{Input:} subdivision graph $G' = (V', E')$; %
a set of active edges $A \subseteq E$; a sequence of $t$ mix-or-move matrices $M_{t},\ldots,M_1$ each of dimension $m\times m$ and supported over at most $O(m\, \polylog{n})$ non-zero coordinates.

\medskip

\textbf{Output:} a set of source edges $A_{S}\subset A$, a set of sink edges $A_{T}\subset A$, and a separation value $\eta$.

\medskip

\textbf{Procedure:}
\begin{enumerate}
\item Sample a unit vector $\rr$ uniformly at random, and project the flow vector of each $e\in A$ 
to $\rr$ to get $\mu_e = \inner{\MM_t\vv_e}{\rr}$, where $\MM_t = M_tM_{t-1}\ldots M_1$. This step is executed by initializing length-$m$ flow vectors with $\vv_{e}$ the indicator vector of $e$, and then running the algorithm from \Cref{lem:flow-vector-mixing}.

\item Assuming w.l.o.g that $\card{\{e\in A \mid \mu_e <\bar{\mu}\}}\leq \card{\{e\in A \mid \mu_e \geq \bar{\mu}\}}$, pick $L=\{e\in A \mid \mu_e <\bar{\mu}\}$, and let $R=A \setminus L$.  %

\item Compute $P_L$ and $P_R$. If $P_L\geq \frac{1}{20} P_{A}$, set $A_{S}$ as the $\card{A}/8$ edges (or all edges in $L$ if there are fewer than $\card{A}/8$ edges in $L$) with the smallest $\mu_{e}$ values from $L$, $A_{T}$ as $R$, and $\eta$ as $\bar{\mu}$.
\item Otherwise:
\begin{enumerate}
\item Let $\ell = \sum_{e\in L} \card{\mu_e - \bar{\mu}}$, and let $\eta=\bar{\mu}+4\ell/\card{A}$.
\item Let $A_T$ be the edges whose $\mu_e$ is at most $\eta$. 
\item Construct new sets $R'=\{e\in A \mid \mu_e \geq \bar{\mu}+6\ell/\card{A}\}$, and let $A_{S}$ be the $|A|/8$ edges with the largest  $\mu_{e}$ values from $R'$.
\end{enumerate}
\end{enumerate}
\end{tbox}

This subroutine is used repeatedly during the cut-matching game in the hierarchical decomposition procedure of \cite{RackeST14}.
We now show this subroutine can be implemented efficiently in the \pram setting. 

\begin{claim}
\label{clm:cut-matching-source-sink}
There is a \pram implementation of the  subroutine \texttt{find-source-and-sinks} that, given an arbitrary sequence of $t$ matrices, each with support over $O(m\, \polylog n)$ nonzero entries, has $O(t \log{n})$ depth and $O(t m\, \polylog{n})$ work.
\end{claim}
\begin{proof}
By \Cref{lem:flow-vector-mixing}, computing the projections $\mu_e$'s takes $O(\log{n})$ depth and $O(m \, \polylog{n})$ work since the support size of each mix-or-move matrix $M_i$ is $O(m\, \polylog{n})$.
The steps that compute the average and the potential all take $O(\log{n})$ depth and $O(m~\polylog{n})$ work. Finally, taking a subset from $A$ (resp. $L$ and $R$) by checking the $\mu$ values also takes $O(\log{n})$ depth and $O(m)$ work. Summarizing the above steps gives the desired $O(\log{n})$ depth and $O(m\, \polylog{n})$ work.
\end{proof}

\subsection{Parallel implementation of the matching player.}
\label{subsec:pram-matching-player}
We now discuss the parallel implementation of 
the matching player. 
The matching player takes as input the 
source edges $A_S$ and the sink edges $A_T$ computed by the cut player, 
and computes a partial fractional 
matching $M$ between the subdivision nodes $X_{A_S}$
and $X_{A_T}$ in the subdivision graph $G'$. 
The matching player uses a $(1-\epsilon)$-approximate 
max-flow algorithm $\mathcal{F}_{\epsilon}$ 
\footnote{We assume that $\mathcal{F}_{\epsilon}$ is such that it also outputs 
an approximate min-cut. Note that Sherman's algorithm \cite{sherman2013nearly} satisfies this assumption.}
to ensure that the 
matching can be routed in $G'$ with constant 
congestion, and uses the $(1-\delta)$-approximate flow-path decomposition algorithm $\mathcal{D}_\delta$ to construct a matching from the output of the flow algorithm.
At this point, we remind the reader that the cut-matching game in \cite{RackeST14}
deviates from that of \cite{KhandekarRV06}.
In \cite{RackeST14}, the matching player chooses to perform either a \emph{matching} step 
or a \emph{deletion} step
and the choice is made by flipping a fair random coin.
The matching step is similar to \cite{KhandekarRV06} where the flow vectors of 
the matched edges are \emph{mixed} 
resulting in a reduction in potential.
The deletion step deletes the flow vectors 
on some edges (potentially \emph{moving} them to new edges) resulting in a new set of active edges which again leads to potential reduction.
The following procedure describes the matching 
player's strategy from \cite{RackeST14} in detail.

\begin{tbox}
\textbf{Subroutine} \texttt{match-or-delete} \cite{RackeST14}

\medskip
	
\textbf{Inputs:} subdivision graph $G' = (V', E')$; a set of candidate edges $F\subseteq E$, active edges $A\subseteq E$, remaining edges $R \subseteq E$; source edges $A_S\subset A$; sink edges $A_T\subset A$; set of edges $B$. 

\textbf{Output:} a $m\times m$ mix-or-move matrix $M$ with support size at most $O(m~\polylog n)$; new candidate edges $F_{\textnormal{new}}$; new active edges $A_{\textnormal{new}}$; new remaining edges $R_{\textnormal{new}}$; set $B_{\textnormal{new}}$.

\medskip

\textbf{Procedure:}
\begin{enumerate}
\item Construct a capacitated graph $G'_{st}$: add a super-source $s$ and connect $s$ to all subdivision vertices $x\in X_{A_{S}}$ with capacity $1$; add a super-sink $t$ and connect $t$ to all subdivision vertices $x\in X_{A_{T}}$ with capacity $\frac{1}{2}$; add capacity $2$ to all edges in $G'$ that are not incident on $s$ or $t$.
\item Run the flow algorithm $\mathcal{F}_\epsilon$ on $G'_{st}$ with $\eps = \frac{1}{3\log^3{n}}$ to obtain a flow $f$; let $C'\in E'$ be the set of edges that correspond to the approximate min \emph{cut}, and let $C\in E$ be the corresponding set of cut edges in $G$.
\item \label{line:flowdecomp} Run the flow-decomposition algorithm $\mathcal{D}_{\delta}$ on $f$ with $\delta = \frac{1}{3\log^3{n}}$, and let $G$ be the data-structure encoding the flow paths (see Lemma \ref{lem:short-cut}). For every edge $(s,x)$ where $x\in X_{A_S}$ with flow $f_{(s,x)}\geq 1/2$, rescale the flow such that $f_{(s,x)}=1$. For every flow path $p$ that uses edge $(s,x)$, rescale the flow on $p$ by $1/f_{(s,x)}$ to make the flow path consistent, which can be done by propagating this rescaling top-down in $G$. Adjust capacities in $G'_{st}$ to make this new flow feasible. 
\item With probability $\frac{1}{2}$, enter the \textbf{matching case}:
\begin{enumerate}
\item Let $M_0$ be the matrix of the fractional matching between $X_{A_S},X_{A_T}$ induced at the top-level of $G$. If $M_0$ is a partial matching, then
make it perfect by adding self loops. 
Then, the mix-or-move matrix is given by $M \leftarrow \frac{1}{2} M_0 + \frac{1}{2} I$ where $I$ is the identity matrix.

\item \textbf{return} mix-or-move matrix $M$, candidate edges $F_{\textnormal{new}} = F = A\uplus R$, set $B_{\textnormal{new}} = B$ 
\end{enumerate}
\item With probability $\frac{1}{2}$, enter the \textbf{deletion case}:
\begin{enumerate}
\item If $((A\cup R)\setminus A_T)\cup C$ induces a balanced clustering in $G$, \textbf{return} new candidate edges $F_{\textbf{new}} = ((A\cup R)\setminus A_T)\cup C$. In this case, $|F_{\textnormal{new}}| \leq (7/8) |F|$ due to which the cut-matching game restarts from scratch, and the remaining return values are not useful. 
\item\label{line:flow-vector-move} Else set $F_{\textbf{new}} = ((A\cup R)\setminus A_S)\cup C$: Let $M_0$ be the fractional matching between $X_{A_S},X_{A_T}$ induced at the top level of $G$. For each matched pair $(x,x')\in (X_{A_S} \times X_{A_T})$ in $M_0$, identify the first edge $y$ in $C$ on the flow path $x \rightsquigarrow x'$ using the data-structure $G$ and construct a mix-or-move matrix $M$ which moves the flow vector (instead of mixing it) from $x$ to $y$. If the total flow received by $y$ from all $x\in A_S$ exceeds $1$, rescale this total flow to have value $1$. Otherwise, zero out this flow, and add this edge $y$ to set $C_B$.
\item Set $A_{\textnormal{new}} = (A\setminus A_S)\cup (C\setminus C_B)$\footnote{Note that this update is mentioned as $A_{\textnormal{new}} = (A\setminus A_T)\cup (C\setminus C_B)$ in \cite{RackeST14} which is a typo.}, $R_{\textnormal{new}} = F_{\textnormal{new}} \setminus A_{\textnormal{new}}$.
\item \textbf{return} mix-or-move matrix $M$, $F_{\textnormal{new}}$, $A_{\textnormal{new}}$, $R_{\textnormal{new}}$, $B_{\textnormal{new}} = B\cup C_{B}$.
\end{enumerate}
\end{enumerate}
\end{tbox}

Note that although the flow decomposition is an approximation version, the total flow value induced by the decomposition is at least $(1-\frac{1}{3\log^3 {n}})^2 \geq (1-\frac{1}{\log^3{n}})$ of the max-flow value (assuming $n\geq 6$), which satisfies the desired requirement as in \cite{RackeST14}. We now show that the matching-or-deletion subroutine can be implemented under the \pram setting efficiently. 

\begin{claim}
\label{clm:cut-matching-match-delete}
There is a \pram implementation of the subroutine \texttt{match-or-delete} using depth 
$O\paren{D(\mathcal{F}_{\eps}, m) + D(\mathcal{D}_{\delta},m) + \polylog{n}}$ and work $O\big(T(\mathcal{F}_{\eps},m)+ T(\mathcal{D}_{\delta},m)+ m \,\polylog{n}\big)$,
where $D(\mathcal{A} , m)$ and $T(\mathcal{A},m)$ are the depth and work required by the algorithm  $\mathcal{A}\in \{\mathcal{F}_\eps,\mathcal{D}_{\delta}\}$ for $\epsilon,\delta = 1/(3\log^3 n)$, respectively.
\end{claim}
\begin{proof}
The first step in the algorithm is the construction of the graph $G'_{st}$ which can be done in $O(1)$ depth and $O(m)$ work. 
The set $C'$ is returned 
by $\mathcal{F}_\eps$,
and $C$ can be constructed from $C'$ in $O(1)$ depth and $O(m\, \polylog{n})$ work. 
There are at most $O(m\, \polylog{n})$ flow paths. Hence, using the data structure $G$ returned by the flow decomposition algorithm $\mathcal{D}_{\delta}$, re-scaling of the flows and adjusting capacities takes $O(\log{n})$ depth and $O(m\, \polylog{n})$ work. %
We now analyze the required depth and work for both the matching and the deletion cases:
\begin{itemize}
\item In the matching case, since the fractional matching matrix has support size  $O(m\, \polylog{n})$, adding self-loops and computing the mixing matrix $M$ takes $O(m\, \polylog{n})$ work 
and $O(1)$ depth.

\item In the deletion case, checking whether an induced clustering is balanced in Line 5a can be done by simply deleting the candidate edges and checking the sizes of the resulting connected components which takes $O(m)$ work and $O(\log n)$ depth.
If we enter Line~\ref{line:flow-vector-move}, since there are at most
$O(m\, \polylog{n})$ matched pairs $(x,x')$, identifying the first edge in 
$C$ on the flow path $x\rightsquigarrow x'$ for 
every matched pair requires $O(\polylog(n))$ depth and $O(m \, \polylog(n))$ work 
using the data structure $G$.
The moving matrix $M$ can then be computed in 
$O(m\, \polylog{n})$ work and $O(1)$ depth.
Rescaling and deleting flows can also be performed 
using $O(m\, \polylog{n})$ work and $O(1)$ depth.
\end{itemize}
Taking the worst-case work and depth among the cases gives us the desired statement.
\end{proof}

\subsection{Putting it together: \texttt{partition-A1}.}
\label{subsec:well-linked-edges}
In this section we combine the 
implementation of the cut and matching players 
to achieve the guarantees of 
Lemma 3.1 in \cite{RackeST14}. This is also the algorithm that achieves the guarantees we claim in \Cref{lem:parta}.
Specifically, given a set of edges $F$ that induces a $3/4$-balanced clustering of the graph,
the goal of Lemma 3.1 is to find a new 
set of edges $\Fnew$ such that (rf. Lemma 3.1 in \cite{RackeST14})
\begin{enumerate}
\item either $\card{\Fnew} \leq \frac{7}{8}\card{F}$;
\item or $\Fnew=A\cup R$ such that $\card{A}\leq \card{F}$, $\card{R}\leq \frac{2}{\log{n}}A$, and the edges in $A$ are $\Omega(1/\log^2{n})$-well-linked.
\end{enumerate}
The following presents  the 
details of the algorithm that is used to prove Lemma 3.1 in \cite{RackeST14}.

\begin{tbox}
\textbf{Subroutine} \texttt{partition-A1}

\medskip
	
\textbf{Input:} Subdivision graph $G' = (V',E')$ of $G=(V,E)$; a set of edges $F\subseteq E$ that induce a $3/4$-balanced partition of $V$.

\textbf{Output:} A new set of edges $F_{\textnormal{new}}$ that also induces a $3/4$-balanced partition of $V$ such that either
\begin{enumerate}
    \item $|F_{\textnormal{new}}|\leq (7/8)|F|$, or
    \item $F_{\textnormal{new}} = A\cup R$ with $A,R$ disjoint such that edges in $A$ are $\Omega(1/\log^2n)$-well-linked, and $|R|\leq 2|A|/\log n$.
\end{enumerate}

\medskip

\textbf{Procedure (cut-matching game):}
\begin{enumerate}
    \item Initialize matching matrix $M_0$ corresponding to self-loops; $A=F$; $R=\emptyset$; $B=\emptyset$.
    \item For $t=1,2,\ldots,O(\log^2 n)$:
    \begin{enumerate}
        \item $(A_S,A_T,\eta) \gets$ \texttt{find-source-and-sinks}$(G',A,\{M_i\}_{0\leq i < t})$.
        \item $(M_t,F_{\textnormal{new}}, A_{\textnormal{new}}, R_{\textnormal{new}}, B_{\textnormal{new}})\gets$\texttt{match-or-delete}$(G',F,A,R,A_S,A_T,B)$.
        \item If $|F_{\textnormal{new}}|\leq (7/8)|F|$: \textbf{return} $F_{\textnormal{new}}$ (return condition 1). 
        \item Else, update $B=B_{\textnormal{new}}$, $A=A_{\textnormal{new}}$, $R=R_{\textnormal{new}}$. 
        \item Compute potential $P_A$ of active edges $A$. 
        \item If $P_A \leq 1/(16n^2)$ (flow vectors in $A$ have mixed): 
        \begin{enumerate}
            \item If $|B|\leq 2|A|/\log n$: \textbf{return} $F_{\textnormal{new}} = A\cup B$ (return condition 2).
            \item Else, we necessarily have $|A\cup R|\leq (7/8)|F|$: \textbf{return} $F_{\textnormal{new}} = A\cup R$ (return condition 1). 
        \end{enumerate}
\end{enumerate}
\end{enumerate}
\end{tbox}

With Claims \ref{clm:cut-matching-source-sink} and \ref{clm:cut-matching-match-delete}, it is not difficult to check that the whole procedure of step 2 can be implemented in poly-logarithmic depth and nearly-linear work. We formalize this as follows.
\begin{lemma}
 \label{lem:racke-connected-edges}
 There is a \pram implementation of the above algorithm that has depth\\ 
$O\paren{(D(\mathcal{F}_{\eps},m)+D(\mathcal{D}_{\delta},m)) 
 \cdot \polylog{n}}$ and work $O\big((m+T(\mathcal{F}_{\eps},m) + T(\mathcal{D}_{\delta},m) )\cdot \polylog{n}\big)$ for $\eps,\delta = O(1/\log^3 n)$.
\end{lemma}
\begin{proof}
Firstly, the number of iterations of 
the cut-matching game is upper bounded by $O(\log^2 n)$. 
Then the  lemma follows immediately from the 
work and depth calculations of the cut and matching 
steps in  
Claims \ref{clm:cut-matching-source-sink} and \ref{clm:cut-matching-match-delete}, respectively.
\end{proof}

\subsection{Extending to the capacitated case.}
\label{subsec:racke-weighted} 

Although \cite{RackeST14} states that their
results hold for capacitated graphs,
they do not detail this extension.
However, for completeness,
we sketch the implementation
of the cut matching game in~\cite{RackeST14} on capacitated graphs,
and discuss how they can be implemented
in \pram.
Throughout this section we assume
graphs have integer capacities bounded by
$\poly(n)$.

We note that the analysis of the cut matching
game follows by viewing each edge $e$
of capacity $c_e$ as $c_e$ uncapacitated parallel copies. It thus remains to show that the cut matching game can be implemented
with $O(m\polylog n)$ work and
$\polylog n$ depth, assuming a
$(1-\eps)$-approximate maxflow algorithm
with the same work and depth.
We first describe a key subroutine that we need throughout the cut matching game,
namely \textit{averaging of flow vectors},
and then describe the changes we make to the cut and matching players.

\paragraph{Averaging flow vectors.}
Note that naively, treating each capacitated edge as uncapacitated copies can potentially
lead to a large number of edges, and maintaining flow vectors on them will be too costly. Therefore
we will always maintain the invariant that the parallel copies of the same capacitated edge carries the
same flow vector. Whenever this variant gets violated (e.g. after a flow mixing or deletion step),
we restore it by averaging out the flow vectors across the parallel copies.
Thereby, we never store more than $O(m\polylog(n))$ flow vectors at any point.
Notice that in the actual implementation of the cut matching game,
we never explicitly store the flow vectors but only store their projections;
and thus we only need to average the projections of these vectors, which
is equivalent to first averaging the flow vectors and then taking projections by linearity.
The averaging of the projections can be done in $O(\log n)$ depth and near-linear work
by parallel summation.

Moreover, as we show
in the claim below, the potential function only decreases after the averaging
of any collection of flow vectors.
Recall that in a graph with $m_0$ uncapacitated edges
$e_1,\ldots,e_{m_0}$ with flow vectors
$f_{e_1},\ldots,f_{e_{m_0}}$ on them,
the potential function in \cite{RackeST14} is defined to be 
\begin{align*}
    \Phi_f :=
    & \min_{c}
    \sum_{i=1}^{m_0} \norm{f_{e_i} - c}^2 =
    \sum_{i=1}^{m_0} \norm{f_{e_i} - \mu}^2,
\end{align*}
where $\mu = \frac{1}{m_0} \sum_{i=1}^{m_0} f_{e_i}$ is the average of all flow vectors.

\begin{claim}
    Let $E' \subseteq E$ be any collection of uncapacitated edges.
    Define new flow vectors $f'_{e_1},\ldots,f'_{e_{m_0}}$ by averaging out
    the flow vectors on edges in $E'$, namely,
    \begin{align*}
        f'_{e} =
        \begin{cases}
            \frac{1}{|E'|} \sum_{e\in E'} f_e & e\in E' \\
            f_e & e\notin E'
        \end{cases}.
    \end{align*}
    Then the potential function can only decrease going from $f$ to $f'$:
    \begin{align*}
        \Phi_f \geq \Phi_f'.
    \end{align*}
\end{claim}
\begin{proof}
    Let $\mu = \frac{1}{m_0} \sum_{i=1}^{m_0} f_{e_i}$ be the average flow vector
    of all edges. Note that this is also the average flow vector with respect to $f'$,
    since averaging the flow vectors of edges in $E'$ does not change the total sum of the flow
    vectors.
    Thus it suffices to compare the contribution of edges in $E'$ to the potential function
    with respect to $f,f'$ respectively.
    To this end, we write the contribution with respect to $f'$ as
    \begin{align*}
        \sum_{e\in E'}
        \norm{f'_e - \mu}^2
        =
        & \sum_{e\in E'}
        \norm{\frac{1}{|E'|} \sum_{e\in E'} f_e- \mu}^2 \\
        =
        & |E'|\cdot
        \norm{\frac{1}{|E'|} \sum_{e\in E'} f_e- \mu}^2.
    \end{align*}
    Letting $\Dcal$ denote the uniform distribution over edges
    in $E'$, we can write the above contribution as
    \begin{align*}
        |E'|\cdot \norm{ \expec{e\in\Dcal}{f_e} - \mu }^2.
    \end{align*}
    Notice that the function $f(x) = \norm{x - \mu}^2$ is quadratic and hence
    also convex. Therefore we can apply Jensen's inequality and obtain
    \begin{align*}
        |E'|\cdot \norm{ \expec{e\in\Dcal}{f_e} - \mu }^2
        \leq & |E'|\cdot \expec{e\in\Dcal}{\norm{ {f_e} - \mu }^2} \\
        = & \sum_{e\in E}\norm{ {f_e} - \mu }^2.
    \end{align*}
    That is, the contribution of edges in $E'$ with respect to $f'$ is at most that
    with respect to $f$, implying that the potentail of $f'$ can only be smaller than that of $f$,
    as desired.
\end{proof}

\paragraph{The cut and matching players.}
The cut player uses the exact same strategy to find the sources and sinks, except that now
they do not explicitly operate all parallel copies, but rather manipulate the parallel copies
of the same edge together, by exploiting the fact that they all have the same flow vectors.
For example, when computing projection of the flow vectors, we only need to do the computation once
for the parallel copies of the same edge; when computing the average of the flow vectors, we just
need to compute a weighted average where each flow vector is weighted by the capacity of the corresponding edge. Notice that when choosing $A_S$, we could end up only choosing a subset of
the parallel copies of the same edge, but leaving the remaining parallel copies out. Then the matching
player will create a flow graph by connecting super source to the split vertex of the edge
with capacity being the number of parallel copies in $A_S$.

The matching player also adopts almost the same strategy as the uncapacitated case,
with the following modifications:
\begin{enumerate}
    \item Most notably, in both matching case and deletion case, where we mix or move flow vectors,
    we average the flow vectors of parallel copies of the same edge afterwards.
    This is because different copies of the same edge could be matched differently, resulting
    in different flow vectors after mixing/moving. However,
    since the matching we found has support size $O(m\polylog(n))$,
    the total number of distinct flow vectors is always bounded by $O(m\polylog(n))$.
    Notice that once again,
    we never explicitly average the flow vectors but only average their projections.
    \item
    At Line~\ref{line:flowdecomp} of \texttt{match-or-delete},
    after we find a flow decomposition,
    we scale the flow paths as follows.
    For every edge $(s,x)$ with flow $f_{(s,x)}\geq (1/2)c_{(s,x)}$, rescale the flow such that $f_{(s,x)}=c_{(s,x)}$; For every flow path $p$ that uses edge $(s,x)$, rescale the flow on $p$ by $c_{(s,x)}/f_{(s,x)}$ to make the flow path consistent, which can be done by propagating this rescaling top-down in the flow decomposition DAG in $O(m\polylog(n))$ work and $\polylog(n)$ depth.
    Adjust capacities in $G'_{st}$ accordingly.
    \item At Line~\ref{line:flow-vector-move} of \texttt{match-or-delete},
    if the total flow received by $y$ exceeds $c_y$, rescale this total flow to have value $c_y$. Otherwise, zero out this flow, and add (the parallel copies of) this edge $y$ to set $C_B$.
    Notice that the rescaling of the flow can again be done by propagating it top-down
    in the flow decomposition DAG in $O(m\polylog(n))$ work and $\polylog(n)$ depth.
\end{enumerate}

%% file: Sherman.tex
\label{sec:shermans}
In this section, we discuss the implementation details of vanilla Sherman's algorithm \cite{sherman2013nearly} in the \pram model, which forms the basis of the near-linear work, polylogarithmic depth approximate max-flow subroutine invoked in our cutting-scheme in Section \ref{sec:new-rst}. This algorithm consists of an outer-algorithm that makes $O(\log m)$ many iterative calls to an inner procedure \texttt{AlmostRoute} that actually performs the gradient descent. At the end of these calls, we are left with a minimum congestion flow that is \emph{almost} feasible, in the sense that there is a negligible residual demand that can be routed in the flow network with $O(1/\poly(m))$ congestion. This outer-algorithm terminates by routing this residual demand along a maximum spanning tree, achieving feasibility of the superimposed flows (i.e. the resultant flow routes the desired demands $b$). This relatively simpler outer algorithm is described below, with the bulk of the technical detail being contained in the \texttt{AlmostRoute} subroutine that implements gradient descent.

\begin{algorithm}{$\texttt{approximate maximum flow}(G,R,b,\alpha,\epsilon)$}\\
    \label{alg:sherman-outer}\nonl\textbf{Input:} Graph $G=(V,E,c)$; $\alpha$-congestion approximator $R$ that is a hierarchical decomposition of $G$; vertex demands $b\in \mathbb{R}^V$; quality of the congestion approximator $\alpha=O(\polylog n)$; precision $\eps >0 $. \\
    \nonl\textbf{Output:} A $(1+\epsilon)$-approximate minimum congestion flow $f\in \mathbb{R}^E$ that routes demands $b$; $(1-\epsilon)$-approximate maximum congested cut $S$.\\
    \textbf{Procedure:}
    \begin{algorithmic}[1]
        \State $b_0 \gets b$; compute $B$, the vertex-edge incidence matrix of $G$.  
        \State $(f_0,S_0) \gets \texttt{AlmostRoute}(G,R,b_0,\alpha,\epsilon)$ 
        \For{$i\gets 1 \ldots \log (2m)$}
            \State $b_i \gets b_{i-1} - Bf_{i-1}$
            \State $(f_i,S_i)\gets \texttt{AlmostRoute}(G,R,b_i,\alpha,1/2)$.\algorithmiccomment{subsequent $S_i$ are not needed}
        \EndFor
        \State Let $t = \log (2m)$ be the final iteration counter of the above loop; set $b_{T}\leftarrow b_t - Bf_t$.
        \State Compute the maximum spanning tree $T$ of $G$.
        \State Let $f_T$ be the flow obtained by routing demands $b_T$ on the tree $T$
        \State \Return flow $f = f_T + \sum_{i=1}^{\log(2m)} f_i$; cut $S = (S_0,\overline{S_0})$.
    \end{algorithmic}
\end{algorithm}

It is easy to see that the above outer-algorithm admits an efficient \pram implementation, assuming that the subroutine \texttt{AlmostRoute} admits a \pram implementation with near-linear work and polylogarithmic depth; there are only $O(\log n)$ many iterations in the outer algorithm, and lines 1,4,6, and 8 can easily be implemented with $O(m)$ work and $O(1)$ depth, and the maximum spanning tree construction in line 7 has a known $O(m)$ work, $O(\polylog n)$ depth \pram algorithm \cite{pram-mst}.

We next discuss the implementation specifics of the subroutine \texttt{AlmostRoute} which actually performs the optimization and is more involved. The key idea behind this subroutine is to transform the constrained optimization problem (for undirected graphs) given in Eqn.~\ref{eq:mincongestioneq} into an unconstrained one using the following \emph{congestion potential} function
\[\phi(f) = \lm(C^{-1}f) + \lm(2\alpha R(b-Bf)),\]
where for any $x\in \mathbb{R}^k$, 
\[\lm(x) := \log \left(\sum_{i=1}^k (e^{x_i} + e^{-x_i})\right)\]
is the symmetric softmax function, a differentiable approximation of $||\cdot||_{\infty}$. Algorithmically, this is achieved via a standard gradient descent which finds a $(1-\epsilon)$-approximate maximum flow $f$ after at most $O(\epsilon^{-3}\alpha^2\log n)$ iterations. Therefore, an efficient \pram implementation of this algorithm effectively reduces to finding an efficient implementation of a single iteration of the descent step within this algorithm, a formal description of which is given below.

\begin{algorithm}{$\texttt{AlmostRoute}(G,R,b,\alpha,\epsilon)$}\\
    \label{alg:sherman-almostroute}\nonl\textbf{Input:} Graph $G=(V,E,c)$; $\alpha$-congestion approximator $R$ that is a hierarchical decomposition of $G$; vertex demands $b\in \mathbb{R}^V$; quality of the congestion approximator $\alpha=O(\polylog n)$; precision $\eps >0 $. \\
    \nonl\textbf{Output:} A $(1+\epsilon)$-approximate minimum congestion flow $f\in \mathbb{R}^E$ that routes demands $b$; $(1-\epsilon)$-approximate maximum congested cut $S$. \\
    \textbf{Procedure:}
    \begin{algorithmic}[1]
        \State Initialize $f \gets 0$; compute  $k_b\leftarrow (16\log n)/(2\alpha\epsilon ||Rb||_{\infty})$; scale $b\leftarrow k_b\cdot b $.  
        \Repeat
            \State Set $k_f\leftarrow 1$; scaling factor $s\leftarrow 17/16$.
            \While {$\phi(f) <16 \epsilon^{-1}\log n$}
                \State Scale $k_f\leftarrow s\cdot  k_f$; $f\leftarrow s\cdot f$; $b\leftarrow s\cdot b$.
            \EndWhile
            \State Set $\delta \leftarrow \sum_{e\in E} \left|c_e \cdot \frac{\partial \phi(f)}{\partial f_e}\right|$.
            \If {$\delta \geq \epsilon/4$}
                \State For each edge $e\in E$, update $f_e\leftarrow f_e - \textnormal{sign}\left(\frac{\partial \phi(f)}{\partial f_e}\right) \cdot \frac{\delta c_e}{1+4\alpha^2}$.
            \Else
                \State Undo scaling $f\leftarrow f/k_f$, $b\leftarrow b/(k_bk_f)$. 
                \State Compute the maximum congested cut $(S,\overline{S})$ from the $(n-1)$ threshold cuts of vertex potentials $\{\pi_v\}_{v\in V}$ induced due to $\partial \phi(f)/\partial f_e$ (described shortly).
                \State \Return flow $f$, cut $S$ .
            \EndIf
        \Until{\bf termination}
    \end{algorithmic}
\end{algorithm}

Observe that implementing the above algorithm requires us to compute $(i)$ the value of the potential function $\phi(f) $, and $(ii)$ the partial derivatives $\partial \phi(f)/\partial f_e$ of the potential with respect to the flow on each edge in the graph. Additionally, we also need to bound the total number of iterations of the above algorithm (lines 2, 4), for which we directly leverage the result of Sherman (Lemma 2.5 in \cite{sherman2013nearly}), which shows that the total number of iterations until termination (line 2) is $O(\alpha^2\epsilon^{-3}\log n)$, and within each iteration, the total number of times we scale the flow and demands (line 4) is $O(\log \alpha)$. We now show how to compute the value of the potential, and its partial derivatives. In order to do so, we shall find it instructive to understand the structure of the congestion approximator $R$.

The congestion approximator $R\in \mathbb{R}^{x\times n}$ is a matrix with each row $i\in [x]$ corresponding to a cut in the graph, and each column corresponding to a vertex. For any cut $i = (S_i,\overline{S_i})$ considered by the congestion approximator, entry $R_{i,v}\in \{0,1\}$ indicates whether vertex $v$ lies on the $S_i$ side of the cut, normalized by the total capacity $c(S_i,\overline{S_i})$ of the cut, i.e. the sum of capacities of all edges crossing this cut. Therefore, the product $[Rb']_i$ of this row of the congestion approximator with any demand vector $b'$ gives the congestion that would be induced by routing these demands across the cut $(S_i,\overline{S_i})$. However, we cannot explicitly construct this matrix due to work and depth constraints, and instead shall use the specific structure of the congestion approximator to efficiently compute these congestion values for all cuts explicitly considered by the approximator. 

The congestion approximator in our case is a $O(\log n)$ depth rooted tree $T$ corresponding to a hierarchical decomposition of the flow instance $G=(V,E)$ upon which Sherman's algorithm is invoked, with the leaves corresponding to the vertices $v\in V$ in the flow network, and the internal nodes corresponding to a cluster consisting of the leaf vertices in the subtree rooted at that internal node. This hierarchical decomposition tree $T$ can equivalently be viewed as a set of cuts in the input graph; each node $i\in T$ in this tree corresponds to a cut $(S_i,\overline{S_i})$, where $S_i$ is the set of vertices corresponding to the leaves in the subtree rooted at node $i$ in the tree $T$. In the following analysis, we shall leverage this view of the congestion approximator in order to efficiently compute the value of the congestion potential, as well as its partial derivatives.

We begin by decomposing the congestion potential into its two components 
\[\phi(f) = \phi_1(f) + \phi_2(f); \text{ where } \phi_1(f) = \lm(C^{-1}f), \text{ and } \phi_2(f) = \lm(2\alpha R(b-Bf)).\]
To compute the first component $\phi_1(f)$, we can simply compute the congestion $f_e/c_e$ of every edge $e\in E$ in parallel, followed by an aggregation step, which can be done with $O(m)$ total work and $O(\log n)$ depth. 

To compute the second component $\phi_2(f)$, we can compute the residual demands $b_v-\sum_{e\in \delta(v)} B_{v,e}f_e$ for every vertex $v\in V$ where $\delta(v)$ corresponds to the set of edges incident on vertex $v$. This also requires $O(m)$ total work and $O(\log n)$ depth (with every vertex first reading the flow values of incident incoming edges followed by those of incident outgoing edges in two separate passes to avoid read collisions). The total demand of any subset of vertices in the congestion approximator (rooted-tree) can then be computed with $\widetilde{O}(n)$ total work and $O(\log n)$ depth using subtree sums. Given the capacity $c(S_i,\overline{S_i})$ of every cut $i=(S_i,\overline{S_i})$ represented by the internal nodes in our congestion approximator, the second term can then be computed via an aggregation, which can be done with $O(m)$ total work and $O(\log n)$ depth.

To compute the partial derivatives, we first consider the component $\phi_1(f)$ in our congestion potential. Then we have that for any edge $e\in E$, the partial derivative 
\[\frac{\partial \phi_1(f)}{\partial f_e} = \frac{\exp(f_e/c_e) - \exp(-f_e/c_e)}{c_e\cdot \exp(\phi_1(f))}\]
which can easily be computed with $O(m)$ work and $O(1)$ depth when the potential $\phi_1(f)$ is known (its computation is described above). 

To compute the partial derivative of the second component $\phi_2(f)$, let $\mathcal{I}$ be the set of all cuts (rows) considered by our congestion approximator, and for any cut $i=(S_i,\overline{S_i})\in \mathcal{I}$, let $y_i = 2\alpha[R(b-Bf)]_i$ be the congestion induced by the residual demands across cut $i=(S_i,\overline{S_i})$. Then we have for any edge $e\in E$, the partial derivative
\[\frac{\partial \phi_2(f)}{\partial f_e} = \sum_{i\in \mathcal{I}} \frac{\partial \phi_2(f)}{\partial y_i}\cdot \frac{\partial y_i}{\partial f_e} = \sum_{i\in \mathcal{I}} \frac{\exp(y_i)-\exp(-y_i)}{\exp(\phi_2(f))}\cdot \frac{2\alpha B_{S_i,e}}{c(S_i,\overline{S_i})},\]
where $c(S_i,\overline{S_i})$ is the capacity of the cut $i=(S_i,\overline{S_i})$ considered in our congestion approximator, and (with some abuse of notation) $B_{S_i,e} = \sum_{v\in S_i}B_{v,e}\in\{-1,0,1\}$ is an indicator of whether in cut $i$, edge $e$ is an incoming edge $(1)$, outgoing edge $(-1)$ or does not cross it $(0)$. The cuts $\mathcal{I}$ are not arbitrary. Rather, they are induced by a single rooted hierarchical decomposition tree $T$, which we can use to efficiently compute this partial derivative for every edge. For an edge $e=(u,v)\in E$, let $T_{u,v}$ be the unique path between $u,v$ in $T$. Then we have that 
\[\frac{\partial \phi_2(f)}{\partial f_e} = \sum_{i\in T_{u,v}} \frac{\exp(y_i)-\exp(-y_i)}{\exp(\phi_2(f))}\cdot \frac{2\alpha B_{S_i,e}}{c(S_i,\overline{S_i})}.\]
Now observe that for any internal node $i$ (corresponding to some cut $(S_i,\overline{S_i})$) that is encountered on the path $T_{u,x}$ between $u$ and the \emph{least-common-ancestor} $x$ of $u,v$ in the rooted tree $T$, we have $B_{S_i,e} = -1$, and for any internal node $i$ that is encountered on the path $T_{x,v}$ between $v$ and $x$ in $\mathcal{T}$, we have $B_{S_i,e} = +1$. Now for any internal node $j$ in $T$, let $T_{j,r}$ denote the unique path in $T$ from the root $r$ of $T$ to node $j$. Then we can define for every internal node $j$ in $T$, a node potential $\pi_j$ as 
\[\pi_j := \sum_{i\in T_{j,r}} \frac{\exp(y_i)-\exp(-y_i)}{\exp(\phi_2(f))}\cdot \frac{2\alpha }{c(S_i,\overline{S_i})},\]
which is easy to compute with $O(n)$ total work and $O(\log n)$ depth through a prefix sum on an Eulerian tour of $T$ that starts and ends at the root $r$ of $T$. This is achieved by setting the weight of the forward edge entering an internal node $i$ from its parent to be $+(\exp(y_i) - \exp(-y_i))/(\exp(\phi_2(f))) \cdot (2\alpha/c(S_i,\overline{S_i}))$ and the reverse edge leaving the internal node $i$ going to its parent to be $-(\exp(y_i) - \exp(-y_i))/(\exp(\phi_2(f))) \cdot (2\alpha/c(S_i,\overline{S_i}))$. Therefore, sum corresponding to subtrees in the prefix sums evaluate to $0$, leaving just the sum of the root $r$ to node $i$ path. Given these node potentials, it is now easy to compute the partial derivatives of any edge $e=(u,v)$ as
\[\frac{\partial \phi_2(f)}{\partial f_e} = \pi_v - \pi_u,\]
which requires just $O(m)$ total work and $O(\log n)$ depth. 

Lastly, Sherman shows that these vertex potentials $\pi_v$ induced by the flow when it is approximately optimal (i.e. when the subroutine terminates) also allow us to efficiently recover the approximate minimum cut (equivalently, the approximate maximum congested cut). Specifically, one of the threshold cuts with respect to the vertex potentials is an approximate min-cut, and this can be computed efficiently in $\widetilde{O}(m)$ total work and $O(\log n)$ depth by sorting the vertices by their potential values and returning the most congested cut from the resulting $n-1$ threshold cuts. Therefore, we have that Sherman's algorithm admits an efficient implementation in the \pram model.

%% file: dp.tex
\label{sec:min-cut-tree}

To compute a hierarchical decomposition on trees in Section \ref{sec:rst-on-trees}, we need to compute an \textit{exact} $s$-$t$ min-cut cut on a congestion approximator tree with the addition of a super-source $s$ and super-sink $t$.
In this section, we show how to compute the exact min-cut when the tree has $O(\log n)$ depth, which is simpler, before extending it to trees with arbitrary depth.

\subsubsection*{The easy case: $O(\log n)$ depth.}
Let $T$ be a tree rooted at $r$ with $O(\log n)$ depth, let $s\not\in T$ be a super-source, and $t\not\in T$ be a super-sink; $s$ and $t$ may be connected arbitrarily to $T$ and these edges may have arbitrary capacity.
For a $s$-$t$ min-cut $(S,\bar{S})$ of $T\cup \{s,t\}$, without loss of generality $s\in S$ and $t\in \bar{S}$.
As such, it remains to determine which nodes of $T$ are in $S$ and which are in $\bar{S}$, and so we consider finding a $s$-$t$ min-cut as a tree problem.
To account for the capacities of edges incident on $s$ or $t$, for each $u\in T$, we set node weights $w_u^{s}=c_{us}$ and $w_u^{t}=c_{ut}$, where $c_{us},c_{ut}$ are the capacities of the $(u,s)$ and $(u,t)$ edges, respectively, or 0 if no such edge exists.

To find the min-cut capacity, we can use dynamic programming.
For each $u\in T$, define $\cut^{s}(u)$ and $\cut^{t}(u)$ as the $s$-$t$ min-cut capacity of the subtree rooted at $u$ with the restriction that $u$ is on the $s$ side of the cut (i.e.~$u\in S$) or $u$ is on the $t$ side of the cut (i.e.~$u\in \bar{S}$), respectively.
The recurrence relations are then $\cut^{s}(u)=w_u^{t}+\sum_{v\in D(u)}\min \{\cut^{s}(v), c_{uv}+\cut^{t}(v)\}$ and $\cut^{t}(u)=w_u^{s}+\sum_{v\in D(u)} \min \{c_{uv}+\cut^{s}(v),\cut^{t}(v)\}$, where $D(u)$ is the set of children of $u$.
Since we assume $T$ has $O(\log n)$ depth, standard dynamic programming techniques allow us to compute $\min \{\cut^{s}(r),\cut^{t}(r)\}$, which is the $s$-$t$ min-cut capacity, in $O(\log n)$ depth and $O(n)$ work.
For brevity, we have presented computing only the capacity of the min-cut, but the actual cut may be found by storing the argmin for each minimum taken in the recurrence relations.

Before continuing to the arbitrary depth case, we introduce a (slightly) generalized problem where some vertices are constrained to be in $S$, or $\bar{S}$, an extension useful when extending to trees of arbitrary depth.
\begin{definition}[$(F_s,F_t)$-Restricted $s$-$t$ Min-Cut]
    \label{lem:low-depth-cut}
    Let $T$ be a tree, let $s$ be a super-source and let $t$ be a super-sink.
    Then, given disjoint subsets $F_s,F_t$ of nodes of $T$, a \textit{$(F_s,F_t)$-Restricted $s$-$t$ Min-Cut} of $T \cup \{s,t\}$ is a minimum $s$-$t$ cut $(S,\bar{S})$ under the restriction that $F_s \cup \{s\}\subseteq S$ and $F_t \cup \{t\} \subseteq \bar{S}$.
\end{definition}

The DP presented before can be easily modified to also solve this extended version when $T$ has $O(\log n)$ depth, using recurrence relations
\begin{align}
    \rcut^{s}(u) &=
    \begin{cases}
        w^{t}_u + \sum_{v\in D(u)} \min \{\rcut^{s}(v),c_{uv}+\rcut^{t}(v)\} &\text{if $u\not\in F_t$}\\
        \infty & \text{if }u\in F_t
    \end{cases} \label{eqn:cuts} \\
    \rcut^{t}(u) &=
    \begin{cases}
        w^{s}_u + \sum_{v\in D(u)} \min \{c_{uv}+\rcut^{s}(v),\rcut^{t}(v)\} &\text{if $u\not\in F_s$} \\
        \infty & \text{if $u\in F_s$} \label{eqn:cutt}
    \end{cases}
\end{align}
where again $D(u)$ is the set of children of $u\in T$.

\begin{lemma}
    Let $T$ be a tree on $n$ nodes rooted at $r$ with depth $O(\log n)$, and let $\rcut^{s}$ and $\rcut^{t}$ be defined as in \eqref{eqn:cuts} and \eqref{eqn:cutt}.
    Then, given disjoint subsets $F_s,F_t$ of vertices of $T$, $\min \{\rcut^{s}(r),\rcut^{t}(r)\}$ is the capacity of an $(F_s,F_t)$-Restricted $s$-$t$ Min-Cut on $T\cup \{s,t\}$.
    Moreover, this value can be computed using an $O(\log n)$ depth and $O(n)$ total work PRAM algorithm.
\end{lemma}
\begin{proof}
    By straightforward dynamic programming, since $T$ has $O(\log n)$ depth, $\rcut^{s}(r)$ and $\rcut^{t}(r)$ can be computed in $O(\log n)$ depth and $O(n)$ work.
    For correctness, first note that the base cases are correct: the cost of any solution where $x\in F_s$ is placed in $\bar{S}$ is infinite (and analogously for $F_{t}$ and $S$) and, by construction, $w_u^{s}$ is the capacity of the edge between $u$ and $s$, if it exists (and similarly for $w_u^{t}$).
    The correctness then follows by induction.
\end{proof}

\subsubsection*{Extending to arbitrary depth trees.}
When $T$ has super-logarithmic depth, we modify the DP and divide into subproblems based on tree separator nodes (see Definition \ref{def:tree-sep}) rather than children.
For a DP subproblem to find the $s$-$t$ min-cut on a tree $T'$, we compute a tree separator node $q$ and recurse on each tree of $T' \setminus \{q\}$.
From the definition of a tree separator node, this results in subproblems on trees which are at most half the size of $T'$.

When combining subproblems, we use the recursive calls to determine the min-cut capacity when $q\in S$ and the min-cut capacity with $q\in \bar{S}$, and return the lower value.
To do this, we must also determine for each $u\in \delta(q)$ (i.e.~each neighbor of $q$ in $T'$) whether $u\in S$ or $u\in \bar{S}$, in order to determine whether to add the capacity of the $(u,q)$ edge to the cut.
As such, for every neighbor $u$ of $q$, we compute 2 subproblems on the connected component of $T' \setminus \{q\}$ containing $u$: one where we constrain $u\in S$, and one where we constrain $u\in \bar{S}$.
Due to recursion, in any given subproblem there might be multiple vertices which are constrained to be in $S$ or $\bar{S}$; we call these the \textit{tracked} vertices.
Each subproblem then takes as input a tree $T'$, a set of tracked vertices $\Acal$, and subset $\Scal \subseteq \Acal$, where the goal of the subproblem is to compute a $s$-$t$ min-cut $(S,\bar{S})$ on $T'$ under the restriction that $\Scal \subseteq S$ and $(\Acal \setminus \Scal) \subseteq \bar{S}$.

However, there are $2^{|\Acal|}$ possible subsets $\Scal \subseteq \Acal$, and so the number of possible subproblems (and thus total work) is exponential in the number of tracked vertices.
As such, we must bound the number of tracked vertices in any subproblem.
To do this, if we ever have a subproblem with 3 tracked vertices, rather than recursing on the components formed by removing a tree separator node, we recurse on the trees formed by removing the LCA of 2 tracked vertices.
By rooting every subtree at a tracked vertex, this results in new subproblems with at most 2 tracked vertices, and so all subproblems have at most 3 tracked vertices.
Importantly, these additional steps to reduce the number of tracked vertices at most double the depth of the recursion, leading to $O(\log n)$ levels of recursion.
Since each level of recursion can be implemented in $O(\log n)$ depth and $O(n)$ work via dynamic programming, we obtain a $O(\log^2 n)$ depth, $\ot{n}$ work PRAM algorithm.

Below we present the full algorithm for general trees.
For brevity, we present computing the capacity of the min-cut; the actual cut may be found by storing the argmin for each min taken in the recurrence relation.
We reuse the notation from the algorithm for $O(\log n)$ depth, with $S$ denoting the side of the cut containing $s$ and $w_u^{s},w_u^{t}$ denoting the capacity of the $(u,s)$ or $(u,t)$ edge, respectively, and 0 if no such edge exists.

\begin{tbox}
    \textbf{\underline{Min Cut on Trees} Subroutine}

    \textbf{Inputs:} Tree $T$ with root $r$, super-source $s\not\in T$ and super-sink $t\not\in T$, both connected arbitrarily to $T$.

    \medskip

    \textbf{Goal:}
    For $T'$ a subtree of $T$, $\Acal \subseteq T'$ a set of tracked vertices and $\Scal\subseteq \Acal$, recursively compute $\cut(T',\Acal,\Scal)$, which is the capacity of a $s$-$t$ min-cut $(S,\bar{S})$ on $T' \cup \{s,t\}$ such that $\Scal \cup \{s\} \subseteq S$ and $(\Acal \setminus \Scal) \cup \{t\} \subseteq \bar{S}$.
    \medskip

    \textbf{Output:} $\cut(T,\emptyset,\emptyset)$ as the $s$-$t$ min-cut capacity.

    \medskip

    \textbf{Computing $\cut(T',\Acal,\Scal)$:}

    If the depth of $T'$ is at most $10\log n$, run the algorithm of Lemma \ref{lem:low-depth-cut} on $T'\cup \{s,t\}$, with $F_s = \Scal$ and $F_{t} = \Acal \setminus \Scal$ (and $T'$ rooted arbitrarily).
    \smallskip

    Otherwise, with $T'$ rooted at any tracked vertex when $|\Acal|\neq \emptyset$ and arbitrarily otherwise:
    \begin{enumerate}
        \item Compute a split vertex $q$:
            \label{itm:split-vertex}
            \begin{enumerate}
                \item If $|\Acal|\leq 2$, set $q$ to be a tree separator node of $T'$, using the algorithm of Lemma \ref{lem:tree-sep}.
                \item If $|\Acal|=3$, set $q$ as the LCA of 2 non-root tracked vertices, using the algorithm of Theorem \ref{thm:lca}.
            \end{enumerate}
        \item Compute the connected components of $T' \setminus \{q\}$, and for each $u\in \delta(q)$\footnote{$\delta(q)$ is the set of neighbors of $q$ in $T'$} define $C^{u}$ as the component containing $u$.
        \item For each $u\in \delta(q)$, update tracked vertices $\Acal^{u}=(\Acal \cap C^{u})\cup \{u\}$ and $\Scal^{u}=\Scal \cap \Acal^{u}$.

        \item Compute the min-cut capacity conditioned on $q\in S$:
            \[\cut_{q\in S}=w_q^{t}+\sum_{u\in \delta(q)}\min \{w_u^{t}+\cut(C^{u},\Acal^{u},\Scal^{u}\cup \{u\}),~w_u^{s}+c_{uq}+\cut(C^{u},\Acal^{u},\Scal^{u})\}\]
        \item Similarly, compute the min-cut capacity conditioned on $q\in \bar{S}$:
            \[\cut_{q\in \bar{S}}=w_q^{s}+\sum_{u\in \delta(q)}\min \{w_u^{t}+c_{uq}+\cut(C^{u},\Acal^{u},\Scal^{u}\cup \{u\}),~w_u^{s}+\cut(C^{u},\Acal^{u},\Scal^{u})\}\]
        \item If $q\not\in\Acal$, return
            \[\cut(T',\Acal,\Scal)=\min \{\cut_{q\in S}, \cut_{q\in \bar{S}}\}\]
        \item If $q\in\Acal$ and $q\in \Scal$, return $\cut(T',\Acal,\Scal)=\cut_{q\in S}$.
        \item If $q\in\Acal$ and $q\not\in \Scal$, return $\cut(T',\Acal,\Scal)=\cut_{q\in \bar{S}}$.
    \end{enumerate}

\end{tbox}

\begin{lemma}[Min Cut on Trees]
    \label{lem:min-cut-tree}
    The algorithm \underline{Min Cut on Trees} computes the $s$-$t$ min-cut capacity of a tree $T$ with the addition of a super-source $s$ and super-sink $t$ in $O(\log^2 n)$ depth and $\ot{n}$ work.
\end{lemma}
\begin{proof}
    To bound the depth and work of the algorithm, we first show that the number of tracked vertices in any subproblem is always at most 3.
    The algorithm begins with a call that has no tracked vertices, and each recursive call adds at most one additional tracked vertex to each subproblem.
    So, to bound the number of tracked vertices, it suffices to show that if a subproblem has 3 tracked vertices, the number of tracked vertices in each recursive call is at most 2.
    Consider the call $\cut(T',\Acal,\Scal)$ with $|\Acal|=3$, and let $q$ be the LCA computed in Line \ref{itm:split-vertex}.
    By the definition of an LCA and the fact that we always root $T'$ at a tracked vertex, it follows that the path in $T'$ between any 2 tracked vertices passes through $q$.
    As such, after removing $q$ from $T'$, each element of $\Acal$ is in a separate connected component.
    Thus, the tracked sets used in the recursive calls each have at most 2 elements: at most one element from $\Acal$, and one neighbor of $q$.

    We claim there are at most $O(\log n)$ recursive levels of the algorithm, where recursive level $i$ consists of all subproblems resulting from $i$ consecutive recursive calls to $\cut(T,\emptyset,\emptyset)$.
    Consider a subproblem to compute $\cut(T',\Acal,\Scal)$ with at most 2 tracked vertices (i.e.~$|\Acal|\leq 2$).
    In this case, we recurse on subtrees which are connected components after the removal of a tree separator node.
    By the definition of a tree separator node (Definition \ref{def:tree-sep}), this removal results in subtrees which are at most half the size of $T'$.
    So, there can be at most $O(\log n)$ such steps before the recursion terminates.
    Now, suppose $|\Acal|=3$ and the subproblem $\cut(T',\Acal,\Scal)$ occurs at level $i$.
    The resulting subproblems in level $i+1$ have at most 2 tracked vertices, and so in level $i+2$, the size of the resulting subtrees is at most half the size of $T'$.
    Thus, there are $O(\log n)$ total levels.
    Each recursive level can be processed in parallel, in $O(\log n)$ depth (to compute connected components, subtree sums, and combining recursive calls), making the total depth of the algorithm $O(\log^2 n)$.

    The subtrees (i.e.~all distinct $T'$ from subproblems) at each level of recursion form a partition of the nodes of $T$.
    Moreover, by rooting trees and computing the split vertex $q$ deterministically, every subproblem on $T'$ has the same set of tracked vertices.
    So, since there are at most 3 tracked vertices in any subproblem, there are at most 8 possible sets $\Scal \subseteq \Acal$, and thus at most 8 subproblems using any one subtree.
    The work to compute a $k$ node subtree, outside of recursion, is at most $O(k)$.
    So, the total work at each level is $O(n)$, and as there are $O(\log n)$ levels, the complete work is $\ot{n}$.

    For subtrees with depth $O(\log n)$, the correctness follows from Lemma \ref{lem:low-depth-cut}.
    For trees with larger depth, we must have either $q\in S$ or $q\in\bar{S}$, where $q$ is the split vertex computed in Line \ref{itm:split-vertex}.
    By induction and the setting of the edge weights $w_u^{s},w_u^{t}$, $\cut_{q\in S}$ is the min-cut with the restrictions on $\Acal$ imposed by $\Scal$ when $q\in S$ (and similarly for $\cut_{q\in\bar{S}}$).
    So, as we take the min of $\cut_{q\in S},\cut_{q\in \bar{S}}$ when $q\not\in\Acal$, $\cut(T',\Acal,\Scal)$ is the correct min-cut capacity.
\end{proof}

%% file: poly-weight.tex
\label{sec:poly-weights}
In this section, we give a $O(\log n)$ depth, $\ot{m}$ work \pram~algorithm which, given $s,t$ and $\varepsilon$ converts any capacitated graph $G$ into one with with $\poly(n/\varepsilon)$ aspect ratio that preserves the $s$-$t$ maximum flow up to a $(1-\varepsilon)$ factor.

\begin{algorithm}{$\polyweights(G,s,t,\varepsilon)$}\\
    \label{alg:poly-weights}\nonl\textbf{Input:} Graph $G$ with arbitrary capacities, $s,t\in V(G)$, error parameter $\varepsilon$. \\
    \nonl\textbf{Output:} Graph $G'$ with $\poly(n/\varepsilon)$ aspect ratio.  Moreover, with $f$ the max $s$-$t$ flow value in $G$ and $f'$ the max $s$-$t$ flow value in $G'$, $(1-\varepsilon)f\leq f' \leq f$.\\
    \textbf{Procedure:}
    \begin{algorithmic}[1]
        \State Initialize $G' \gets G$. 
        \State Let $T$ be a maximum spanning tree of $G$, computed using the algorithm of \cite{pram-mst}.
        \State Compute $c'$ as the capacity of the lowest capacity edge on the unique $s$-$t$ path in $T$.
        \State For any edge $e$ in $G'$ with capacity larger than $mc'$, reduce its capacity to $mc'$.
        \State For any edge $e$ in $G'$ with capacity less than $\varepsilon c'/m$, delete $e$.
        \State \Return{$G'$}
    \end{algorithmic}
\end{algorithm}

The algorithm of \cite{pram-mst} has $O(\log n)$ depth and $O(m)$ work and finding all edges on the $s$-$t$ path in $T$ can be done in $O(\log n)$ depth and $\ot{m}$ work using subtree sums, so computing $c'$ and modifying the capacities can be done in $O(\log n)$ depth and $\ot(m)$ work.
Moreover, by construction, the modified graph $G'$ has aspect ratio $m^2/\varepsilon=\poly(n/\varepsilon)$ and no capacities have been increased, so it remains to show that the maximum $s$-$t$ flow does not reduce by more than a $(1-\varepsilon)$ factor.

For this, we first need the following simple lemma about maximum spanning trees.
\begin{lemma}
    \label{lem:mst-capacity}
    Let $G$ be a capacitated graph and let $T$ be a maximum capacity spanning tree of $G$.
    Let $P$ be the unique $s$-$t$ path in $T$ (which is also a path in $G$), and let $P'$ be any other $s$-$t$ path in $G$.
    Then,
    \[\min_{e\in P} c_e \geq \min_{e\in P'} c_e\]
    where $c_e$ is the capacity of edge $e$.
\end{lemma}
\begin{proof}
    Let $c'=\min_{e\in P}c_e$, and suppose for contradiction there exists in $G$ a $s$-$t$ path $P'$ such that for all $e\in P'$, $c_e > c'$.
    Let $e'$ be an edge on $P$ with capacity $c'$.
    Removing $e'$ from $T$ results in exactly two connected components $S$ and $V \setminus S$ (with $s\in S$ and $t\in V \setminus S$).
    Since $P'$ is an $s$-$t$ path, it must contain an edge $h=(u,v)$ such that $u\in S$ and $v\in V \setminus S$, so removing $e'$ from $T$ while adding $h$ results in a spanning tree $T'$.
    But $c_h > c_{e'}$ by assumption, and so the capacity of $T'$ is greater than that of $T$, contradicting that $T$ is a maximum spanning tree of $G$.
\end{proof}

This then allows us to show the procedure decreases the maximum flow by at most a $\varepsilon$ factor.
\begin{lemma}
    Let $G$ be a capacitated graph and let $G'=\polyweights(G,s,t,\varepsilon)$.
    Suppose it possible to route $f$ units of flow from $s$ to $t$ in $G$.
    Then, it is possible to route $(1-\varepsilon)f$ units of flow from $s$ to $t$ in $G'$.
\end{lemma}
\begin{proof}
    Let $T$ be a maximum spanning tree of $G$, and let $c'$ be the capacity of the minimum capacity edge on the unique $s$-$t$ path in $T$.
    By Lemma~\ref{lem:mst-capacity}, all other $s$-$t$ paths in $G$ have minimum capacity at most $c'$; thus, every $s$-$t$ path can support at most $c'$ units of flow.
    There can be at most $m$ disjoint paths from $s$ to $t$, and so it follows that $f\leq c'm$.
    Thus, reducing the capacity of all edges with capacity greater than $mc'$ to $mc'$ does not affect the maximum flow.
    Similarly, clearly $f\geq c'$, as there is a $s$-$t$ path with minimum capacity $c'$.
    Thus, as the sum flow on edges of capacity at most $\varepsilon c'/m$ is at most $\varepsilon c'$, deleting edges of capacity at most $\varepsilon c'/m$ reduces the flow by at most $\varepsilon f$.
    It then follows that the max flow in $G'$ must have value at least $(1-\varepsilon)f$.
\end{proof}